\documentclass[aps,
  prx,           
  reprint,       
  amsmath,       
  amssymb,       
  superscriptaddress, 
  twocolumn,10pt,
  showpacs, 
  nofootinbib    
]{revtex4-2}
\usepackage{epsfig,amssymb,amsmath,psfrag,epstopdf,color}
\pdfoutput=1
\usepackage{graphicx}
\usepackage[export]{adjustbox}

\usepackage{graphicx}          
\usepackage{hyperref}          
\usepackage{xcolor}            
\usepackage{booktabs}          
\usepackage{multirow}          
\usepackage{amsfonts}          
\usepackage{bm}                
\usepackage[separate-uncertainty=true]{siunitx} 
\usepackage{paralist}
\usepackage{hyperref}
\usepackage{placeins}          
\usepackage{orcidlink}

\allowdisplaybreaks

\usepackage[T1]{fontenc}
\usepackage{tikz}
\usetikzlibrary{
  positioning,
  arrows.meta,
  shapes.geometric,
  fit,
  backgrounds,
  calc,
  decorations.pathreplacing,
  decorations.markings
}

\definecolor{cpurplefill}{HTML}{EEEDFE}
\definecolor{cpurplestroke}{HTML}{534AB7}
\definecolor{cpurpletext}{HTML}{3C3489}

\definecolor{cpinkfill}{HTML}{FBEAF0}
\definecolor{cpinkstroke}{HTML}{993556}
\definecolor{cpinktext}{HTML}{72243E}

\definecolor{ctealfill}{HTML}{E1F5EE}
\definecolor{ctealstroke}{HTML}{0F6E56}
\definecolor{ctealtext}{HTML}{085041}

\definecolor{cbluefill}{HTML}{E6F1FB}
\definecolor{cbluestroke}{HTML}{185FA5}
\definecolor{cbluetext}{HTML}{0C447C}

\definecolor{ccoralfill}{HTML}{FAECE7}
\definecolor{ccoralstroke}{HTML}{993C1D}
\definecolor{ccoraltext}{HTML}{712B13}

\definecolor{cgrayfill}{HTML}{F1EFE8}
\definecolor{cgraystroke}{HTML}{5F5E5A}
\definecolor{cgraytext}{HTML}{444441}

\definecolor{camberfill}{HTML}{FAEEDA}
\definecolor{camberstroke}{HTML}{854F0B}
\definecolor{cambertext}{HTML}{633806}

\definecolor{cgreenfill}{HTML}{EAF3DE}
\definecolor{cgreenstroke}{HTML}{3B6D11}
\definecolor{cgreentext}{HTML}{27500A}

\definecolor{orcidlogocol}{HTML}{A6CE39}

\definecolor{lundbg}{HTML}{F8F7F4}
\definecolor{lundborder}{HTML}{D3D1C7}
\definecolor{hotnodeA}{HTML}{E24B4A}
\definecolor{hotnodeB}{HTML}{F09595}
\definecolor{hotnodeC}{HTML}{F7C1C1}
\definecolor{coldnode}{HTML}{85B7EB}
\definecolor{truthnode}{HTML}{97C459}
\definecolor{truthstroke}{HTML}{3B6D11}
\definecolor{treebrown}{HTML}{854F0B}
\definecolor{treeamber}{HTML}{EF9F27}
\definecolor{treelight}{HTML}{FAC775}
\definecolor{treedark}{HTML}{BA7517}
\definecolor{divgray}{HTML}{D3D1C7}
\definecolor{textmuted}{HTML}{5F5E5A}
\definecolor{textdark}{HTML}{2C2C2A}

\definecolor{OIblue}{HTML}{56B4E9}       
\definecolor{OIorange}{HTML}{E69F00}     
\definecolor{OIvermillion}{HTML}{D55E00} 
\definecolor{OIgreen}{HTML}{009E73}      
\definecolor{OIyellow}{HTML}{F0E442}     
\definecolor{OIpink}{HTML}{CC79A7}       
\definecolor{OIblack}{HTML}{000000}      

\definecolor{OIbluefill}{HTML}{DCEEF9}
\definecolor{OIorangefill}{HTML}{FBF0D0}
\definecolor{OIvermillionfill}{HTML}{F6DDD0}
\definecolor{OIgreenfill}{HTML}{D0EDE4}

\definecolor{structfill}{HTML}{F2F2F2}
\definecolor{structstroke}{HTML}{555555}
\definecolor{structtext}{HTML}{333333}
\definecolor{mutedtext}{HTML}{666666}
\definecolor{divline}{HTML}{CCCCCC}
\definecolor{lundbg}{HTML}{FAFAFA}
\definecolor{lundborder}{HTML}{CCCCCC}

\definecolor{nodeimportH}{HTML}{E69F00}  
\definecolor{nodeimportM}{HTML}{F0C860}  
\definecolor{nodeunimport}{HTML}{BBBBBB} 
\definecolor{truthblue}{HTML}{56B4E9}    
\definecolor{truthstroke}{HTML}{0072B2}  

\tikzset{
  basebox/.style={
    rounded corners=4pt, minimum height=12mm, minimum width=36mm,
    align=center, font=\small, inner sep=4pt, line width=0.5pt
  },
  purplebox/.style={basebox, fill=cpurplefill, draw=cpurplestroke,
    text=cpurpletext},
  pinkbox/.style={basebox, fill=cpinkfill, draw=cpinkstroke,
    text=cpinktext},
  tealbox/.style={basebox, fill=ctealfill, draw=ctealstroke,
    text=ctealtext},
  bluebox/.style={basebox, fill=cbluefill, draw=cbluestroke,
    text=cbluetext},
  coralbox/.style={basebox, fill=ccoralfill, draw=ccoralstroke,
    text=ccoraltext},
  graybox/.style={basebox, fill=cgrayfill, draw=cgraystroke,
    text=cgraytext, minimum width=26mm, minimum height=9mm},
  amberbox/.style={basebox, fill=camberfill, draw=camberstroke,
    text=cambertext},
  greenbox/.style={basebox, fill=cgreenfill, draw=cgreenstroke,
    text=cgreentext},
  pinkqbox/.style={basebox, fill=cpinkfill, draw=cpinkstroke,
    text=cpinktext},
  arr/.style={-{Stealth[length=4pt,width=3pt]}, textmuted, line width=0.6pt},
  darr/.style={-{Stealth[length=4pt,width=3pt]}, textmuted, line width=0.8pt},
  dashdiv/.style={divgray, line width=0.5pt, dash pattern=on 3pt off 3pt},
  stagelab/.style={font=\footnotesize\bfseries, text=textdark},
  mutedlab/.style={font=\scriptsize, text=textmuted},
  axislab/.style={font=\scriptsize, text=textmuted},
  lundpanel/.style={rounded corners=2pt, fill=lundbg, draw=lundborder,
    line width=0.5pt},
}

\DeclareRobustCommand{\Sec}[1]{Sec.~\ref{#1}}

\DeclareRobustCommand{\App}[1]{App.~\ref{#1}}
\DeclareRobustCommand{\Tab}[1]{Table~\ref{#1}}

\DeclareRobustCommand{\Fig}[1]{Fig.~\ref{#1}}

\DeclareRobustCommand{\Eq}[1]{Eq.~\ref{#1}}

\DeclareRobustCommand{\pt}{p_{\mathrm{T}}}
\DeclareRobustCommand{\DeltaR}{\Delta{\mathrm{R}}}
\DeclareRobustCommand{\kt}{k_{\mathrm{T}}}
\DeclareRobustCommand{\tauNN}[2]{\tau_{#1#2}}   
\DeclareRobustCommand{\zg}{z_{\mathrm{g}}}
\DeclareRobustCommand{\nsd}{n_{\mathrm{SD}}}

\DeclareRobustCommand{\lundnet}{\text{LundNet}}
\DeclareRobustCommand{\parnet}{\text{ParticleNet}}
\DeclareRobustCommand{\part}{\text{Particle Transformer}}
\DeclareRobustCommand{\gnnexplainer}{\text{GNNExplainer}}
\DeclareRobustCommand{\gnnshap}{\text{GNNShap}}
\DeclareRobustCommand{\gradcam}{\text{GradCAM}}
\DeclareRobustCommand{\G}{\mathcal{G}}
\DeclareRobustCommand{\Gs}{\mathcal{G}_{\mathrm{expl}}}
\DeclareRobustCommand{\fidp}{\mathrm{Fid}^{+}}
\DeclareRobustCommand{\fidm}{\mathrm{Fid}^{-}}

\DeclareRobustCommand{\GeV}{\mathrm{GeV}}

\begin{document}

\title{Explainable AI for Jet Tagging: A Comparative Study of GNNExplainer,
       GNNShap, and GradCAM for Jet Tagging in the Lund Jet Plane}

\author{Pahal D. Patel\orcidlink{0009-0001-9343-9164}}
\email{pahaldp22@iitk.ac.in}
\affiliation{Department of Physics, Indian Institute of Technology, Kanpur\\
Uttar-Pradesh-208016, India}

\author{Sanmay Ganguly\orcidlink{0000-0003-1285-9261}}
\email{sanmay@iitk.ac.in}
\affiliation{Department of Physics, Indian Institute of Technology, Kanpur\\
Uttar-Pradesh-208016, India}


\begin{abstract}
Graph neural networks such as \parnet{} and transformer based networks on point clouds such as \part{} achieve state-of-the-art
performance on jet tagging benchmarks at the Large Hadron Collider, yet the
physical reasoning behind their predictions remains opaque.  We present different 
methods, i.e. perturbation-based (\gnnexplainer{}), Shapley-value-based
(\gnnshap{}), and gradient-based (\gradcam{}); adapted to operate on
\lundnet{}'s Lund-plane graph representation.  Leveraging the fact that each
node in the Lund plane corresponds to a physically meaningful parton
splitting, we construct Monte Carlo truth explanation masks and introduce a
physics-informed evaluation framework that goes beyond standard fidelity
metrics.  We perform the analysis in three transverse-momentum 
bins
($\pt \in [500,700]$, $[800,1000]$, and the inclusive region $[500,1000]$ ${\GeV}$), revealing how
explanation quality and focus shift between non-perturbative and perturbative
regimes.  We further quantify the correlation between explainer-assigned node
importance and classical jet substructure observables---$N$-subjettiness
ratios $\tauNN{2}{1}$ and $\tauNN{3}{2}$ and the energy correlation functions---establishing the degree to which the model has learned known QCD
features.  
We find that overall the weight assigned by explainability methods has a correlation with analytic observables, with expected shift across different phase space regimes, indicating that a trained neural network indeed learns some aspects of jet-substructure moments.
Our open-source implementation enables reproducible explainability studies
for graph-based jet taggers.

\end{abstract}

\maketitle

\section{Introduction}
\label{sec:intro}
At the Large Hadron Collider (LHC), proton--proton collisions routinely produce hard scattered partons (quarks and gluons), along with other heavy particles like top quark, $W/Z$ bosons etc., that fragment and hadronize into collimated sprays of stable particles, such as hadrons, photon and leptons. A subset of such particles, collected through a sequential clustering algorithm,  known as jets~\cite{Salam:2010nqg}. Jets are a relic of quantum properties carried by the partons and thus a major probe of short distance dynamics, taking place inside a hadron--hadron collision processes.
The internal structure of hadronic jets is driven by the fundamental branching probabilities of Quantum Chromodynamics (QCD). The evolution of a parton from a hard scale down to the hadronization scale is governed by a cascade of soft and collinear emissions, a process successfully described by parton shower algorithms and analytical resummation techniques \cite{Larkoski:2017jix}. The Lund Jet Plane (LJP) \cite{Dreyer:2018nbf} has successfully emerged as the definitive phase space description for characterizing this radiation pattern. By projecting jet constituents onto a phase space defined by the natural logarithm of emission transverse momentum $\ln(\kt)$ and $\ln(1/\Delta)$, where  $\Delta$ is the splitting angle. The LJP provides a direct visualization of the factorization properties of QCD, separating non-perturbative effects, wide-angle soft radiation, and hard collinear splittings into distinct kinematic regions.

Identifying whether a
given jet originated from the hadronic decay of a 
boosted heavy particle---a
top quark, a $W$ or $Z$ boson, or a Higgs boson---or
from a generic quantum
chromodynamics (QCD) process, viz. a hard light quark or gluon fragmentation, is one of the central 
experimental challenges
at the LHC~\cite{Larkoski:2017jix, Kogler:2018hem, Marzani:2019hun}.  This
identification, commonly referred to as \emph{jet tagging}, directly
underpins searches for physics beyond the Standard
Model (BSM)~\cite{Kasieczka:2019dbj} as well as precision measurements of different Standard Model (SM)
processes such as
electroweak and Higgs boson couplings~\cite{CMS-PAS-HIG-23-012}.

Over the past decade, jet tagging has been transformed by machine learning (ML).
Early approaches treated jets as images and applied convolutional neural
networks~\cite{deOliveira:2015xxd, Komiske:2016rsd, Kasieczka:2017nvn},
while more recent architectures exploit the point-cloud or graph structure
of jets directly. Graph neural networks (GNN)
have proven particularly well-suited to this
problem because the relational structure of jets---particles 
connected by
angular proximity or clustering history---maps naturally onto 
graph representations~\cite{Shlomi:2020gdn, Thais:2022iok}. ParticleNet~\cite{Qu:2019gqs} introduced dynamic graph
convolutions over particle clouds; JEDI-net~\cite{Moreno:2019neq} modelled
explicit pairwise particle interactions; LorentzNet~\cite{Gong:2022lye}
enforced Lorentz equivariance; and the Particle
Transformer~\cite{Qu:2022mxj} adapted attention mechanisms with learned
pairwise interaction biases to achieve state-of-the-art performance. There are a subset of partons, viz. b-quarks and c-quarks, for which the corresponding jets have a peculiar property of having secondary vertex \cite{Piacquadio:2008zza} inside a jet. Classification of such sub-classes of jet using 
GNNs have turned out to have major 
implications \cite{Shlomi:2020ufi, ATLAS:2025dkv} in terms of 
performance gain over other competing algorithms. 


The rapid adoption of these deep-learning taggers in experimental analyses
by ATLAS~\cite{ATLAS:2025dkv} and CMS~\cite{CMS-PAS-HIG-23-012} raises a
question that goes beyond raw performance: \emph{what has the model learned,
and can it be trusted?}  Unlike classical substructure
observables---$N$-subjettiness ratios
$\tauNN{2}{1}$  \cite{Thaler:2010cxa,Thaler:2011gf}, energy correlation functions 
\cite{Larkoski:2013eya, 
Moult:2016cvt}, or the soft-drop
momentum fraction 
$\zg$ \cite{Larkoski:2014wba}---
which are constructed
from first-principles QCD 
reasoning and whose 
discriminating power can be
calculated analytically, neural network taggers operate as black boxes.
Their internal decision boundaries are distributed across millions of
learned parameters with no transparent connection to the underlying
physics.  This opacity is not merely an aesthetic concern.  If a tagger
learns to exploit features that are artifacts of the Monte Carlo (MC)
simulation---such as details of the hadronization model, the parton
shower cutoff, or the detector response simulation---its performance may
degrade when applied to real collision
data 
\cite{Barnard:2016qma, Dreyer:2020brq}.

A jet can have a graph representation in the two dimensional feature space spanned by 
pseudo-rapidity ( \(\eta\)) 
and azimuthal angle (\(\phi\)). The same jet can have a graph representation in LJP, capturing the 
unique feature that each node carries the information of parton fragmentation and thus has an unique 
representation in a particular region of LJP phase space, 
associated with an unique physics process. The correspondence between a physics process and a sub-region of LJP is shown in
\Fig{fig:ljp_schematic_diagram}.
Details of the graph representation in LJP, 
of a hadronic jet, is described in Section~\ref{sec:ljp}. From the
known methods of explainable artificial intelligence (XAI) on 
graphs, if one can associate an importance score to each of the 
nodes and vertices of a graph, then it's possible to trace down what
kind of physics features
are more emphasized by a trained neural network model. We have designed our analysis strategy, based on this broader principle.

The remainder of this paper is organised as follows.
Section~\ref{sec:ljp} reviews the Lund jet plane, the \lundnet{}, \parnet{} and \part{}
architecture, the three explainability methods, and the jet substructure
observables used in our correlation analysis.
Section~\ref{sec:methods} describes our adaptations of each explainer to
LJP, the construction of physics-informed ground-truth explanation
masks, and the evaluation framework including the substructure correlation
methodology.  
Section~\ref{sec:results} presents our results: the method comparison,
Lund-plane explanation heatmaps, the $\pt$-dependent analysis, and the
substructure correlation studies.  Section~\ref{sec:discussion} discusses
method trade-offs, physics insights, robustness considerations, and
limitations.  We conclude in Section~\ref{sec:outlook}, with an outlook towards application of the method stidies and it's utility in future jet physics program.

\FloatBarrier
\section{Uniqueness of Lund Jet Plane for Explainability}
\label{sec:ljp}

\subsection{The Lund jet plane}
\label{sec:lund_plane}

The Lund jet plane~\cite{Dreyer:2018nbf} provides a two-dimensional
representation of the radiation pattern within a jet that is both
theoretically calculable and experimentally measurable.  Starting from a
jet clustered with the Cambridge/Aachen (C/A)
algorithm \cite{Dokshitzer:1997in, Wobisch:1998wt}, one iteratively
undoes the clustering: at each step the softer of the two branches is
recorded as an emission, characterized by its transverse momentum
fraction $z = \pt^{\text{soft}}/(\pt^{\text{hard}}+\pt^{\text{soft}})$,
its opening angle \( \DeltaR = \sqrt{( \eta_{\text{hard}} - \eta_{\text{soft}})^2 + (\phi_{\text{hard}} - \phi_{\text{soft}})^2} \) relative to the harder branch, and the
transverse momentum scale of the splitting
$\kt = z\,\pt\,\DeltaR$. Here, \(\pt\) denotes the transverse momentum. Each emission is mapped to a point in the
$(\ln \DeltaR,\,\ln\kt)$ plane.  The procedure is applied recursively
along the harder branch to define the \emph{primary} Lund plane, and
optionally along secondary branches to build the full Lund tree.

\begin{figure}
    \includegraphics[width=\columnwidth]{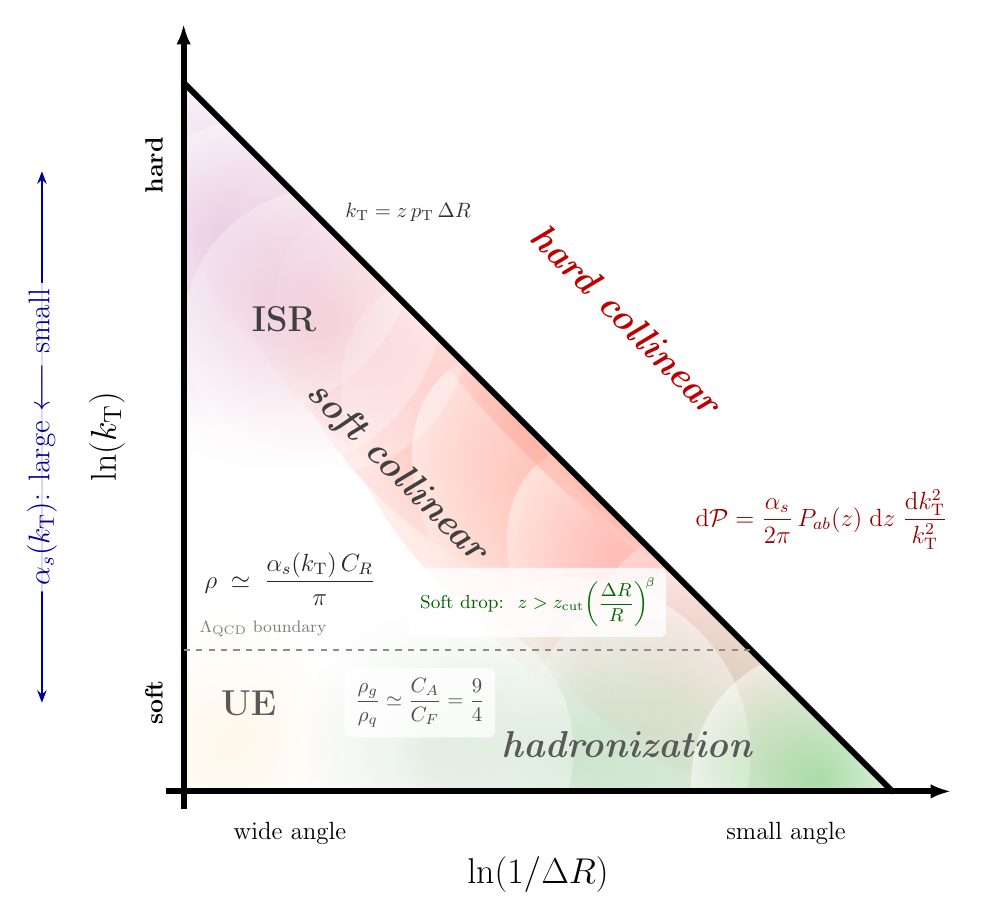}
    \caption{A schematic description of Lund Jet Plane, showing individual quantum effects in different parts of the phase space. The parton splitting functions
    \(P_{ab}(z)\) and the color factors \(C_A\,,C_F\) can be found in any standard QCD literature \cite{Salam:2010zt}.}
    \label{fig:ljp_schematic_diagram}
\end{figure}

In the soft-collinear limit of QCD, the primary Lund plane density at
leading-logarithmic accuracy is approximately
uniform~\cite{Dreyer:2018nbf, Lifson:2020gua}:
\begin{equation}
  \label{eq:lund_density}
  \rho(\kt\,,\DeltaR)
  \;\simeq\;
  \frac{2}{\pi} \alpha_s(k_\mathrm{T})\,C_R\,,
\end{equation}
where $C_R$ is the colour factor of the initiating parton ($C_F = 4/3$
for quarks, $C_A = 3$ for gluons).  The $C_A/C_F$ 
ratio is what makes
quark/gluon discrimination possible at the Lund-plane
level.  Deviations
from uniformity arise from running coupling effects, hard collinear
splittings, and non-perturbative contributions at low $\kt$. These phenomena are shown in the schematic diagram \Fig{fig:ljp_schematic_diagram}.

The ATLAS
experiment and the CMS experiment have measured this density in $\sqrt{s}=\SI{13}{\TeV}$
data \cite{ATLAS:2020bbn, CMS:2023lpp}, validating the perturbative predictions and
constraining the parton shower models. The CMS experiment has also proposed a method to correct multi-prong jet substructure using LJP \cite{CMS:2025eyd}. On the theoretical frontiers, there are proposal of searching for beyond standard model (BSM) physics using LJP \cite{Cohen:2023mya}.

\subsection{Jet tagging in the Lund Jet Plane : Towards a correlation between analytic structure and ML methods}
\label{sec:lund_plane_ml}
Machine Learning based jet tagging and jet mass regression algorithms are now state of the art performer within the mainstream experimental collaborations \cite{CMS-PAS-HIG-23-012,ATLAS:2026vyw}.  The need for interpretable machine learning in the physical sciences has
been broadly articulated \cite{Grojean:2022mef, Wetzel:2025review,
Murdoch:2019pnas, 
BarredoArrieta:2020xaif}.  In high-energy physics
specifically, recent work has begun to address this gap through attention
map analysis for Particle Transformers \cite{Wang:2024rup}, 
and physics-informed architectures such as 
PELICAN \cite{Bogatskiy:2023nnw}
and E-PCN \cite{Islam:2025kjf},
the latter of which used Gradient-weighted
Class Activation Mapping (GradCAM) to attribute classification decisions to
specific kinematic variables.  Shapley value based feature importance studies for quark-vs.-gluon tagging
have been recently carried out in the article \cite{Vent:2025ddm}. Interpretable jet physics using infrared and collinear safe observables has been looked into as well \cite{Konar:2025vts}.
However, these studies have mostly 
focused on
a single explainability method, for a targeted architectures other than
Lund-plane-based GNNs.
In the broader machine learning community, a rich 
ecosystem of explainability methods has been developed specifically for
GNNs \cite{Yuan:2022taxonomy, Li:2022graphxai}.
\gnnexplainer{} \cite{Ying:2019gnnexplainer}, a 
perturbation-based
approach, identifies compact 
subgraph structures and node 
feature subsets
that maximize the mutual 
information with the model's 
prediction.
\gnnshap{} 
\cite{Akkas:2024gnnshap}, 
grounded in cooperative game 
theory,
assigns Shapley values to edges in the computational graph 
through
GPU-parallelized coalition 
sampling, providing 
explanations with axiomatic
fairness guarantees. 
Graph-adapted \gradcam{} \cite{Pope:2019gradcam}
extends gradient-weighted class activation mapping to the graph domain,
producing node-level importance scores via a single backward pass.  These
three methods span the major paradigm families, i.e. perturbation-based,
game-theoretic, and gradient-based and have been systematically compared
on synthetic and molecular benchmarks~\cite{Amara:2022graphframex,
Li:2022graphxai, Agarwal:2023evaluating}.  Yet, in the context of High Energy Physics (HEP),
no study has systematically applied and
compared all three, to a
domain-specific GNN architecture operating on a physically grounded graph
representation where the ground truth for explanations can be derived from
first-principles theory.

GNN's on LJP, such as \lundnet{} \cite{Dreyer:2020brq} provide a uniquely powerful testbed
for filling this gap.  Unlike generic particle-cloud representations,
\lundnet{} builds its input graph from the Lund jet
plane \cite{Dreyer:2018nbf}, a theoretically motivated representation of
the radiation phase space within a jet.  The Lund plane is constructed by
iteratively declustering a jet using the Cambridge/Aachen
algorithm 
\cite{Dokshitzer:1997in, 
Wobisch:1998wt} and recording 
the
transverse momentum $\kt$, 
opening angle $\DeltaR$, and 
momentum fraction
$z$ at each step. Each 
splitting is mapped to a point in the
two-dimensional \( \Big(\ln (1/\DeltaR),\,\ln(\kt) \Big) \) plane, producing a set of nodes
with direct physical meaning: each node corresponds to an individual
parton emission in the jet's showering history.  The primary Lund plane
density is calculable in perturbative QCD \cite{Lifson:2020gua}, and 
the
ATLAS experiment has measured it in
data \cite{ATLAS:2020bbn}. Thus, it is worthy to redeploy \parnet{} and \part{} algorithms on graphs
constructed from LJP and explain their performance.

The Lund-plane representation offers three decisive advantages for
explainability studies.  First, because each node corresponds to a
specific splitting, an explainer that highlights certain nodes is
directly stating which emissions matter for the classification---a
statement that can be checked against QCD expectations (e.g., that top
tagging should rely on the three-prong hard splitting structure).
Second, classical substructure observables such as $\tauNN{2}{1}$,
$\tauNN{3}{2}$, and energy correlation function
ratios \(C_2\,, C_3\) \cite{Larkoski:2013eya}, and the iterated soft-drop multiplicity
$\nsd$ \cite{Larkoski:2014wba, Dreyer:2018nbf} can each be decomposed
into contributions from specific Lund-plane nodes, providing a common
language in which both the explainer output and the physicist's
observables can be expressed and directly compared.  Third, the structure
of the LJP changes systematically with jet transverse momentum
$\pt$: at low boost, the plane is sparsely populated and
non-perturbative effects (hadronization, underlying event) contribute
significantly to soft, wide-angle emissions, while at high boost the
plane is dense and perturbative hard splittings dominate the
substructure \cite{Dreyer:2018nbf, Adams:2015hiv}.  Studying
explainability as a function of $\pt$ therefore probes whether the
model's reasoning shifts between these physically distinct regimes.

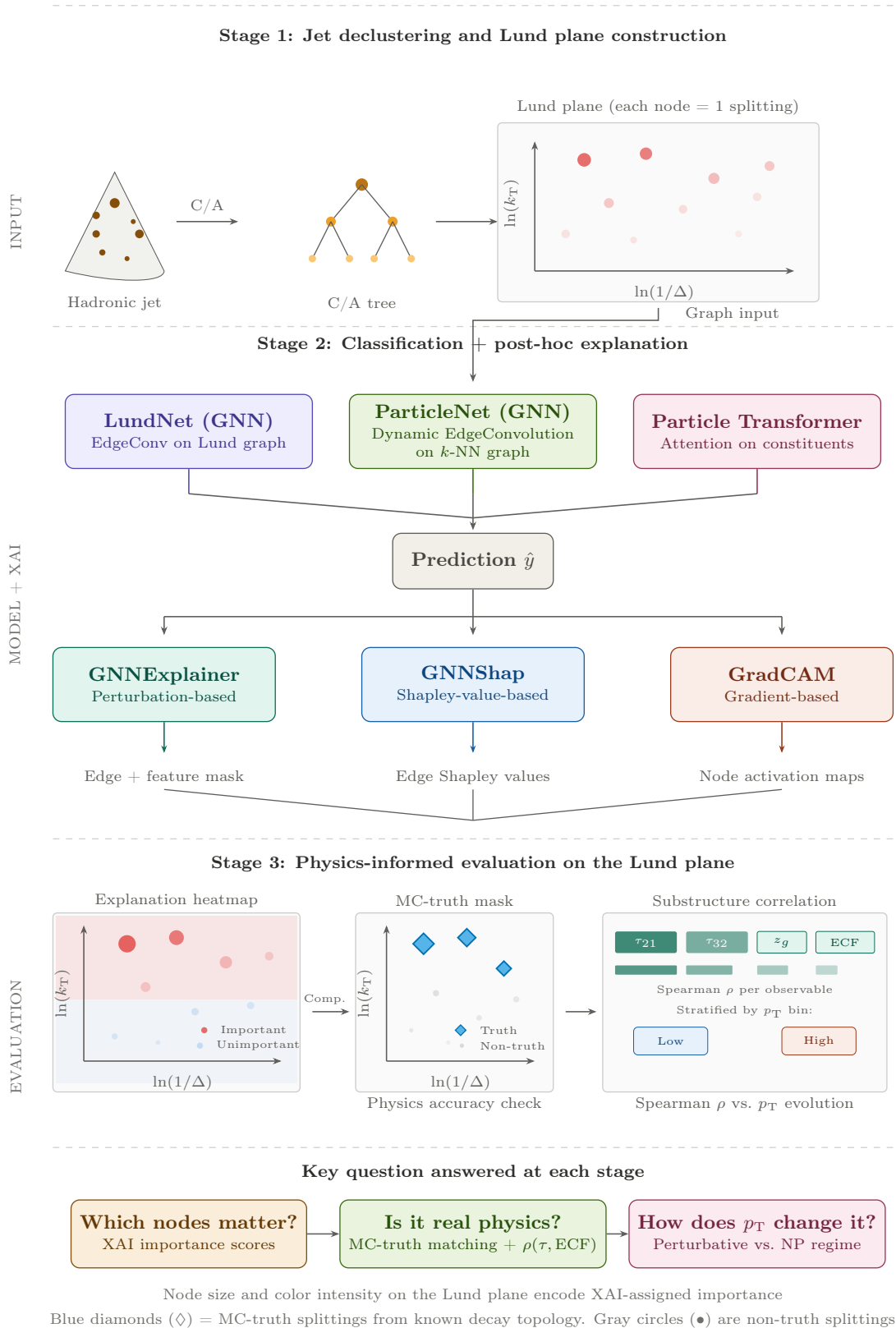
\begin{figure*}[!tp]
\centering

\begin{tikzpicture}[x=1mm, y=1mm]

\draw[dashdiv] (0,7) -- (136,7);
\node[stagelab] at (68, 2) {Stage 1: Jet declustering and Lund plane construction};

\begin{scope}[shift={(0,-18)}]
  \fill[divgray, opacity=0.3]
    (10,-2) -- (2,-18) .. controls (10,-20) and (18,-20) .. (18,-18) -- cycle;
  \draw[textmuted, line width=0.4pt]
    (10,-2) -- (2,-18) .. controls (10,-20) and (18,-20) .. (18,-18) -- cycle;
  \foreach \x/\y/\r in {7/-12/0.6, 10/-7/0.8, 14/-12/0.7,
    8/-15/0.5, 12/-16/0.4, 7/-9/0.6, 13/-10/0.4}{
    \fill[treebrown] (\x,\y) circle (\r);
  }
  \node[mutedlab, below] at (10,-21) {Hadronic jet};
\end{scope}

\draw[arr] (20,-28) -- node[above, mutedlab]{C/A} (30,-28);

\begin{scope}[shift={(40,-18)}]
  \fill[treedark] (10,-4) circle (1);
  \draw[textmuted, line width=0.5pt] (10,-4) -- (5,-10);
  \draw[textmuted, line width=0.5pt] (10,-4) -- (15,-10);
  \fill[treeamber] (5,-10) circle (0.8);
  \fill[treeamber] (15,-10) circle (0.8);
  \draw[textmuted, line width=0.5pt] (5,-10) -- (2,-16);
  \draw[textmuted, line width=0.5pt] (5,-10) -- (8,-16);
  \draw[textmuted, line width=0.5pt] (15,-10) -- (12,-16);
  \draw[textmuted, line width=0.5pt] (15,-10) -- (18,-16);
  \foreach \x in {2,8,12,18} \fill[treelight] (\x,-16) circle (0.6);
  \node[mutedlab, below] at (10,-21) {C/A tree};
\end{scope}

\draw[arr] (62,-28) -- (72,-28);

\begin{scope}[shift={(72,-12)}]
  \draw[lundpanel] (0,0) rectangle (52,-29);
  \draw[arr] (6,-24) -- (48,-24);
  \draw[arr] (6,-24) -- (6,-3);
  \node[axislab] at (27,-27.5) {$\ln(1/\Delta)$};
  \node[axislab, rotate=90] at (2.2,-13) {$\ln(k_\mathrm{T})$};
  \fill[hotnodeA, opacity=0.8]  (14,-6) circle (1.1);
  \fill[hotnodeA, opacity=0.7]  (24,-5) circle (1.0);
  \fill[hotnodeB, opacity=0.6]  (35,-9) circle (0.9);
  \fill[hotnodeB, opacity=0.5]  (18,-13) circle (0.8);
  \fill[hotnodeC, opacity=0.5]  (30,-14) circle (0.7);
  \fill[hotnodeC, opacity=0.4]  (11,-18) circle (0.7);
  \fill[hotnodeC, opacity=0.4]  (22,-19) circle (0.55);
  \fill[hotnodeC, opacity=0.4]  (42,-12) circle (0.7);
  \fill[hotnodeB, opacity=0.5]  (44,-7) circle (0.8);
  \fill[hotnodeC, opacity=0.3]  (39,-18) circle (0.55);
  \node[mutedlab] at (26,2.5) {Lund plane (each node = 1 splitting)};
\end{scope}

\node[mutedlab, rotate=90, opacity=0.9] at (-6,-28) {INPUT};

\draw[dashdiv] (0,-45) -- (136,-45);

\node[stagelab] at (68,-48) {Stage 2: Classification + post-hoc explanation};

\draw[arr] (98,-42) -- (98,-44) -| (68,-54);
\node[mutedlab] at (110,-43) {Graph input};

\node[purplebox, minimum width=40mm] (lundnet) at (22,-62)
  {\textbf{LundNet (GNN)}\\[-1pt]{\scriptsize EdgeConv on Lund graph}};

\node[greenbox, minimum width=40mm] (parnet) at (68,-62)
  {\textbf{ParticleNet (GNN)}\\[-1pt]{\scriptsize Dynamic EdgeConvolution}\\[-2pt]{\scriptsize on $k$-NN graph}};

\node[pinkbox, minimum width=40mm] (part) at (114,-62)
  {\textbf{Particle Transformer}\\[-1pt]{\scriptsize Attention on constituents}};

\draw[cpurplestroke, line width=0.6pt] (lundnet.south) -- ++(0,-4);
\draw[cgreenstroke,  line width=0.6pt] (parnet.south)  -- ++(0,-4);
\draw[cpinkstroke,   line width=0.6pt] (part.south)    -- ++(0,-4);
\draw[textmuted, line width=0.6pt]
  ($(lundnet.south)+(0,-4)$) -- (68,-76);
\draw[textmuted, line width=0.6pt]
  ($(parnet.south)+(0,-4)$)  -- (68,-76);
\draw[textmuted, line width=0.6pt]
  ($(part.south)+(0,-4)$)    -- (68,-76);
\draw[arr] (68,-76) -- (68,-78);

\node[graybox] (pred) at (68,-83)
  {\textbf{Prediction $\hat{y}$}};

\draw[textmuted, line width=0.6pt] (pred.south) -- (68,-92);
\draw[textmuted, line width=0.6pt] (18,-92) -- (118,-92);
\foreach \x in {18,68,118}{
  \draw[arr] (\x,-92) -- (\x,-95);
}

\node[tealbox] (gnnexp) at (18,-103)
  {\textbf{GNNExplainer}\\[-1pt]{\scriptsize Perturbation-based}};

\node[bluebox] (gnnshap) at (68,-103)
  {\textbf{GNNShap}\\[-1pt]{\scriptsize Shapley-value-based}};

\node[coralbox] (gradcam) at (118,-103)
  {\textbf{GradCAM}\\[-1pt]{\scriptsize Gradient-based}};

\draw[ctealstroke, line width=0.6pt, -{Stealth[length=3.5pt]}]
  (gnnexp.south) -- ++(0,-5);
\draw[cbluestroke, line width=0.6pt, -{Stealth[length=3.5pt]}]
  (gnnshap.south) -- ++(0,-5);
\draw[ccoralstroke, line width=0.6pt, -{Stealth[length=3.5pt]}]
  (gradcam.south) -- ++(0,-5);

\node[mutedlab] at (18,-118) {Edge + feature mask};
\node[mutedlab] at (68,-118) {Edge Shapley values};
\node[mutedlab] at (118,-118) {Node activation maps};

\draw[textmuted, line width=0.5pt] (18,-120) -- (68,-125);
\draw[textmuted, line width=0.5pt] (118,-120) -- (68,-125);
\draw[textmuted, line width=0.5pt] (68,-120) -- (68,-125);

\node[mutedlab, rotate=90, opacity=0.9] at (-6,-90) {MODEL + XAI};

\draw[dashdiv] (0,-128) -- (136,-128);

\node[stagelab] at (68,-132) {Stage 3: Physics-informed evaluation on the Lund plane};

\node[mutedlab, rotate=90, opacity=0.9] at (-6,-160) {EVALUATION};

\begin{scope}[shift={(0,-140)}]
  \node[mutedlab] at (20,2) {Explanation heatmap};
  \draw[lundpanel] (0,0) rectangle (40,-29);
  \fill[hotnodeA, opacity=0.12] (0.5,-0.5) rectangle (39.5,-14);
  \fill[coldnode, opacity=0.08]  (0.5,-14) rectangle (39.5,-27.5);
  \draw[arr] (5,-24) -- (37,-24);
  \draw[arr] (5,-24) -- (5,-2);
  \node[axislab] at (21,-27.5) {$\ln(1/\Delta)$};
  \node[axislab, rotate=90] at (1.5,-13) {$\ln(k_\mathrm{T})$};
  \fill[hotnodeA, opacity=0.85] (12,-5) circle (1.4);
  \fill[hotnodeA, opacity=0.70] (20,-4) circle (1.2);
  \fill[hotnodeB, opacity=0.60] (28,-8) circle (1.0);
  \fill[hotnodeB, opacity=0.45] (15,-12) circle (0.8);
  \fill[coldnode, opacity=0.30] (23,-16) circle (0.6);
  \fill[coldnode, opacity=0.25] (10,-20) circle (0.5);
  \fill[coldnode, opacity=0.25] (32,-15) circle (0.6);
  \fill[hotnodeB, opacity=0.40] (35,-7) circle (0.7);
  \fill[coldnode, opacity=0.20] (17,-21) circle (0.4);
  \fill[hotnodeA, opacity=0.8] (24.5,-19) circle (0.5);
  \node[axislab, anchor=west] at (26.0,-19) {\tiny Important};
  \fill[coldnode, opacity=0.4] (23.8,-21.5) circle (0.5);
  \node[axislab, anchor=west] at (25.3,-21.5) {\tiny Unimportant};
\end{scope}

\node[mutedlab] at (44,-154) {\tiny Comp.};
\draw[arr] (42,-156) -- (48,-156);


\begin{scope}[shift={(49,-140)}]
  \node[mutedlab] at (16,2) {MC-truth mask};
  \draw[lundpanel] (0,0) rectangle (32,-29);
  \draw[arr] (5,-24) -- (29,-24);
  \draw[arr] (5,-24) -- (5,-2);
  \node[axislab] at (17,-27.5) {$\ln(1/\Delta)$};
  \node[axislab, rotate=90] at (1.5,-13) {$\ln(k_\mathrm{T})$};
  \node[draw=truthstroke, fill=truthblue, line width=0.7pt,
    diamond, inner sep=1.2pt, minimum size=3.5mm] at (11,-5) {};
  \node[draw=truthstroke, fill=truthblue, line width=0.7pt,
    diamond, inner sep=1.0pt, minimum size=3mm] at (18,-4) {};
  \node[draw=truthstroke, fill=truthblue, line width=0.7pt,
    diamond, inner sep=0.8pt, minimum size=2.5mm] at (24,-9) {};
  \fill[nodeunimport, opacity=0.4] (13,-13) circle (0.5);
  \fill[nodeunimport, opacity=0.3] (20,-17) circle (0.45);
  \fill[nodeunimport, opacity=0.3] (9,-19)  circle (0.35);
  \fill[nodeunimport, opacity=0.3] (26,-14) circle (0.5);
  \fill[nodeunimport, opacity=0.2] (15,-21) circle (0.35);

  \node[draw=truthstroke, fill=truthblue, line width=0.5pt,
    diamond, inner sep=0.5pt, minimum size=1.8mm] at (17,-19) {};
  \node[axislab, anchor=west] at (19,-19) {\tiny Truth };
  \fill[nodeunimport, opacity=0.5] (17.2,-21.5) circle (0.35);
  \node[axislab, anchor=west] at (19,-21.5) {\tiny Non-truth };
  \node[mutedlab] at (16,-31) { Physics accuracy check};
\end{scope}

\draw[arr] (83,-156) -- (88,-156);

\begin{scope}[shift={(89,-140)}]
  \node[mutedlab] at (23,2) {Substructure correlation};
  \draw[lundpanel] (0,0) rectangle (46,-29);
  \fill[ctealstroke, opacity=0.8, rounded corners=1pt] (2,-3) rectangle (12,-6.5);
  \node[font=\tiny, white] at (7,-4.8) {$\tau_{21}$};
  \fill[ctealstroke, opacity=0.55, rounded corners=1pt] (13.5,-3) rectangle (23.5,-6.5);
  \node[font=\tiny, white] at (18.5,-4.8) {$\tau_{32}$};
  \fill[ctealfill, rounded corners=1pt, draw=ctealstroke, line width=0.3pt]
    (25,-3) rectangle (33,-6.5);
  \node[font=\tiny, ctealtext] at (29,-4.8) {$z_g$};
  \fill[ctealfill, rounded corners=1pt, draw=ctealstroke, line width=0.3pt]
    (34.5,-3) rectangle (44,-6.5);
  \node[font=\tiny, ctealtext] at (39.2,-4.8) {ECF};
  \fill[ctealstroke, opacity=0.7, rounded corners=0.5pt] (2,-8.5) rectangle (12,-10);
  \fill[ctealstroke, opacity=0.5, rounded corners=0.5pt] (13.5,-8.5) rectangle (21,-10);
  \fill[ctealstroke, opacity=0.35, rounded corners=0.5pt] (25,-8.5) rectangle (30,-10);
  \fill[ctealstroke, opacity=0.25, rounded corners=0.5pt] (34.5,-8.5) rectangle (38,-10);
  \node[axislab] at (23,-12.5) {\tiny Spearman $\rho$ per observable};
  \node[axislab] at (23,-16) {\tiny Stratified by $p_\mathrm{T}$ bin:};
  \fill[cbluefill, rounded corners=2pt] (5,-18.5) rectangle (17,-23);
  \draw[cbluestroke, line width=0.3pt, rounded corners=2pt]
    (5,-18.5) rectangle (17,-23);
  \node[font=\tiny, cbluetext] at (11,-20.8) {Low};
  \fill[ccoralfill, rounded corners=2pt] (29,-18.5) rectangle (41,-23);
  \draw[ccoralstroke, line width=0.3pt, rounded corners=2pt]
    (29,-18.5) rectangle (41,-23);
  \node[font=\tiny, ccoraltext] at (35,-20.8) {High};
  \node[mutedlab] at (23,-31) {Spearman $\rho$ vs.\ $p_\mathrm{T}$ evolution};
\end{scope}

\draw[dashdiv] (0,-178) -- (136,-178);
\node[stagelab] at (68,-182) {Key question answered at each stage};

\node[amberbox, minimum width=38mm, minimum height=11mm] (q1) at (22,-192)
  {\textbf{Which nodes matter?}\\[-1pt]{\scriptsize XAI importance scores}};

\node[greenbox, minimum width=38mm, minimum height=11mm] (q2) at (68,-192)
  {\textbf{Is it real physics?}\\[-1pt]{\scriptsize MC-truth matching + $\rho(\tau,\mathrm{ECF})$}};

\node[pinkqbox, minimum width=38mm, minimum height=11mm] (q3) at (114,-192)
  {\textbf{How does $p_\mathrm{T}$ change it?}\\[-1pt]{\scriptsize Perturbative vs.\ NP regime}};

\draw[arr] (q1.east) -- (q2.west);
\draw[arr] (q2.east) -- (q3.west);

\node[mutedlab, opacity=0.9] at (68,-202)
  {Node size and color intensity on the Lund plane encode XAI-assigned importance};
\node[mutedlab, opacity=0.9] at (68,-206)
  {Blue diamonds ($\Diamond$) = MC-truth splittings from known decay topology. Gray circles ($\bullet$) are non-truth splittings};

\end{tikzpicture}
\caption{The figure summarizes the analysis strategy of explainability of a ML based jet tagger in the Lund Jet Plane. We study the combinations of jet taggers viz. \lundnet{}, \parnet{} and \part{} along with three explainability methods on GNN's viz. \gnnexplainer{}, \gnnshap{} and \gradcam{}. The importance weight assigned by each of these nine combinations in the LJP phase space is cross verified with first principle analytical computations -- the MC-truth. A comprehensive correlation study across $\pt$ bins are performed to understand which part of the XAI predictions are in agreement with theoretical observables like jet substructure moments. A cross correlation between the explainability scores are also studied in order to understand the degree of mutual agreement among the different XAI methods deployed. }
\label{fig:analysis_strategy}
\end{figure*}

In this paper, we present a systematic, multi-method
explainability study of a Lund-plane-based GNN for jet tagging.  Our
contributions are threefold:

\begin{enumerate}
  \item \textbf{Multi-method comparison on a physics-grounded GNN.}  We
        adapt \gnnexplainer{}, \gnnshap{}, and \gradcam{} to  \lundnet{}, \parnet{} and \part{}  to conduct
        a rigorous comparison across three tagging benchmarks (top, $H$,
        , and quark/gluon), using not only standard fidelity and
        sparsity metrics~\cite{Amara:2022graphframex} but also a new
        physics accuracy metric based on the intersection-over-union
        with Monte Carlo truth explanation masks.

  \item \textbf{$\bm{\pt}$-dependent explainability analysis.}  We
        perform the full analysis in two transverse-momentum bins
        ($\pt \in [500,700]$ {GeV} and $[800,1000]$ {GeV} ), 
        revealing how explanation quality and
        focus evolve from the lower-energetic to the deeply
        perturbative regime.  This exposes whether the model exploits
        infrared-and-collinear (IRC) unsafe features at low $\pt$---a
        direct diagnostic for simulation dependence---and how the
        computational cost of each explainer scales with the denser
        Lund-plane graphs at high $\pt$.

  \item \textbf{Quantitative correlation with substructure
        observables.}  We introduce a three-level framework viz. visual
        overlay, Spearman rank correlation, and importance-weighted
        observable reconstruction, to directly compare
        explainer-assigned node importance with per-node contributions
        to $\tauNN{2}{1}$, $\tauNN{3}{2}$, energy correlation
        functions \(C_2\,,C_3\).  This establishes the degree to
        which \lundnet{} has learned the same discriminating features
        that decades of analytical QCD research have
        identified~\cite{Dasgupta:2013ihk,
        Larkoski:2017jix}, and reveals whether the model has also
        discovered novel features beyond classical observables.
\end{enumerate}
The overall strategy towards a systematic construction of Explainable AI (XAI) for jet taggers are shown in \Fig{fig:analysis_strategy}

\FloatBarrier
\section{Explainability Methods For Jet Taggers}
\label{sec:methods}

\subsection{Explainer adaptation to Lund Jet Plane}
\label{sec:adaptation}

Each explainability method requires non-trivial adaptation to \lundnet{}'s
graph structure.  We summarise the key design choices here; implementation
details and hyperparameter sensitivity studies are in \App{app:xai_math}.

\paragraph{Adapting \gnnexplainer{}.}
The edge mask $\mathbf{M}$ is optimised over all EdgeConv layers
simultaneously.  For \lundnet{}5, the $k$NN graph is recomputed at each
layer from learned embeddings, so we freeze the graph topology to that of
the trained model and optimise masks on the resulting static edge set.
The working principle of \gnnexplainer{} are given in \App{app:gnnexplainer}.

\paragraph{Adapting \gnnshap{}.}
We define the player set as the edges in the computational graph of the
target node (for node-level tasks) or the full graph (for graph
classification).  For \lundnet{}, each edge corresponds directly to a
parent--child splitting relation, making the Shapley attribution
physically transparent.  The rest of the working details of \gnnshap{} can be found in \App{app:gnnshap}.

\paragraph{Adapting \gradcam{}.}
We compute activation maps at each EdgeConv layer and aggregate via
gradient weighting as in \Eq{eqn:gradcam_w}.  To obtain
per-feature importance, we additionally compute the gradient of output with respect to the input feature as 
$\partial y^c / \partial x_{i,d}$ for each input dimension $d \in
\{\ln z,\,\ln\Delta,\,\ln\kt,\ldots\}$, producing a feature-resolved
attribution alongside the node-level score. These attributions are then used to compute the interpretability scores. The details of \gradcam{} implementations are discussed in \App{app:gradcam}.

\subsection{Physics-informed ground truth}
\label{sec:ground_truth}

Lund Jet Plane gives us an analytic handle to trace down jets in appropriate regimes in phase space \cite{Lifson:2020gua}.
We construct Monte Carlo (MC) based truth explanation masks by
tracing the
hard-scattered partons through the showering and hadronization stages and
identifying which Lund-plane nodes descend from the primary parton decay chain.

For \emph{top jets} ($\mathrm t \to bW \to bq\bar{q}'$), the ground-truth
graph $\G^{\mathrm{top}}$ consists of nodes on the Lund tree that
are ancestral to the three hardest splittings corresponding to 
the $\mathrm b$,
$ \mathrm q\,, \bar{q}'$ prongs.  
For \emph{Higgs ($H$) tagging} ($\mathrm H \to q\bar{q}$),
$\Gs^H$ is the single primary splitting that carries the $W$ mass, plus
its immediate descendants.  

\begin{figure*}[!t]
\includegraphics[width=0.32\textwidth]{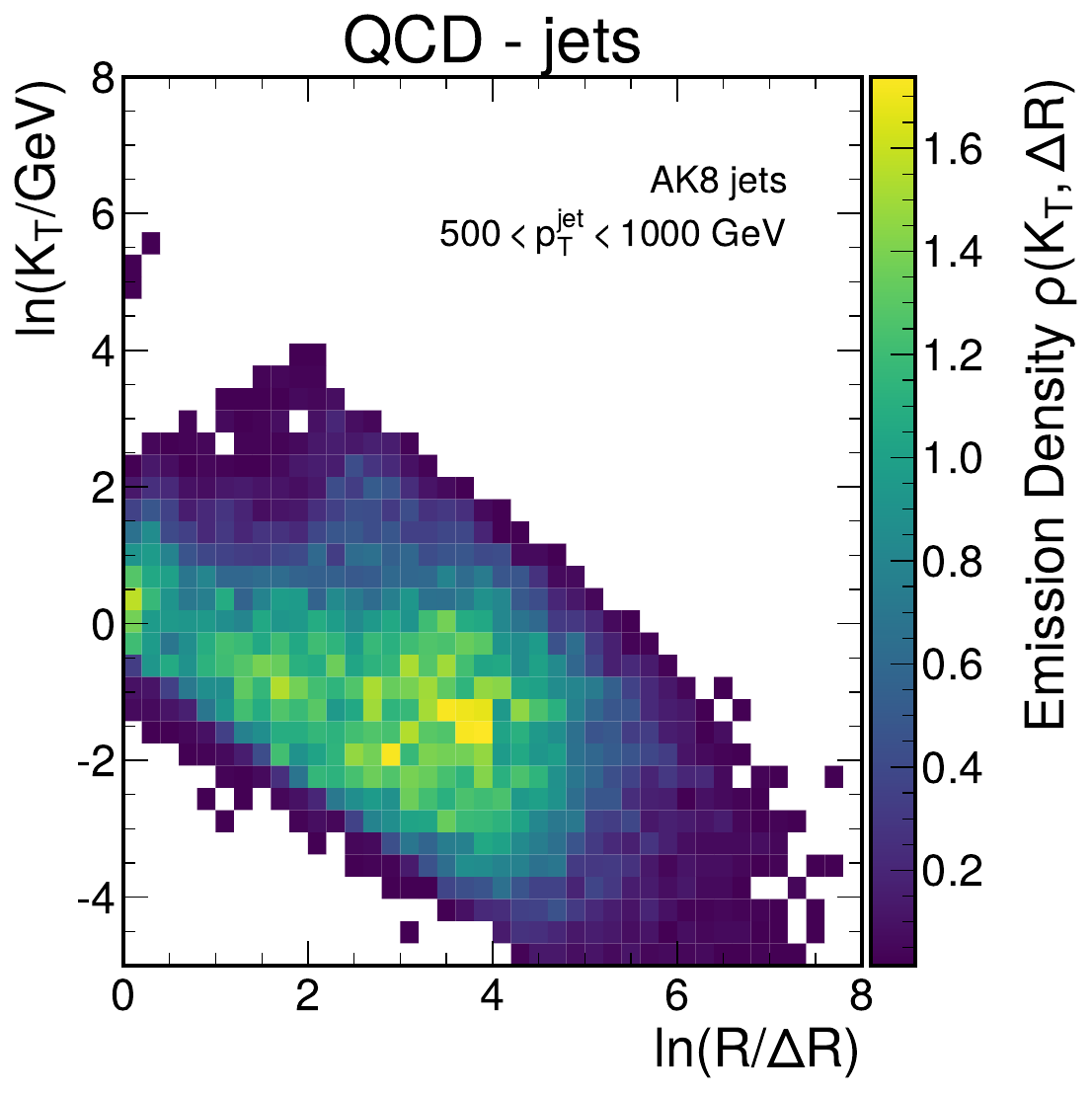}
\includegraphics[width=0.32\textwidth]{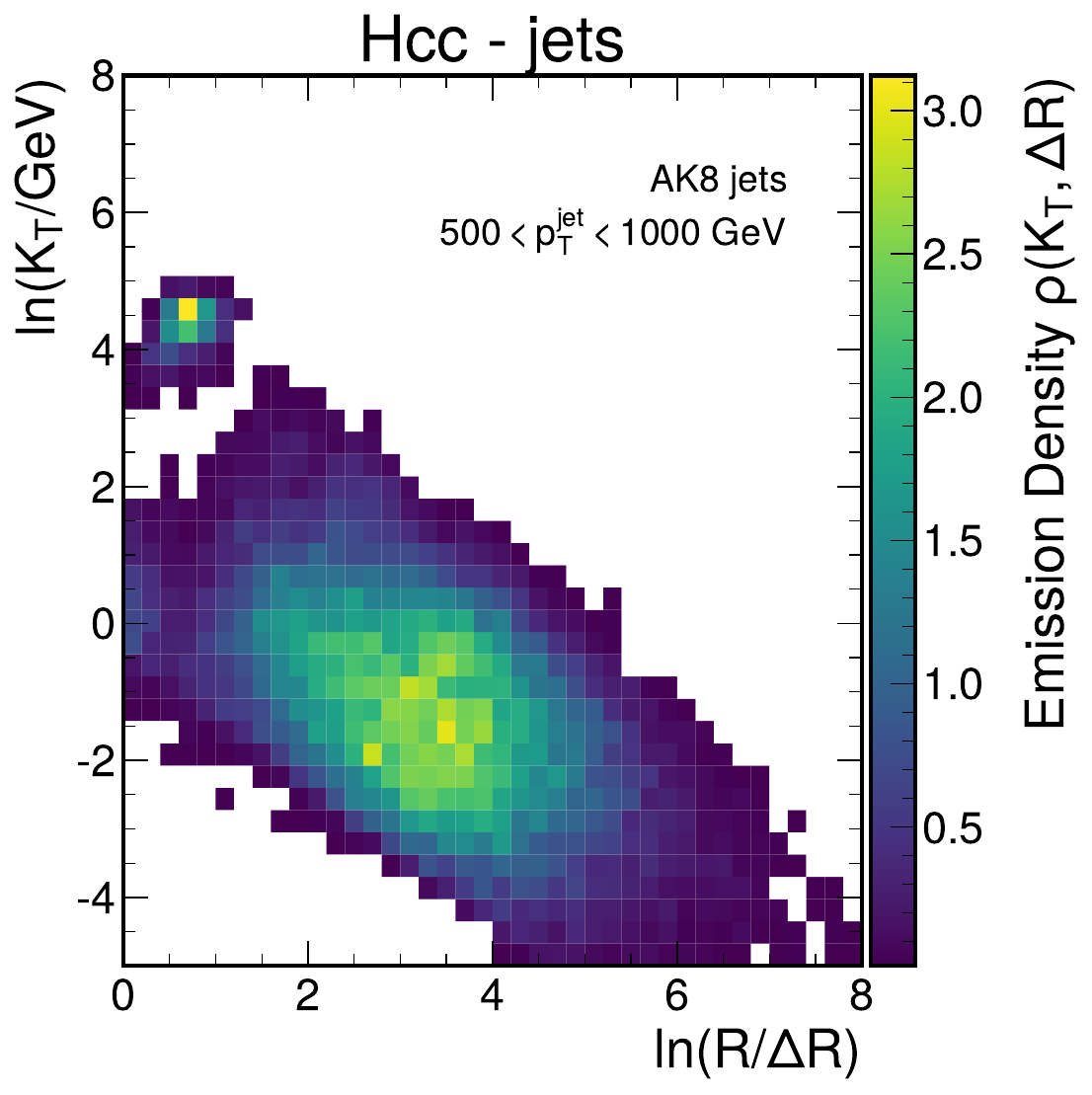}
\includegraphics[width=0.32\textwidth]{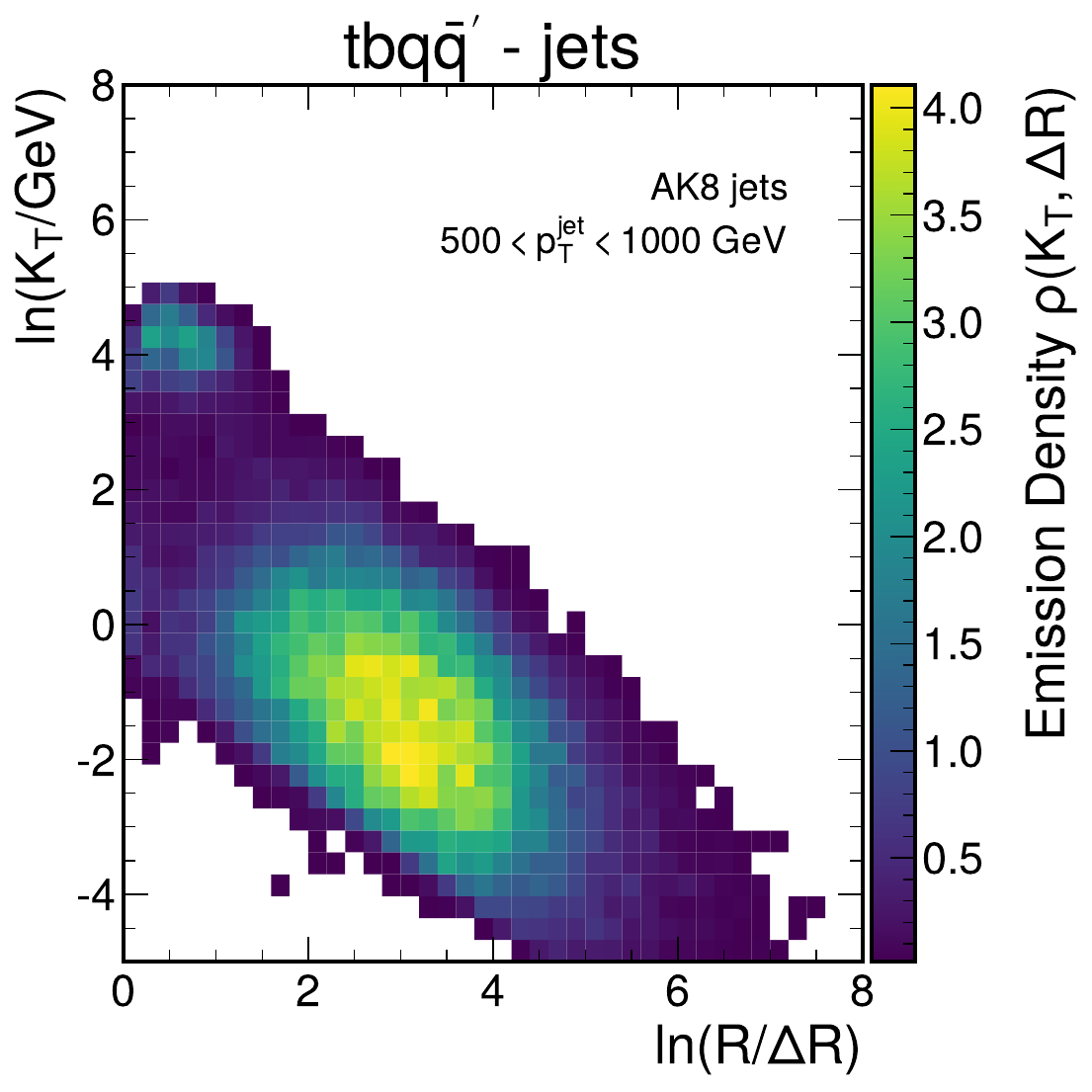}
    \caption{
    Distribution of average LJP emission density 
    \(\rho(k_T\,,\DeltaR)\) for 1-prong, 2-prong and 3-prong 
    jets, reconstructed with anti-$\kt$ algorithm \cite{Cacciari:2008gp} and jet radius parameter \(\mathrm R = 0.8 \). The 1-prong jets are populated by parton splitting of
    light partons only and hence low $\kt$ region 
    ( $ \mathrm ln(\kt) \le 0$) have an over density. 
    It's important to note that for 1-prong jets, the regions
    with lower values of $\mathrm ln(R/\DeltaR)$ has a higher
    density as well, originating from soft wide angled radiation. 
    On the other hand, the middle and right panel shows, jets originating 
    from hadronic decay of $\mathrm H \rightarrow c\bar{c}$ -- 
    the 2-prong 
    jets and the jets originating from hadronic decay of  
    $\mathrm{t \rightarrow b q\bar{q'}}$ -- the 3-prong jets have an observable 
    density overpopulation in high-$\kt$ -- high-$\DeltaR$ regime of phase space. 
    This feature indicates that hadronic showers within these jets are originating 
    from fragmentation of a massive particle.
    }
    \label{fig:raw_ljp_density}
\end{figure*}

We deploy the analytic understanding to retrieve the average LJP density per jet 
\begin{equation}
\label{eqn:ljp_density_form}
\rho(\kt\,,\DeltaR) = \frac{1}{N_{\mathrm jets}}
\frac{d^2 N_{\mathrm emissions}}{d ~ ln(k_T)  ~~d 
~ ln(\frac{R}{\DeltaR})  } \,,
\end{equation}
for 1-prong ($\mathrm{QCD}$), 2-prong ($\mathrm{H \rightarrow c\bar{c}}$) and 3-prong ($\mathrm{t \rightarrow b q\bar{q'}}$) jets. The same quantity, at leading order, is analytically described by \Eq{eq:lund_density}. The density distributions are shown in \Fig{fig:raw_ljp_density} which captures the statistical pattern of parton emission densities of 1-prong, 2-prong and 3-prong jets.
When each emission \(i\) (\(i \in [1\,,2\,, \ldots\,,N_{\mathrm emissions}]\)) in the LJP, for a given jet, has an associated weight \(w_i\), then the weighted LJP density is defined as 

\begin{equation}
\label{eqn:ljp_density_form_wt}
\rho_{W}(\kt\,,\DeltaR) = \frac{1}{N_{\mathrm jets}}
\frac{d^2 \Big( \sum_{i = 1}^{N_{\mathrm emissions} } w_{i} \Big)}{d ~ ln(k_T)  ~~d 
~ ln(\frac{R}{\DeltaR})  } \,.
\end{equation}

A large part of the higher density regions are populated with low energetic parton splitting, as expected from first principle analytic understanding of QCD. The jets originating from light partons, a.k.a QCD jets have a visible density enhancement in the lower $\mathrm ln(R/\DeltaR)$ region, corresponding to soft wide-angled emissions. This pattern is well understood from the Sudakov double logarithm structure of parton branching amplitudes \cite{Larkoski:2024uoc}. Once we train a neural network for downstream tagging task and pass it through an explainable method, we try to see if the explainer associates an importance weight to different regions of LJP and wether that weight factor is matching our analytic understanding of LJP density.

\begin{figure*}[!t]
\includegraphics[width=0.32\textwidth]{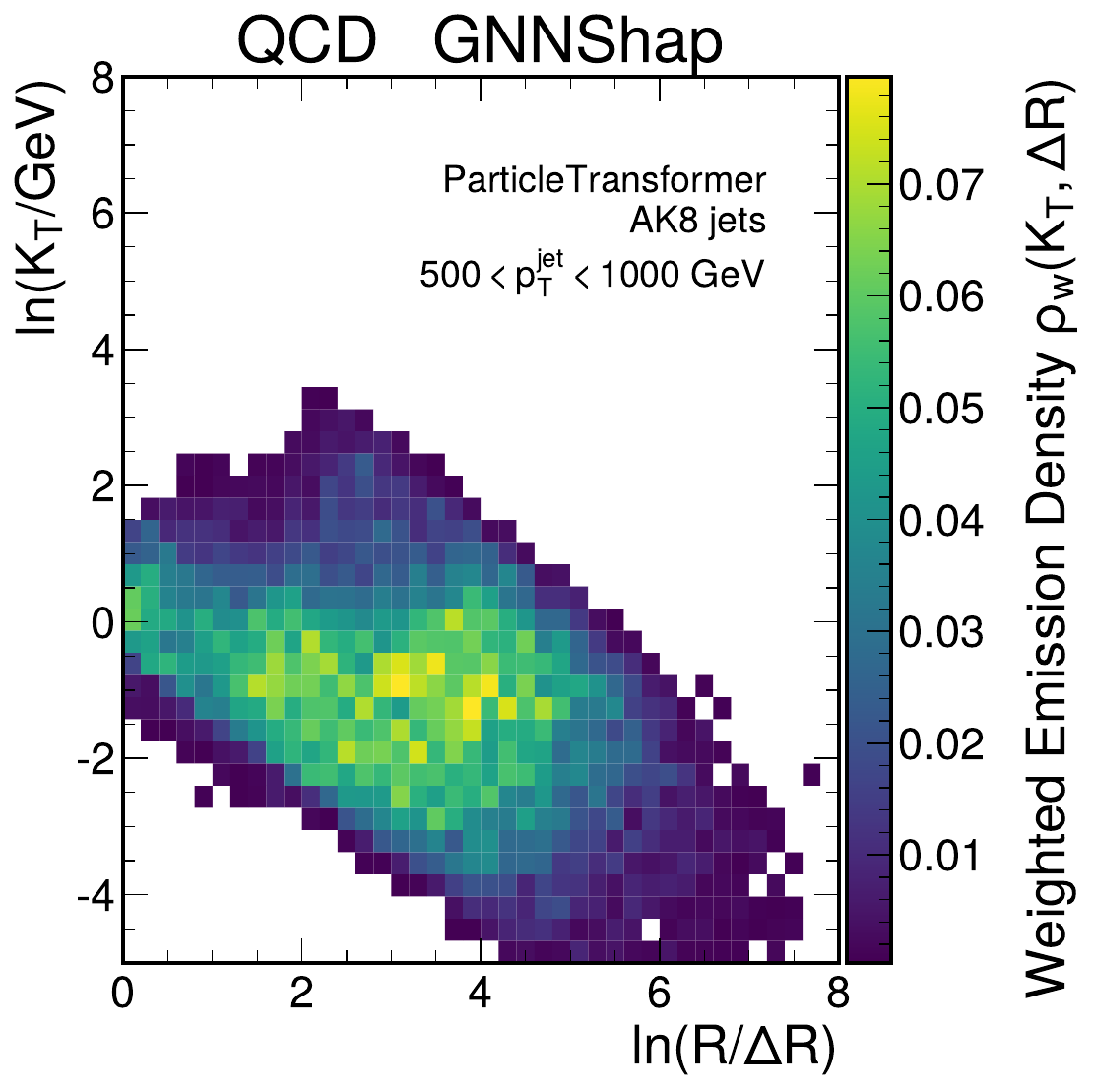}
\includegraphics[width=0.32\textwidth]{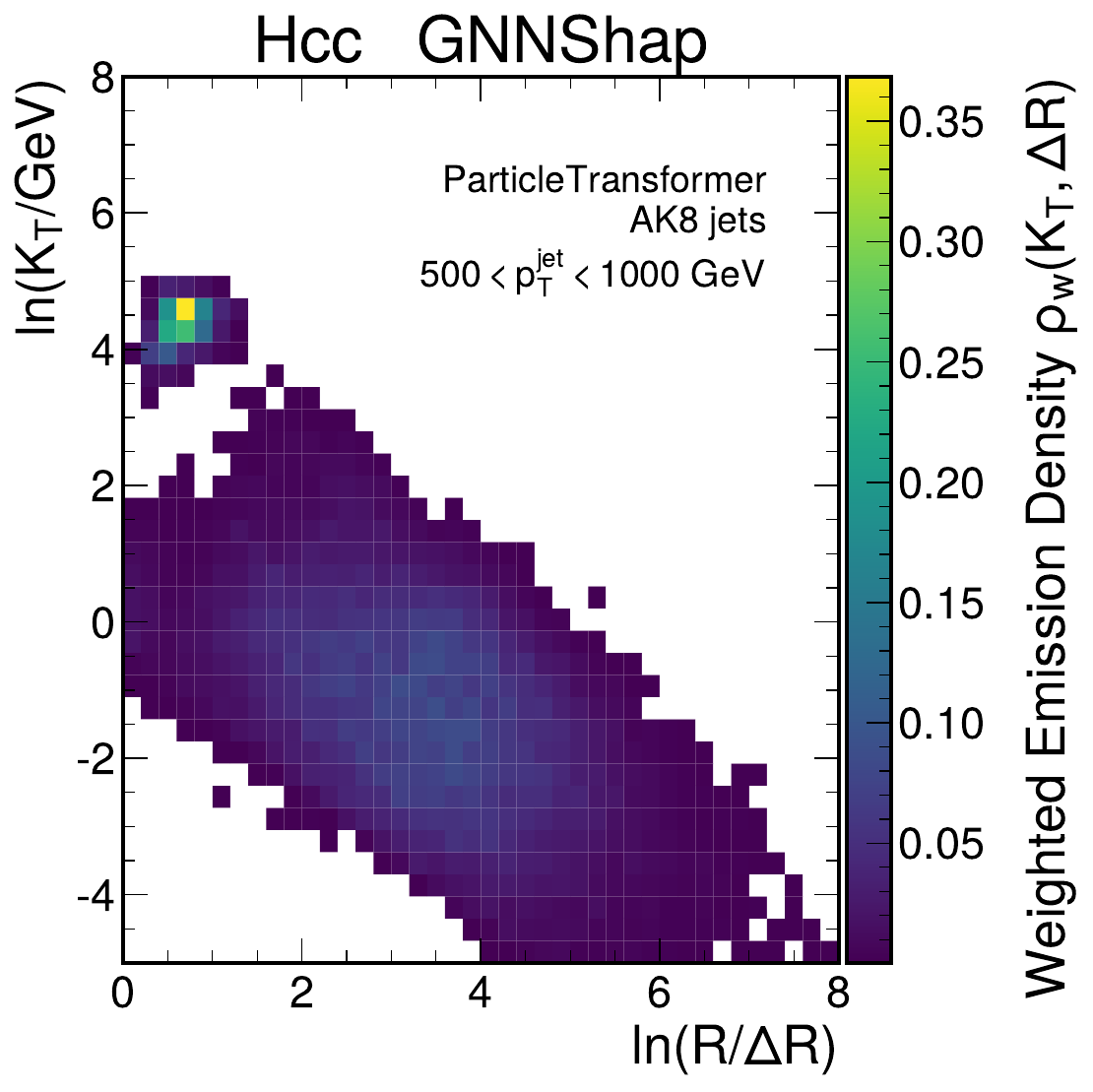}
\includegraphics[width=0.32\textwidth]{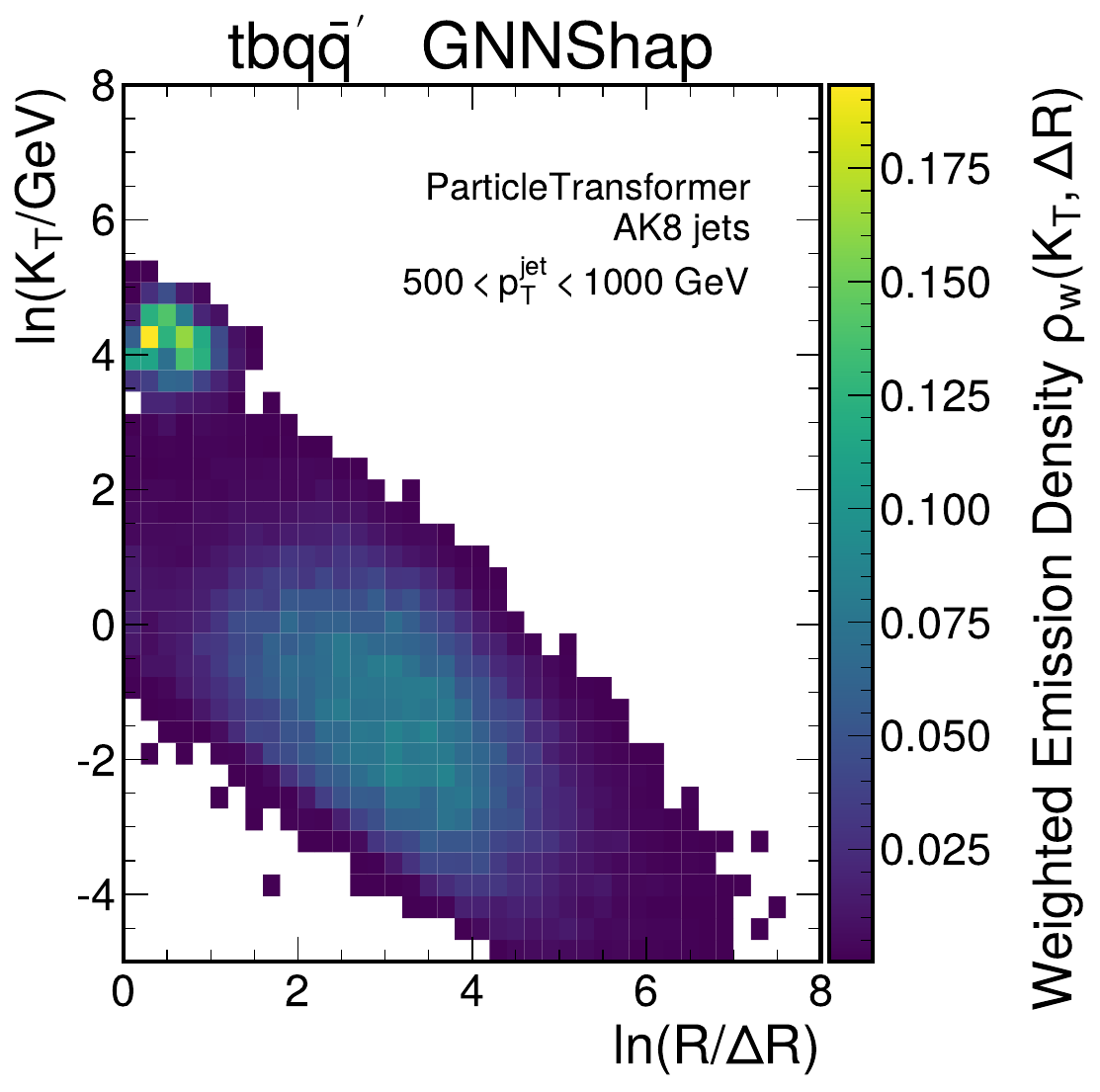}
\includegraphics[width=0.32\textwidth]{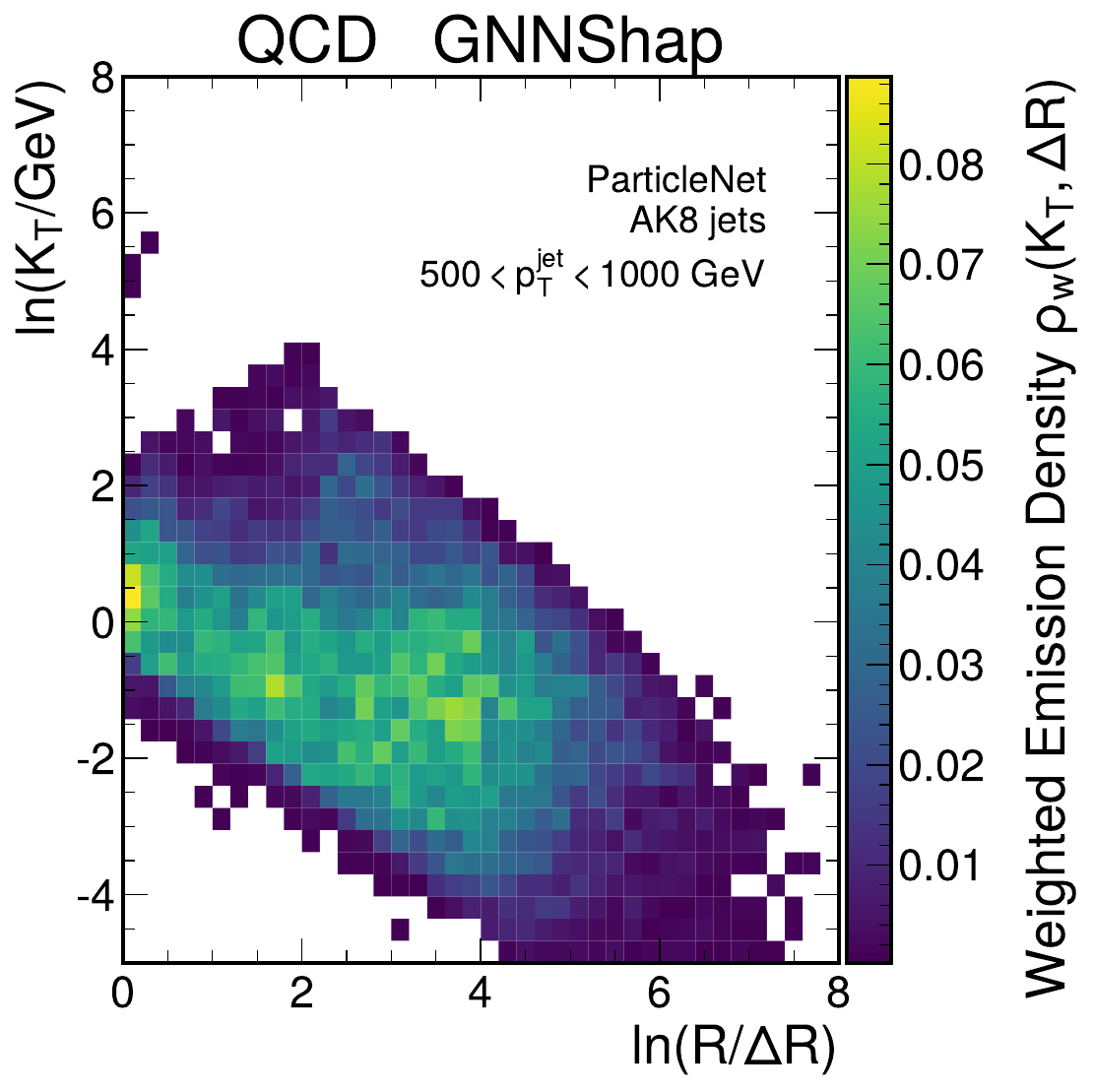}
\includegraphics[width=0.32\textwidth]{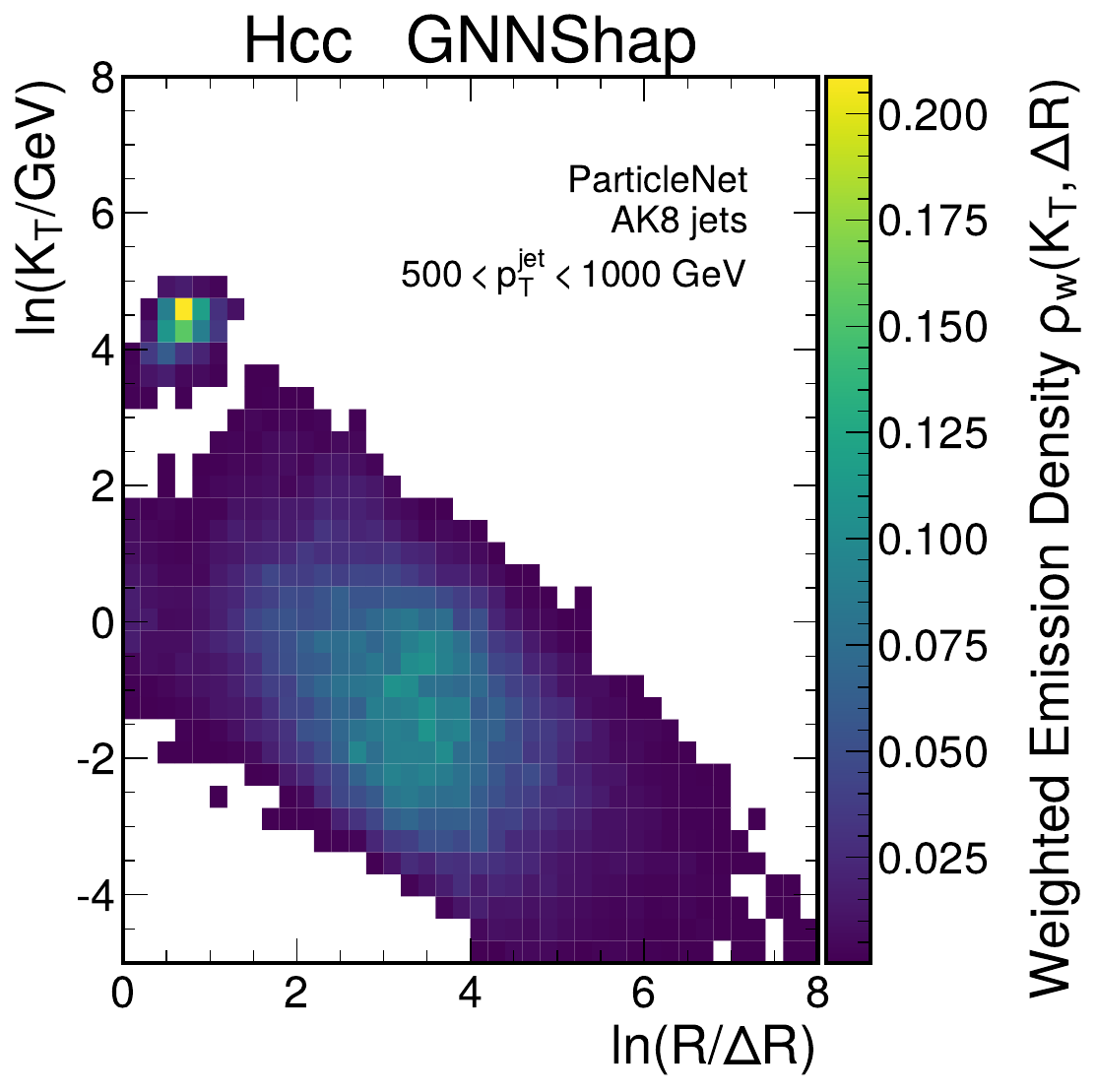}
\includegraphics[width=0.32\textwidth]{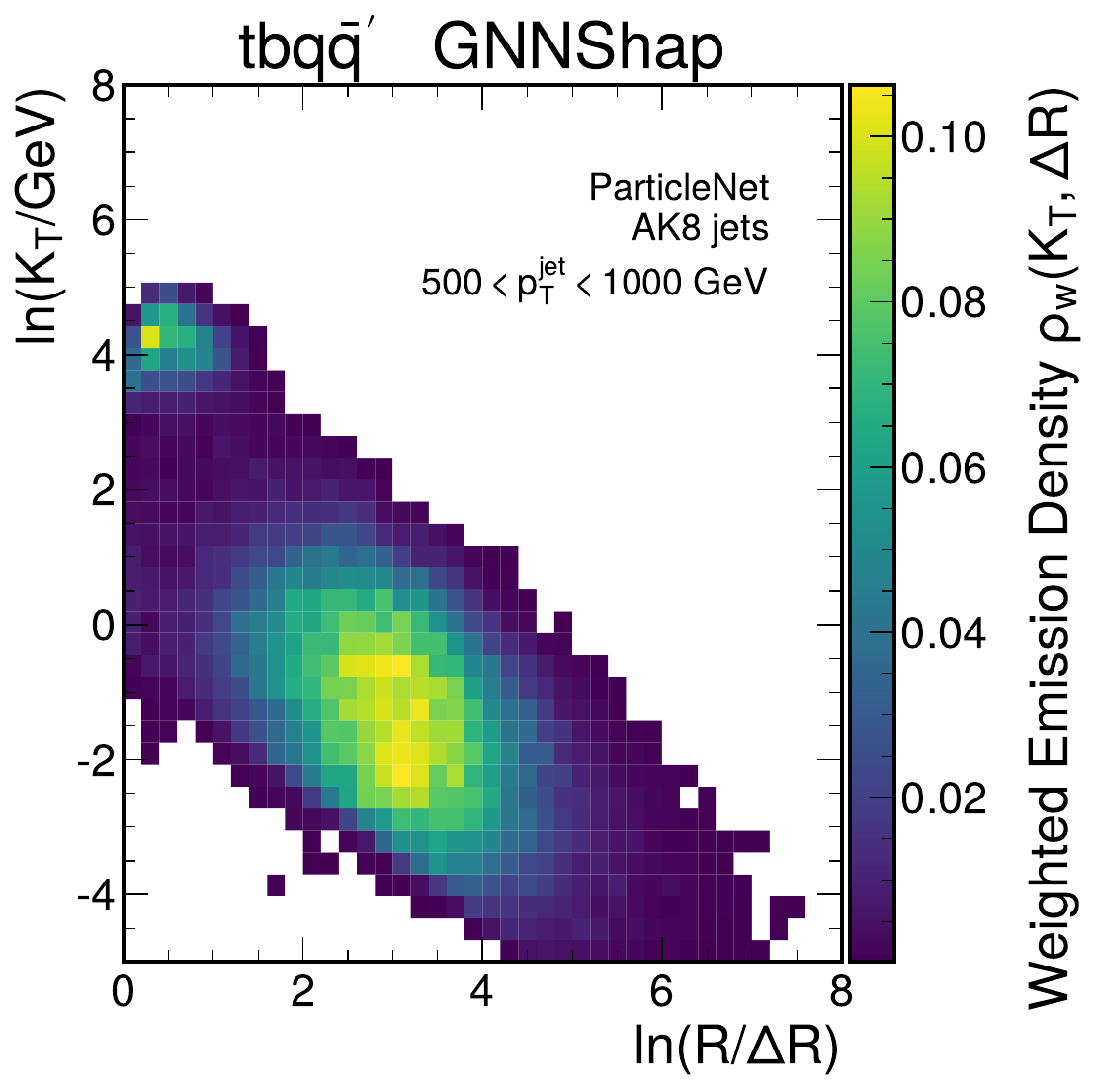}
    \caption{
    Distribution of average weighted LJP emission 
    density 
     \(\rho_{W}(k_T\,,\DeltaR)\) for 1-prong, 2-prong 
     and 3-prong jets, identified as true-positive category by the respective classifiers. The weighted density diagrams 
     here are displayed only for \gnnshap{} explainer 
     for \part{} and \parnet{} architectures only 
     (since both \parnet{} and \lundnet{} uses 
     Edge-Convolution, we have just restricted the 
     comparison between two different type of 
     architectures). When we compare the weighted 
     distributions with the raw LJP density 
     distributions, shown in 
     \Fig{fig:raw_ljp_density}, we see that for 
     1-prong jets (first column of the above figure) 
     both the architectures capture the low $k_{T}$ 
     radiation pattern of the QCD emission over a wide
     angular range. The weighted density distribution for 2-prong \(H \rightarrow c \bar{c}\) decay, we find that both the networks associate larger weights to the hard wide angle splitting, thus indicating it is learning the sub-structure pattern of the jets. The third column on the right, displaying the weighted phase space distribution of 3-prong hadronic top-quark decay, shows that both architectures capture the jet substructure pattern as well as the fragmentation of light partons, populating the low $k_{T}$ regions. In \App{app:full_results}, we further investigate the correlation of learned weights (in terms of weighted LJP features defined in \Eq{eq:feat_avg}) and the jet substructure moments. 
    }
    \label{fig:weighted_ljp_density_gnnshap_benchmark}
\end{figure*}

\begin{figure*}[!t]
\includegraphics[width=0.32\textwidth]{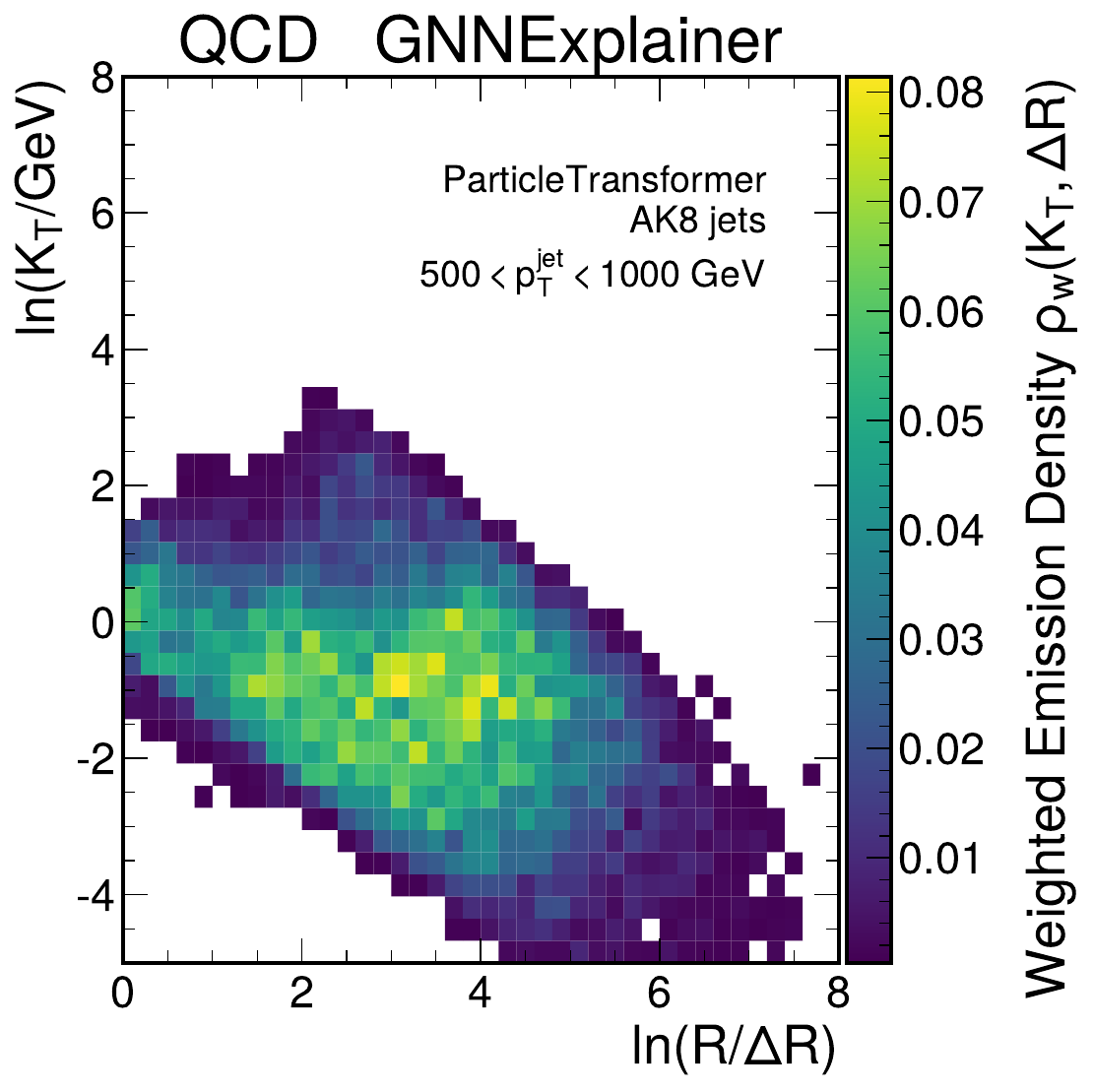}
\includegraphics[width=0.32\textwidth]{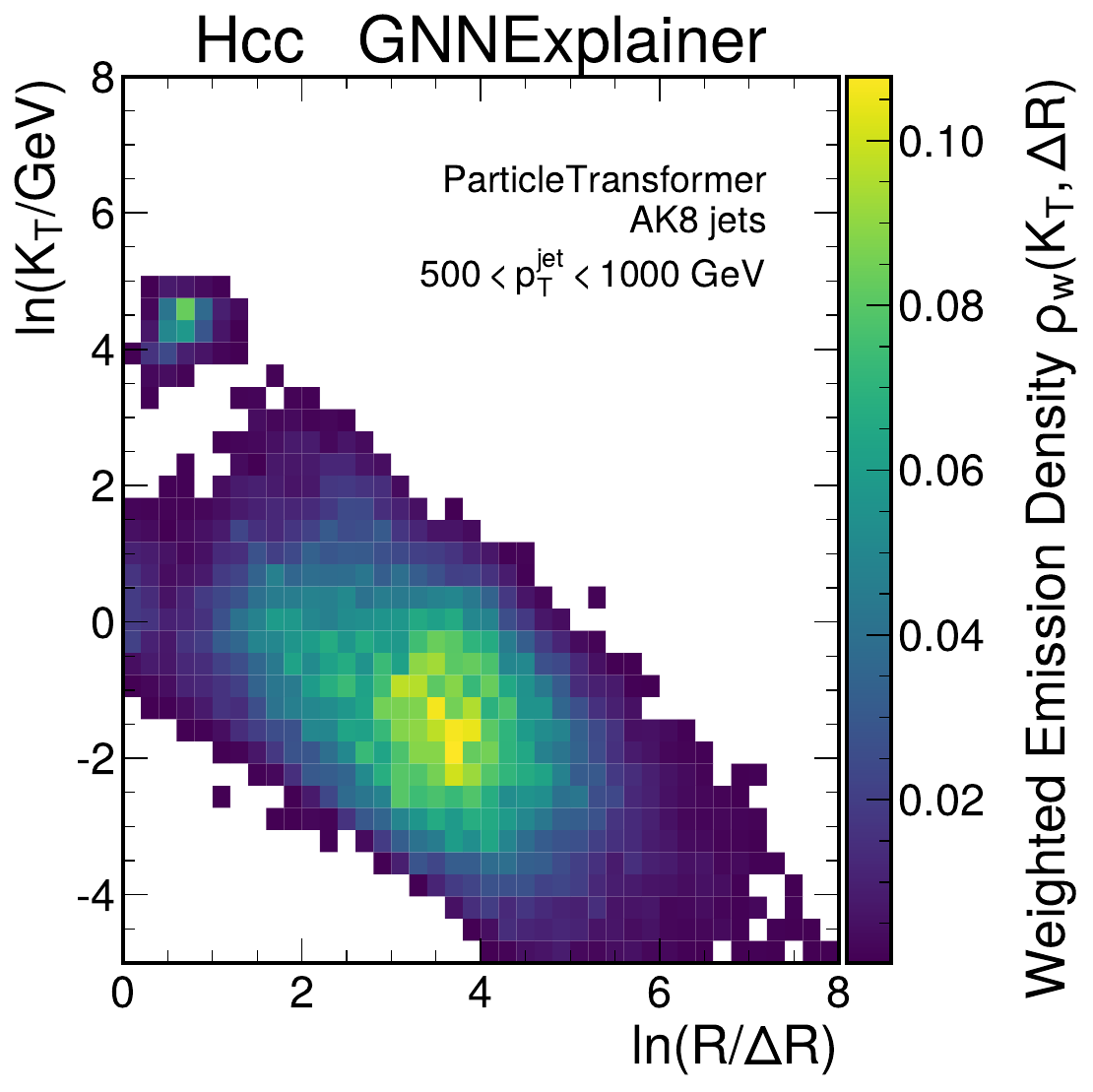}
\includegraphics[width=0.32\textwidth]{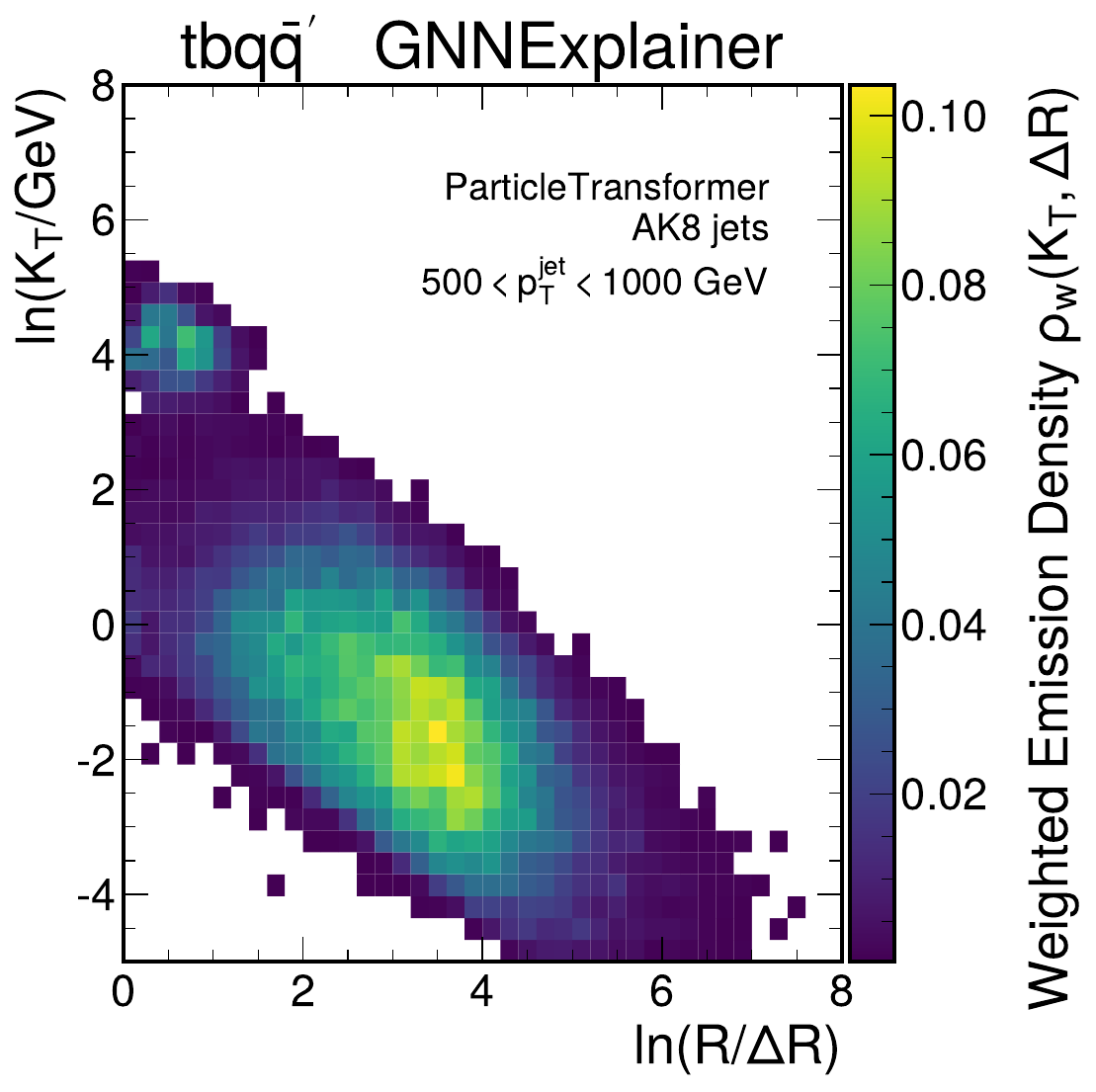}
\includegraphics[width=0.32\textwidth]{FIGURES/particle_transformer/pt_500_1000/lund_plane/gnnshap_ensemble_qcd.pdf}
\includegraphics[width=0.32\textwidth]{FIGURES/particle_transformer/pt_500_1000/lund_plane/gnnshap_ensemble_hcc.pdf}
\includegraphics[width=0.32\textwidth]{FIGURES/particle_transformer/pt_500_1000/lund_plane/gnnshap_ensemble_tbqq.pdf}
\includegraphics[width=0.32\textwidth]{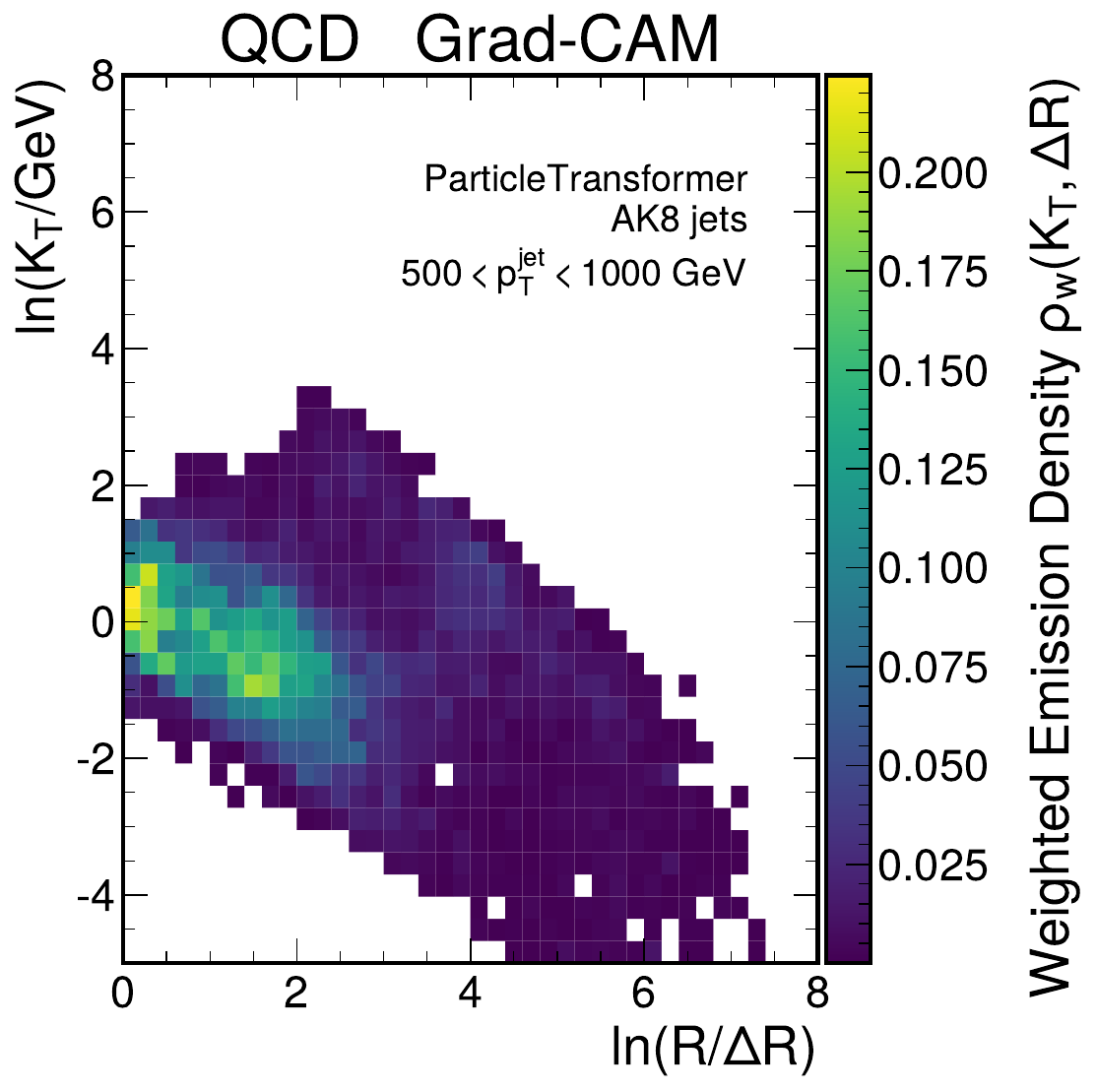}
\includegraphics[width=0.32\textwidth]{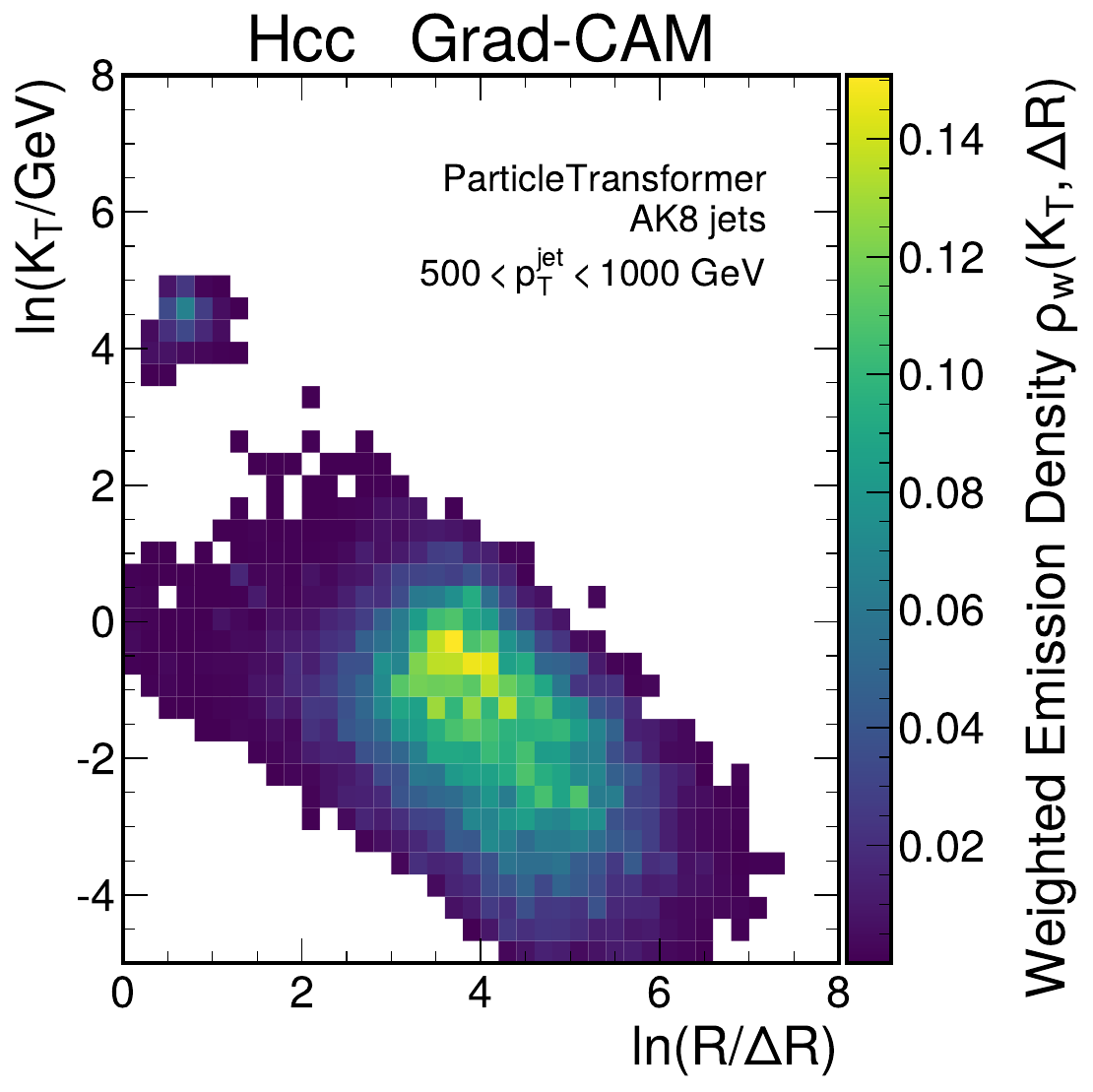}
\includegraphics[width=0.32\textwidth]{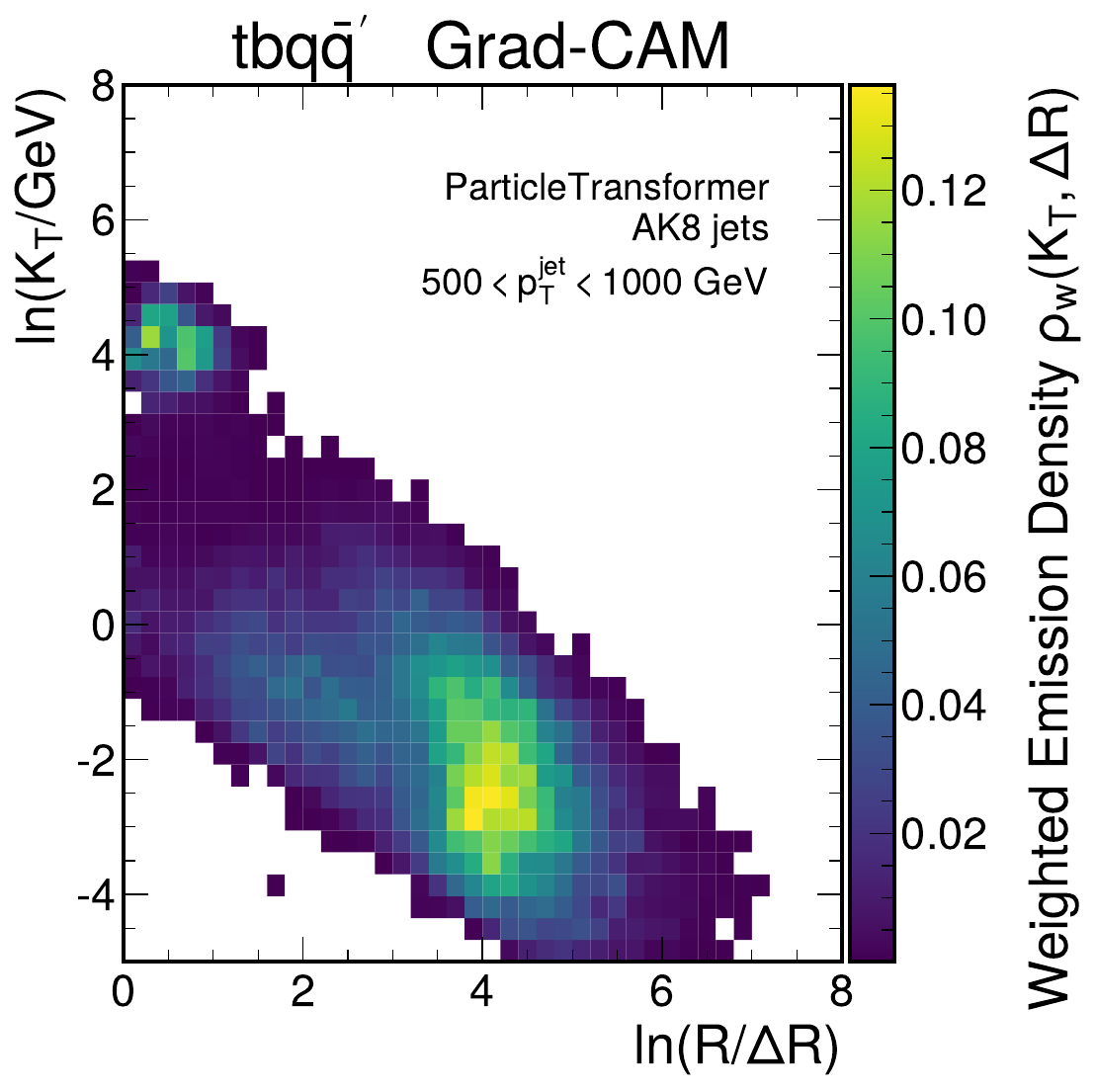}
    \caption{
    Distribution of average weighted LJP emission density 
    \(\rho_{W}(k_T\,,\DeltaR)\) for 1-prong, 2-prong and 3-prong 
    jets, which are classified in the true-positive category by the \part{} algorithm. The top row shows the weighted distribution, when the output of the  trained network is passed through the \gnnexplainer{} algorithm. The middle and bottom row show similar weighted distributions in the LJP, when the explainer weights are computed through \gnnshap{} and \gradcam{} algorithms, respectively. When we compare with a similar \Fig{fig:weighted_ljp_density_gnnshap_benchmark}, we find that for 1-prong case \gradcam{} has certain distinct features compared to the other two explainer and emphasizes on wider angle radiation pattern more. In the middle column we see that while \gnnshap{} exclusively emphasizes on two hard jet substructures, the behavior of the other two explainer are somewhat in contrast. \gradcam{} puts highest weights on soft and small angle radiations. For 3-prong hadronic top quark decay, \gnnshap{} still relies on the hard jet substructure only. However \gnnexplainer{} and \gradcam{}, like the 2-prong case, relies heavily on the low \(\kt\) fragmentation features of the hard sub-partons. These comparative features (shown for the entire $\pt$ range only, while we explicitly study them for low and high $\pt$ bins separately ) tend to demonstrate that \gnnshap{} is least sensitive to soft radiations when combined with \part{} architecture. 
    }
    \label{fig:weighted_ljp_density_part_benchmark}
\end{figure*}

\subsection{Evaluation framework}
\label{sec:evaluation}

To successfully transit neural networks from predictive black boxes
to reliable engines of physical discovery, we must be able to validate 
the interpretable models derived from them. In this section we study three different metrics viz. Fidelity metrics, Weighted Feature Average and Jet substructure correlation, which provide this
crucial validation by statistically quantifying how faithfully an 
interpretable approximation maps the exact decision boundaries and 
latent dynamics of the original network.

\begin{figure*}[!t]
    \includegraphics[width=0.32\textwidth]{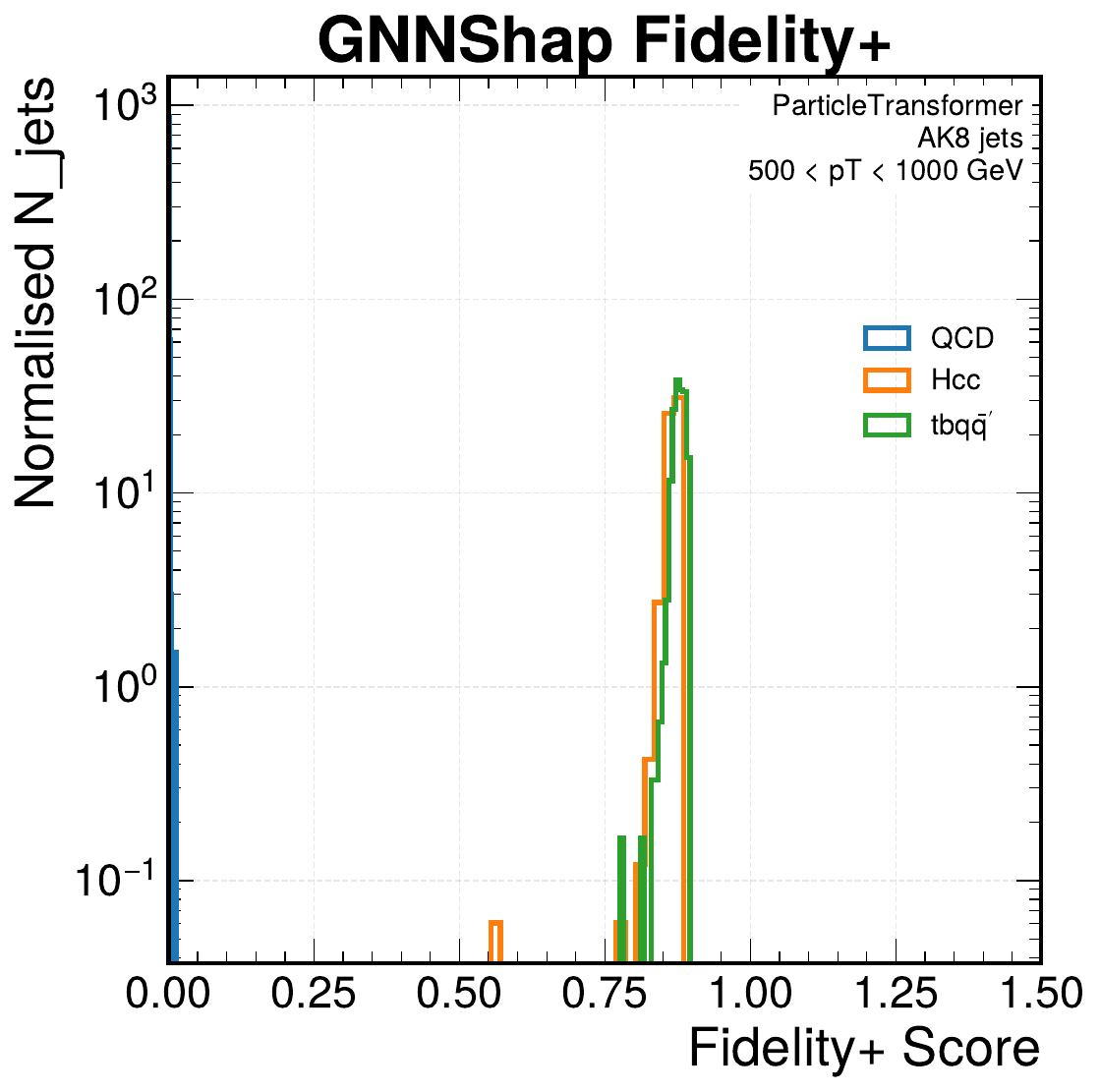}
     \includegraphics[width=0.32\textwidth]{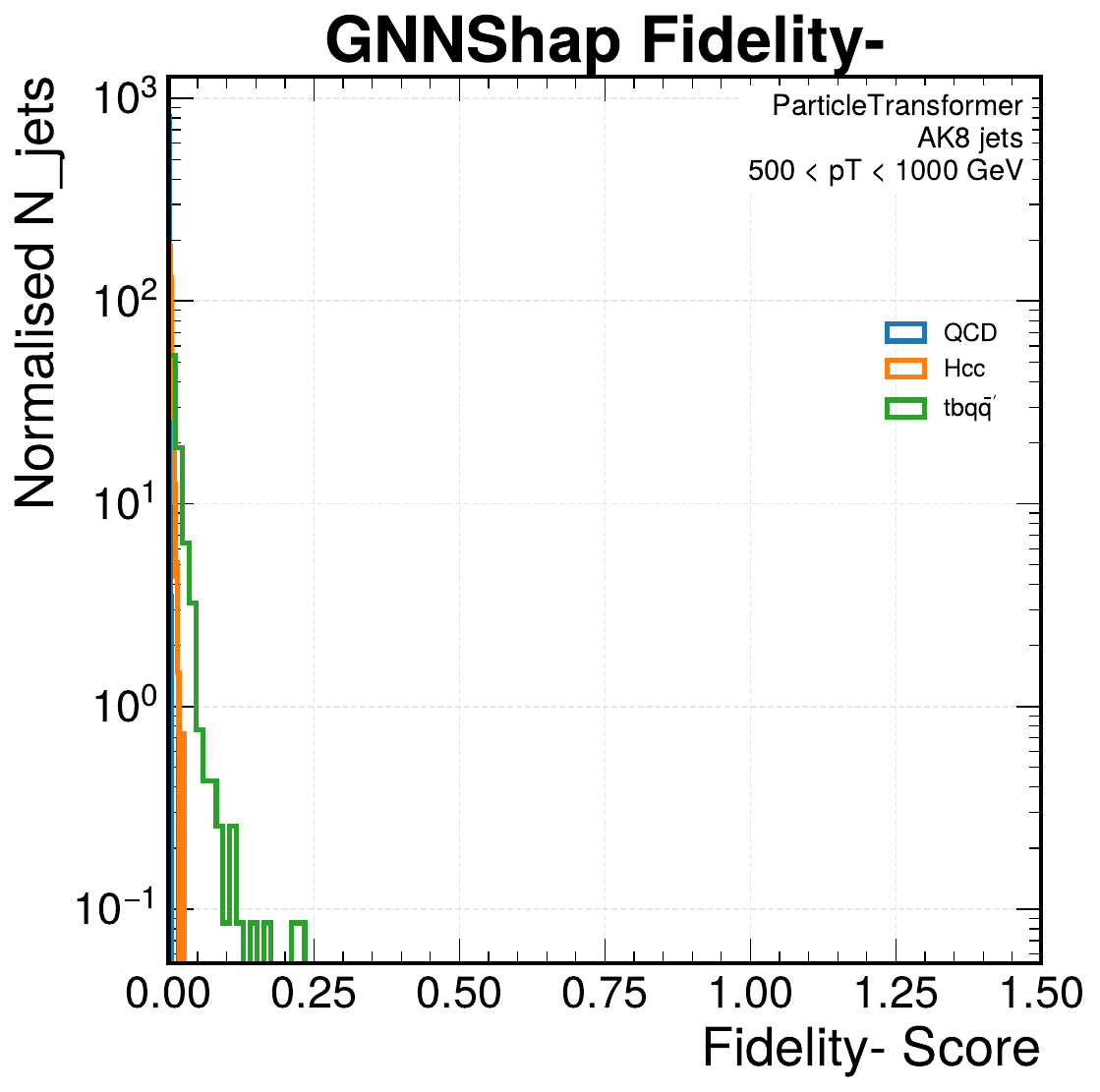}
    \caption{The distribution of the $\fidp$ and $\fidm$ score for the \part{} network along with \gnnshap{} explainability method. 
      We see that for 2-prong and 3-prong jets the shift of the $\fidp$ score is as expected on the higher side where as for 1-prong jets the $\fidp$ score is populated around zero, contrary to the expectation. Explanation of such anomalous behavior lies in the LJP density distribution of 1-prong jets, for which the importance is not restricted to top-$k=15$ sub-graph only, as seen by comparing \Fig{fig:raw_ljp_density} and \Fig{fig:weighted_ljp_density_gnnshap_benchmark}\,, respectively. 
       The $\fidm$ distribution is peaked around zero for all the three cases, as expected.}
    \label{fig:gnnshap_fidplusminus}
\end{figure*}

\begin{figure*}[!t]
    \includegraphics[width=0.32\textwidth]{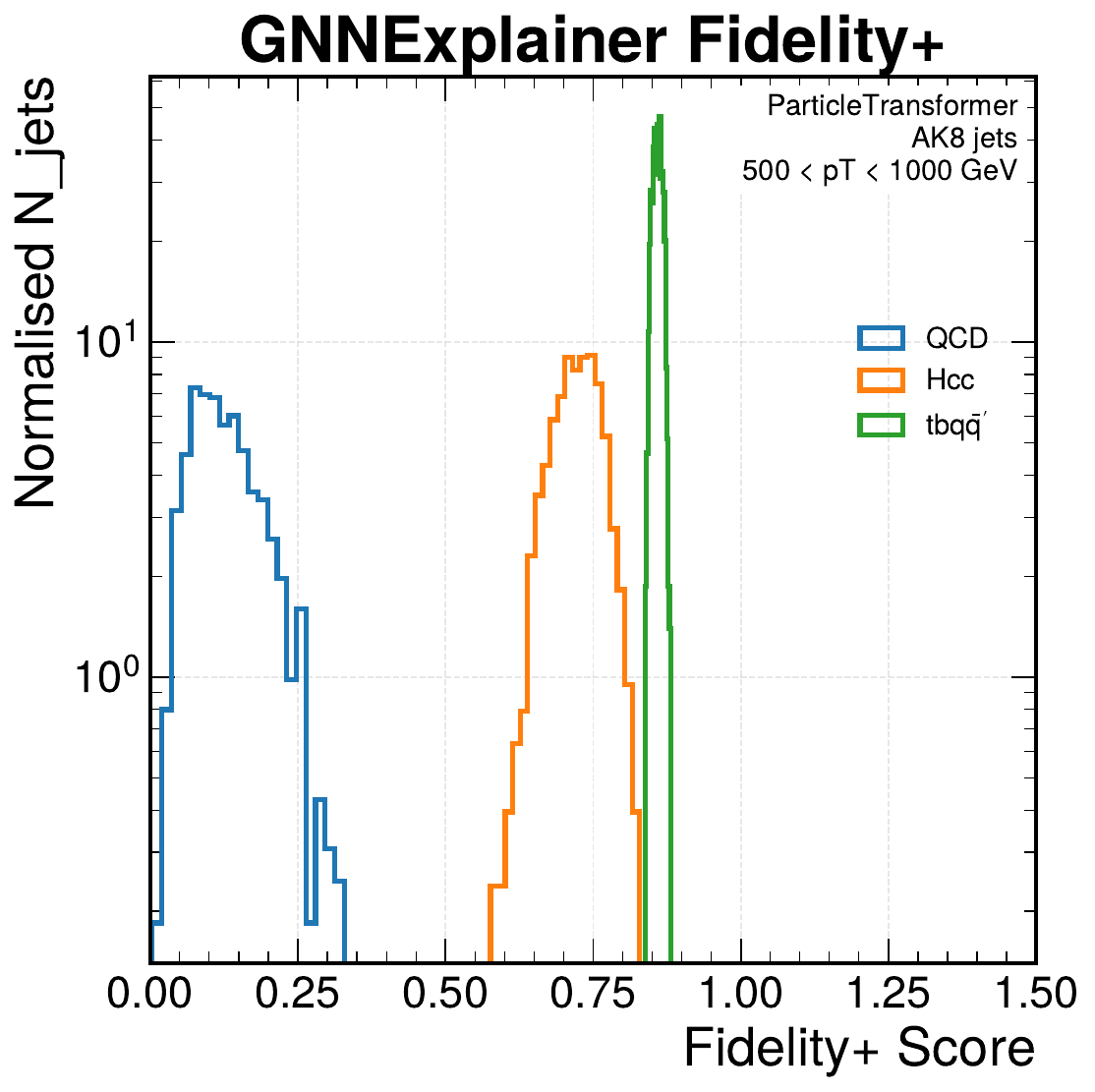}
    \includegraphics[width=0.32\textwidth]{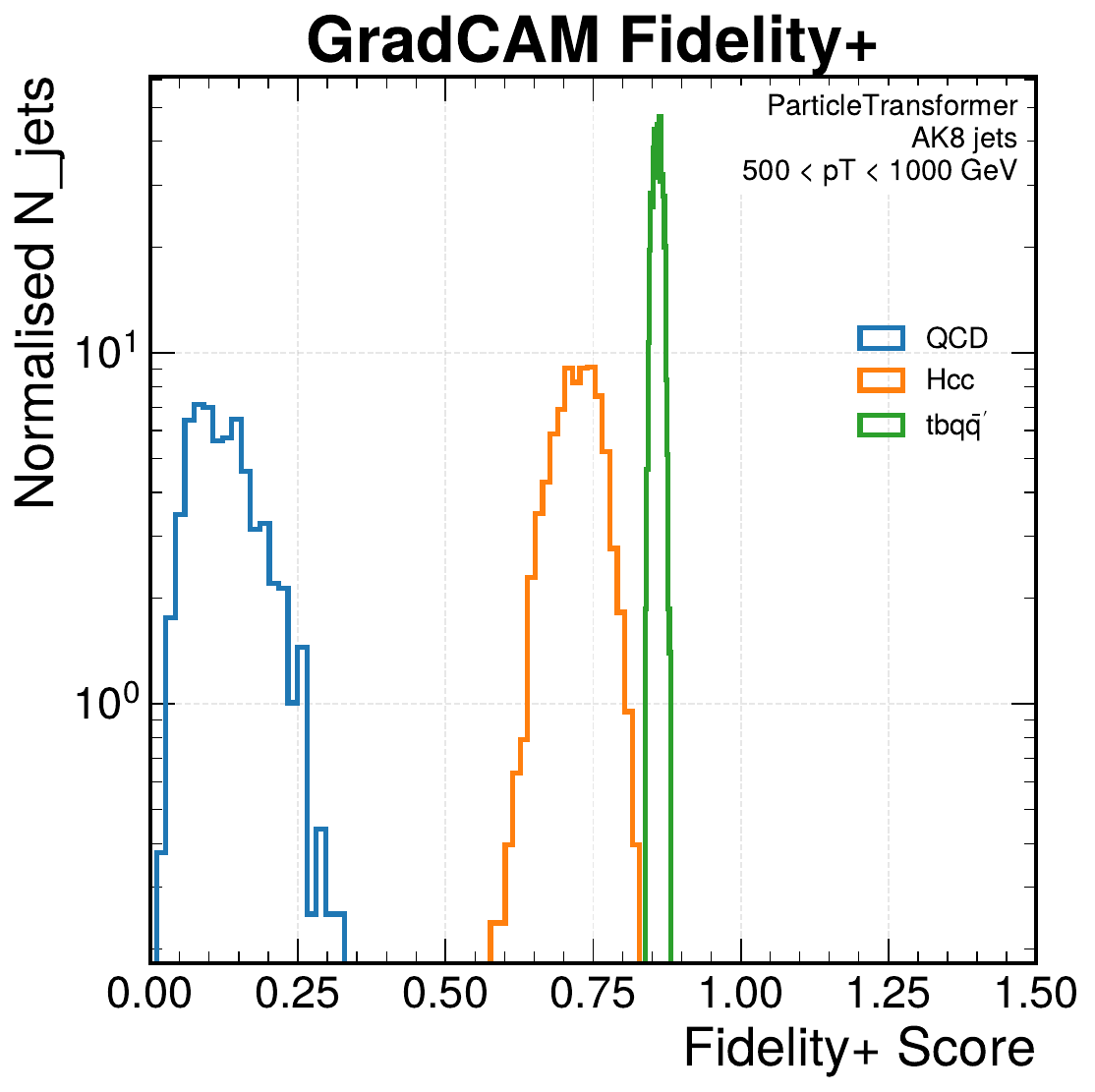}
    \caption{The distribution of the $\fidp$ score for the \part{} network along with \gnnexplainer{} and \gradcam{} explainability method. The shift close to one for 2-prong and 3-prong jets are originating from the fact that top-$k$ nodes are concentrated on the higher $\kt$ regions, corresponding to the hard jet substructure. The partial shift for 1-prong jets
      is again traced down to concentrated explanation 
      weights in the low $\kt$  regions of LJP, as demonstrated by \Fig{fig:raw_ljp_density} and  \Fig{fig:weighted_ljp_density_gnnshap_benchmark}}
    \label{fig:gnnshap_fidplus}
\end{figure*}

\vspace{0.5cm}
\paragraph{Fidelity metrics.}
Fidelity metrics on the  \(\G\) are designed to capture wether a 
sub-graph of the entire graph has more importance towards the 
downstream learning task. To test this, we first identify 
the edges whose explainability weights are more than 0.5\footnote{The threshold 
value of 0.5 may require a further optimization on per class basis. In the current 
literature, we apply it as an uniform threshold to identify important edges}. It is important to clarify here that while \gnnshap{} returns the edge importance score directly, the \gnnexplainer{} and \gradcam{} explainability methods return an importance score \(w_i\), for the \(i\)-th node, of the graph. To compare all the explainability methods on the same footing, for the later two explanation methods, we convert the node assigned weights to edge weights by taking an average value \(w_{ij} = \frac{1}{2}(w_i + w_j)\).  The 
graph, formed by this subset of edges, is identified as explanation subgraph \(\Gs\) 
and the mutually orthogonal part of the graph 
\(  \G \setminus \Gs \), which likely has much less impact 
towards downstream decision making.  \\
Following Ref.~\cite{Amara:2022graphframex}, we define:
\begin{align}
  \label{eq:fidelity}
  \fidm &= f(\G) - f(\Gs)\,,\\
  \fidp &= f(\G) - f(\G \setminus \Gs)\,,
\end{align}
where $f(\cdot)$ denotes the model's predicted probability for the
correct class.  $\fidm$ measures the prediction drop when only the
explanation subgraph is retained. It's worth noting that ideally if the exact explanation subgraph \(\Gs\) can be identified, then \(f(\G)\) and \(f(\Gs)\) should return very close by values and thus a distribution of such scores are expected to populate a region around zero.  
On the other hand, $\fidp$ measures
the drop when the explanation subgraph
\(\Gs\) is removed. The network is
evaluated on the remaining subgraph \(  \G \setminus \Gs \). We 
expect a large shift on the network evaluation score and thus 
\(\fidp\) is expected to populate regions closer to one. The 
distribution of $\fidm$ and $\fidp$ variables are shown in 
\Fig{fig:gnnshap_fidplusminus}. The distributions for 2-prong 
and 3-prong jets show that the $\fidp$ scores are as expected 
on the higher side where as for 1-prong jets the $\fidp$ score is 
populated around zero, showing a counter intuitive feature. Such an
opposite behavior can be traced back to the LJP density diagrams 
\Fig{fig:raw_ljp_density} and 
\Fig{fig:weighted_ljp_density_gnnshap_benchmark}. The first column on both 
the figures, showing 1-prong QCD emission pattern features, demonstrate that
the node importance in this case is spread over the entire LJP graph and the 
weight is not concentrated among a subgraph \(\Gs\) only. Hence when we evaluate 
the network score on \(f(\G)\) and \( f(\G \setminus \Gs) \), the two numbers 
are not much different from each other as \(\Gs\) and \(  \G \setminus \Gs \)
has similar level of importance in this case. This point is discussed in details in Sec.~\ref{sec:disc_anomaly}.
In the \Fig{fig:gnnshap_fidplus} we see the $\fidp$ distribution for 
\gnnexplainer{} and \gradcam{} for \part{}. The shift close to one for 2-prong
and 3-prong jets are self explanatory where as a partial shift of 1-prong jets
is again traced down to concentrated explanation weights in
the low $\kt$ 
regions of LJP, as seen from comparison of \Fig{fig:raw_ljp_density} and  \Fig{fig:weighted_ljp_density_gnnshap_benchmark}. 

\vspace{0.5cm}
\paragraph{Weighted Feature Average}
The combination of three different networks (\part{} 
\begin{figure*}[!t]
\includegraphics[width=0.32\textwidth]{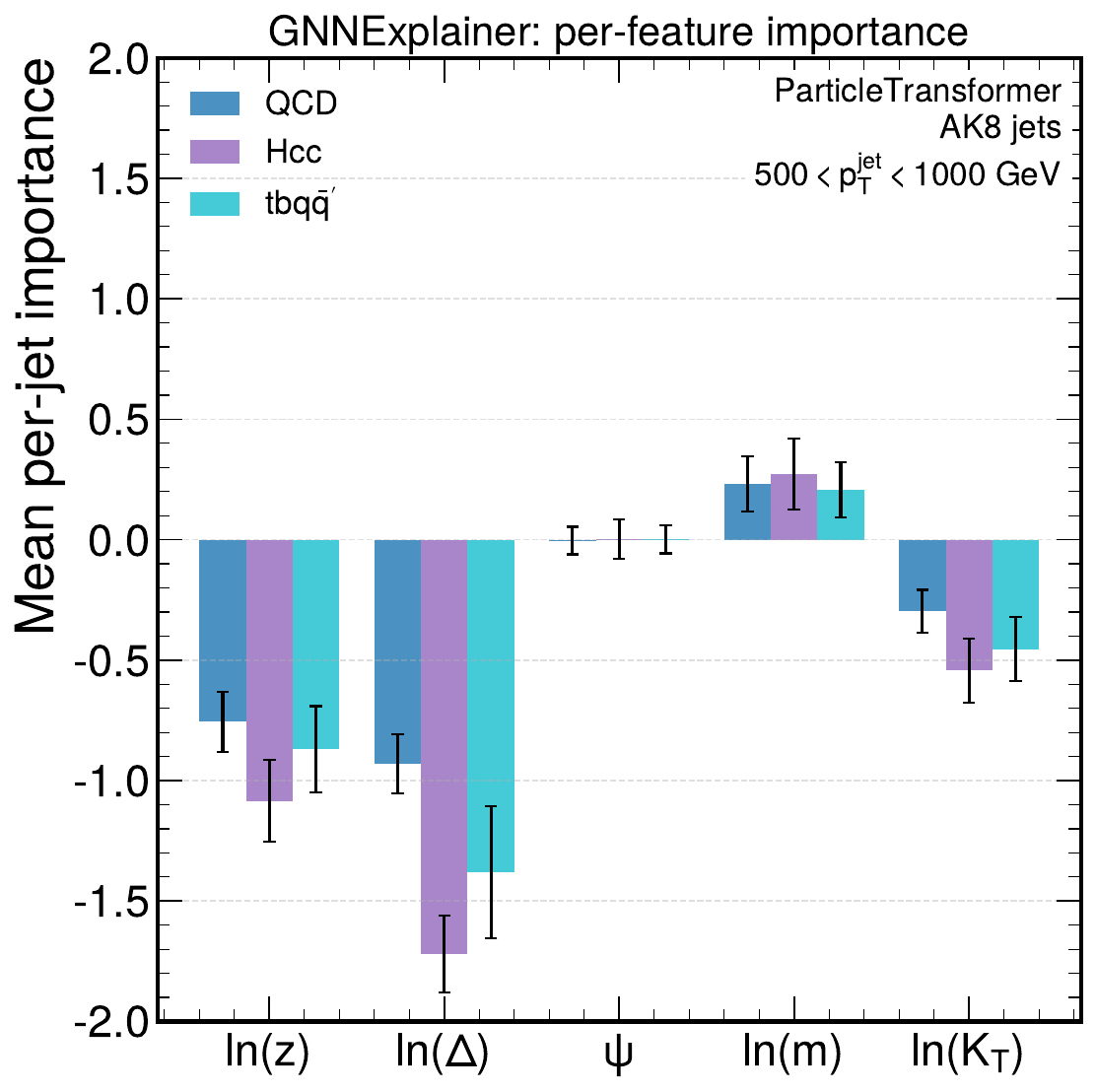}
\includegraphics[width=0.32\textwidth]{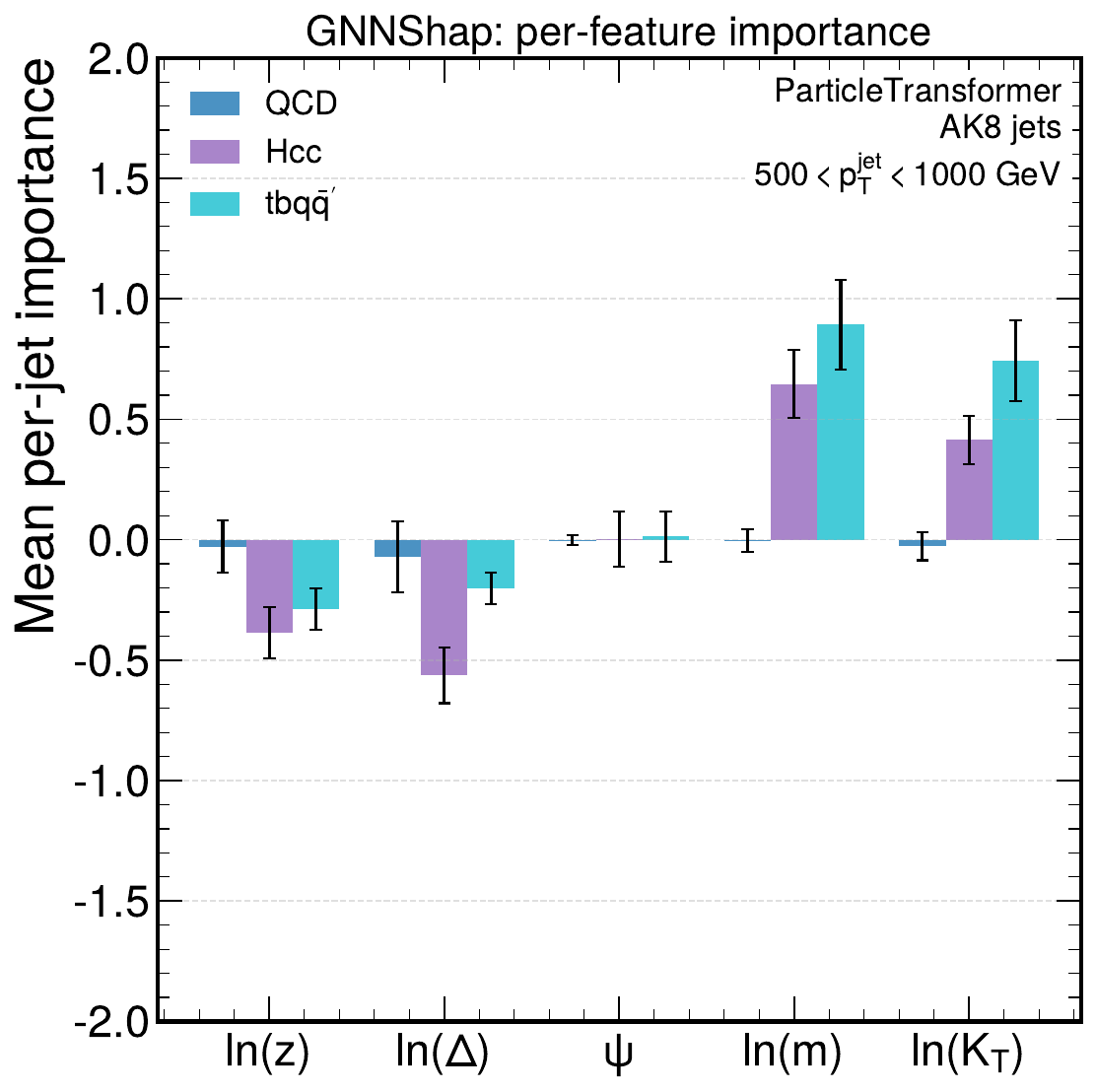}   
\includegraphics[width=0.32\textwidth]{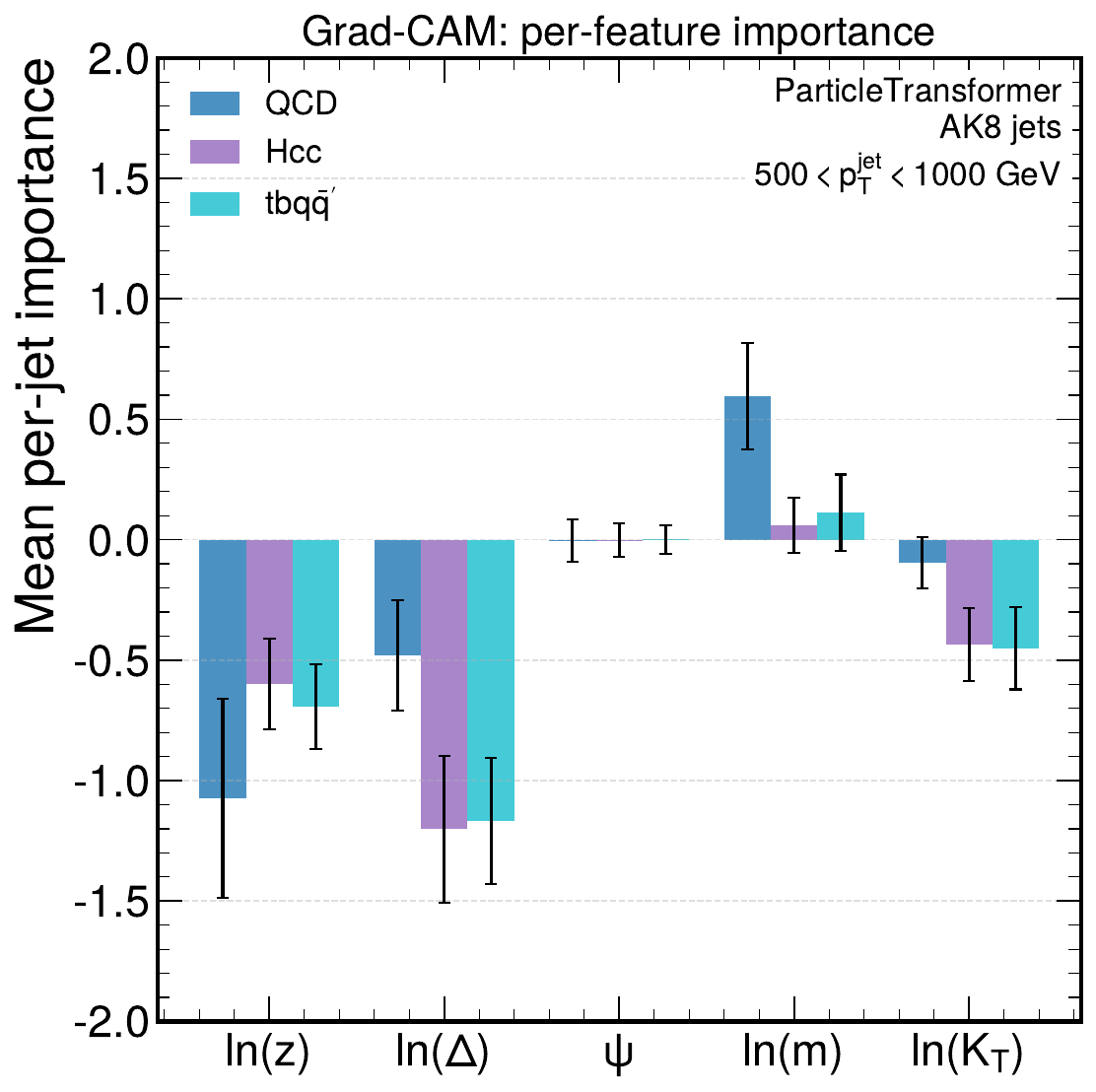}
\includegraphics[width=0.32\textwidth]{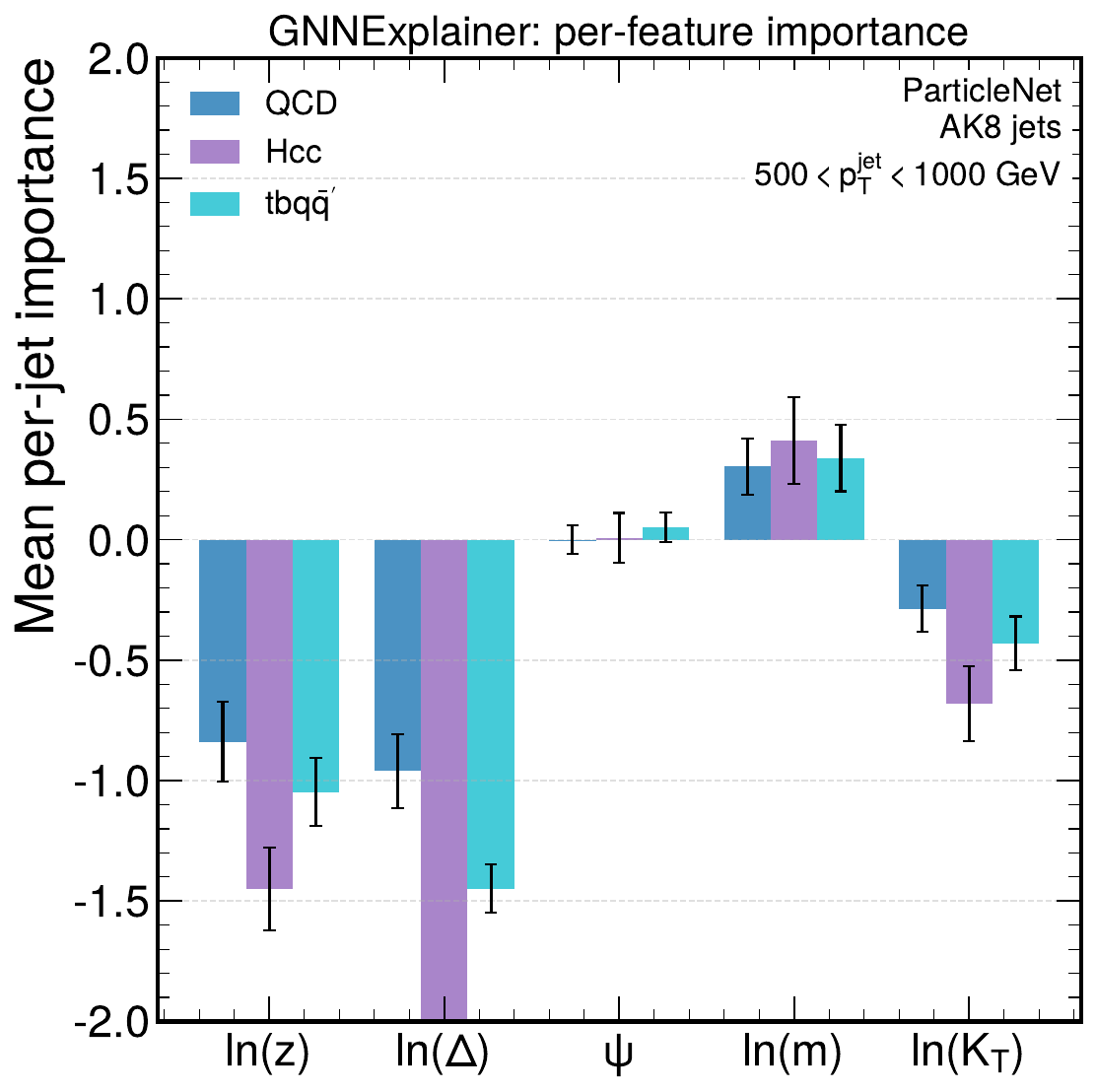}
\includegraphics[width=0.32\textwidth]{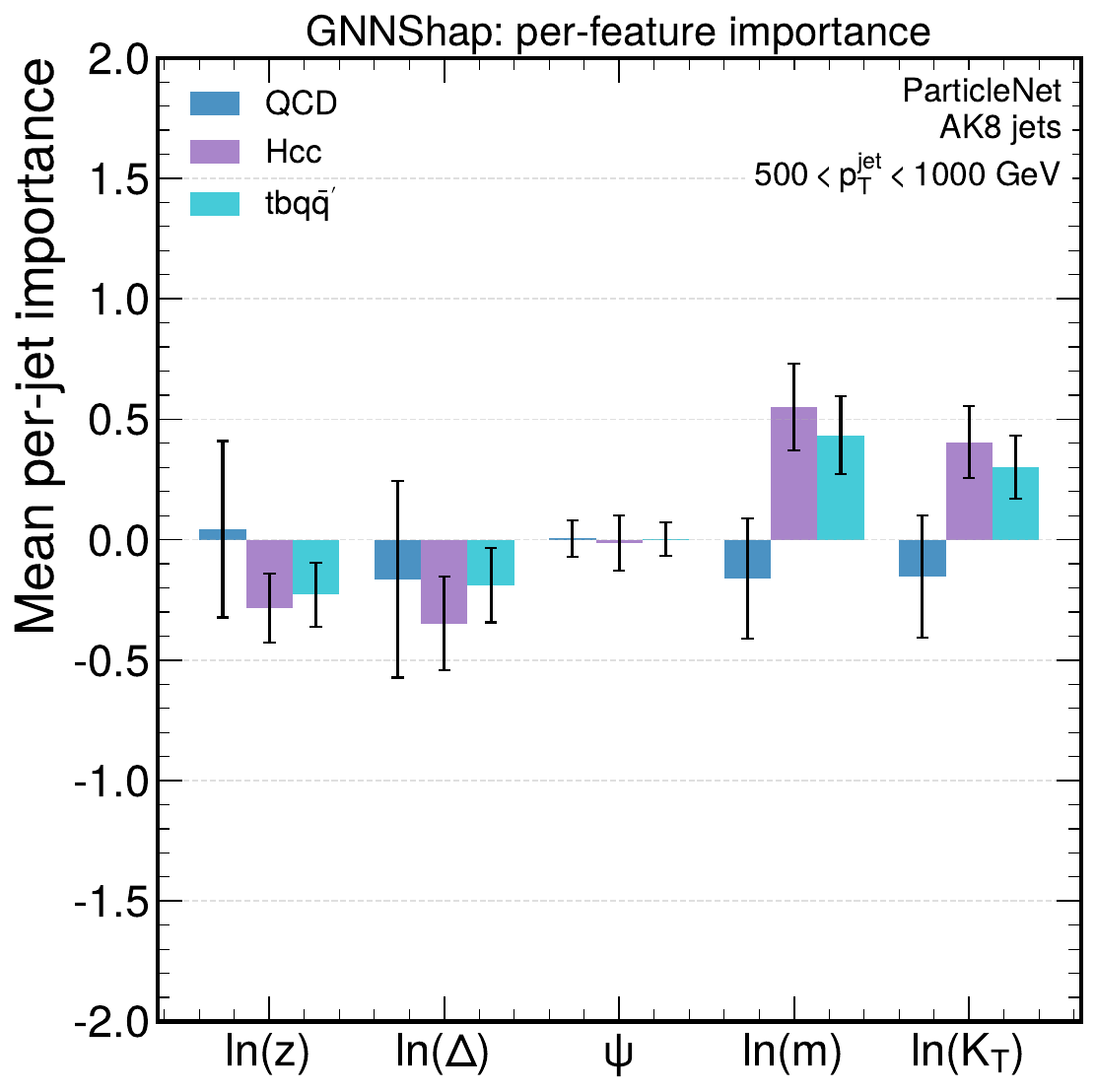}   
\includegraphics[width=0.32\textwidth]{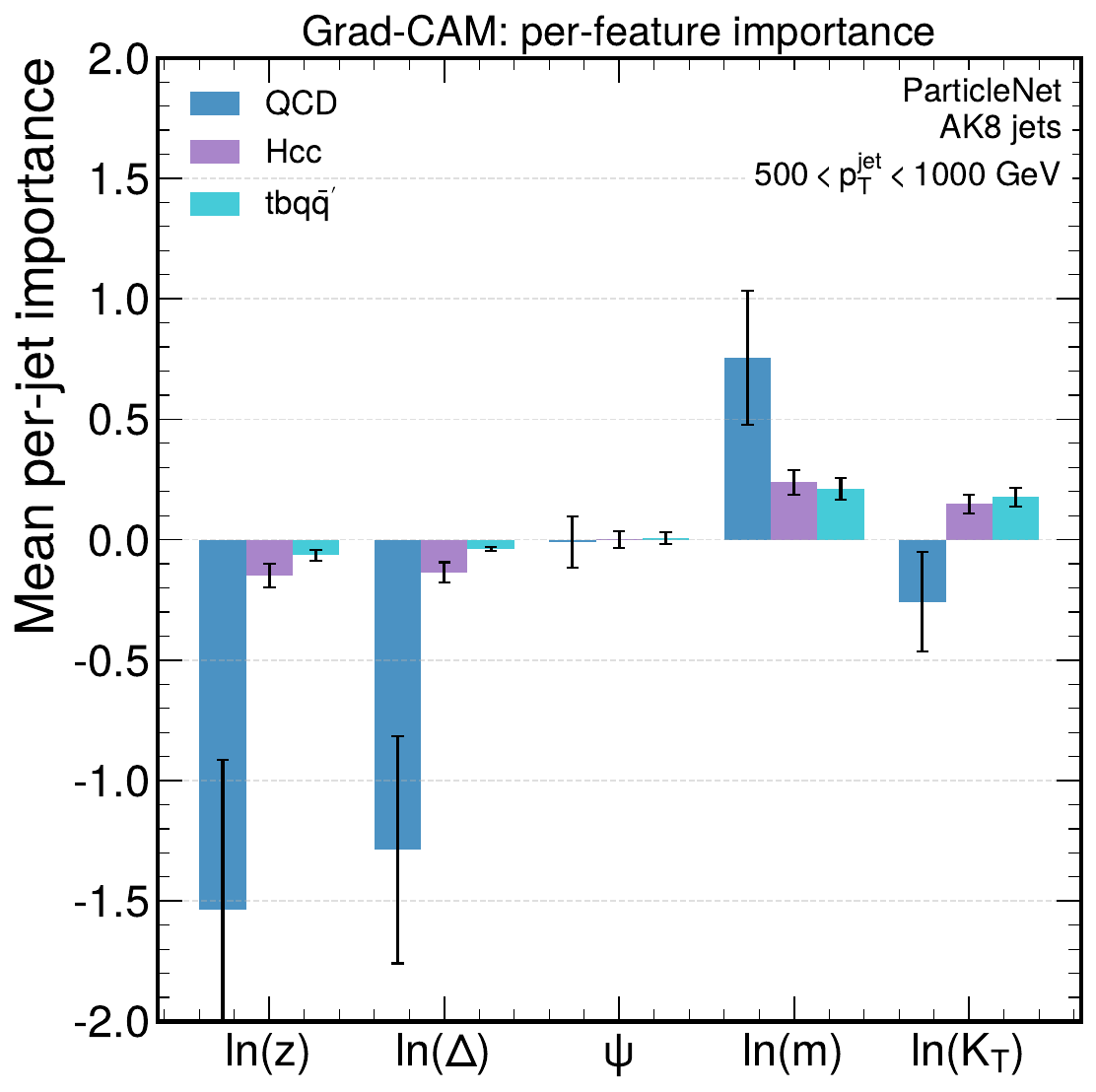}
\includegraphics[width=0.32\textwidth]{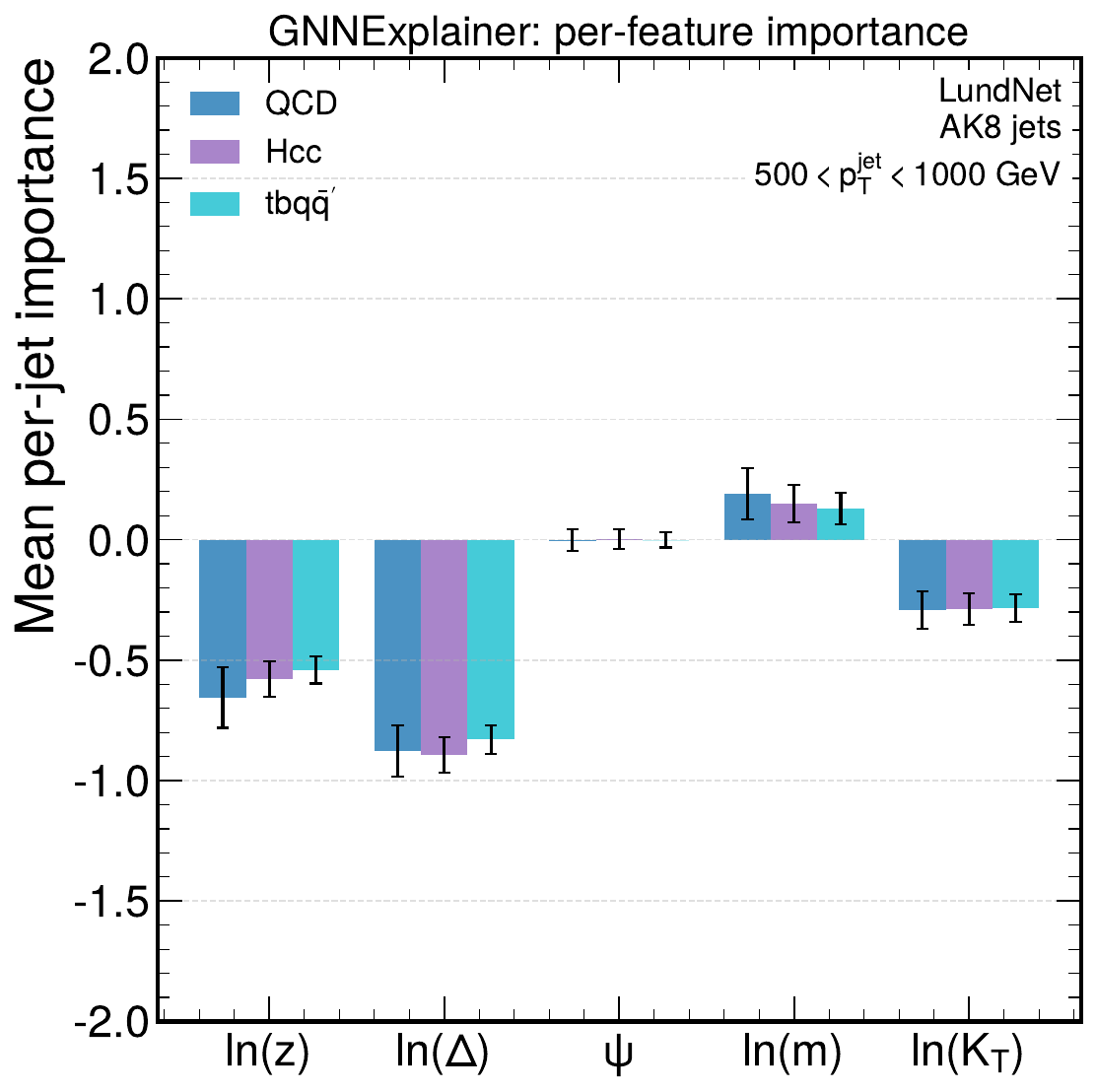}
\includegraphics[width=0.32\textwidth]{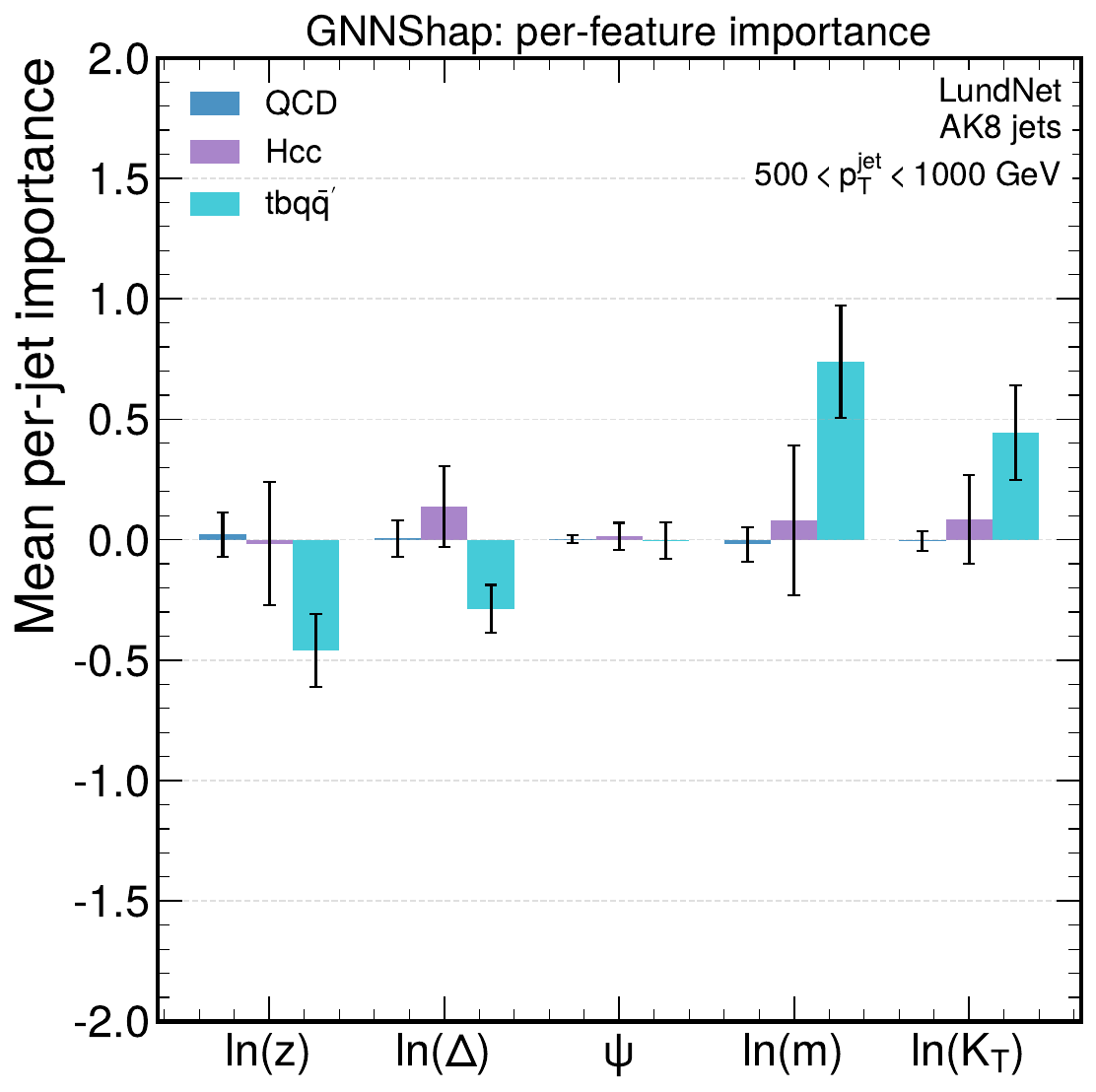}   
\includegraphics[width=0.32\textwidth]{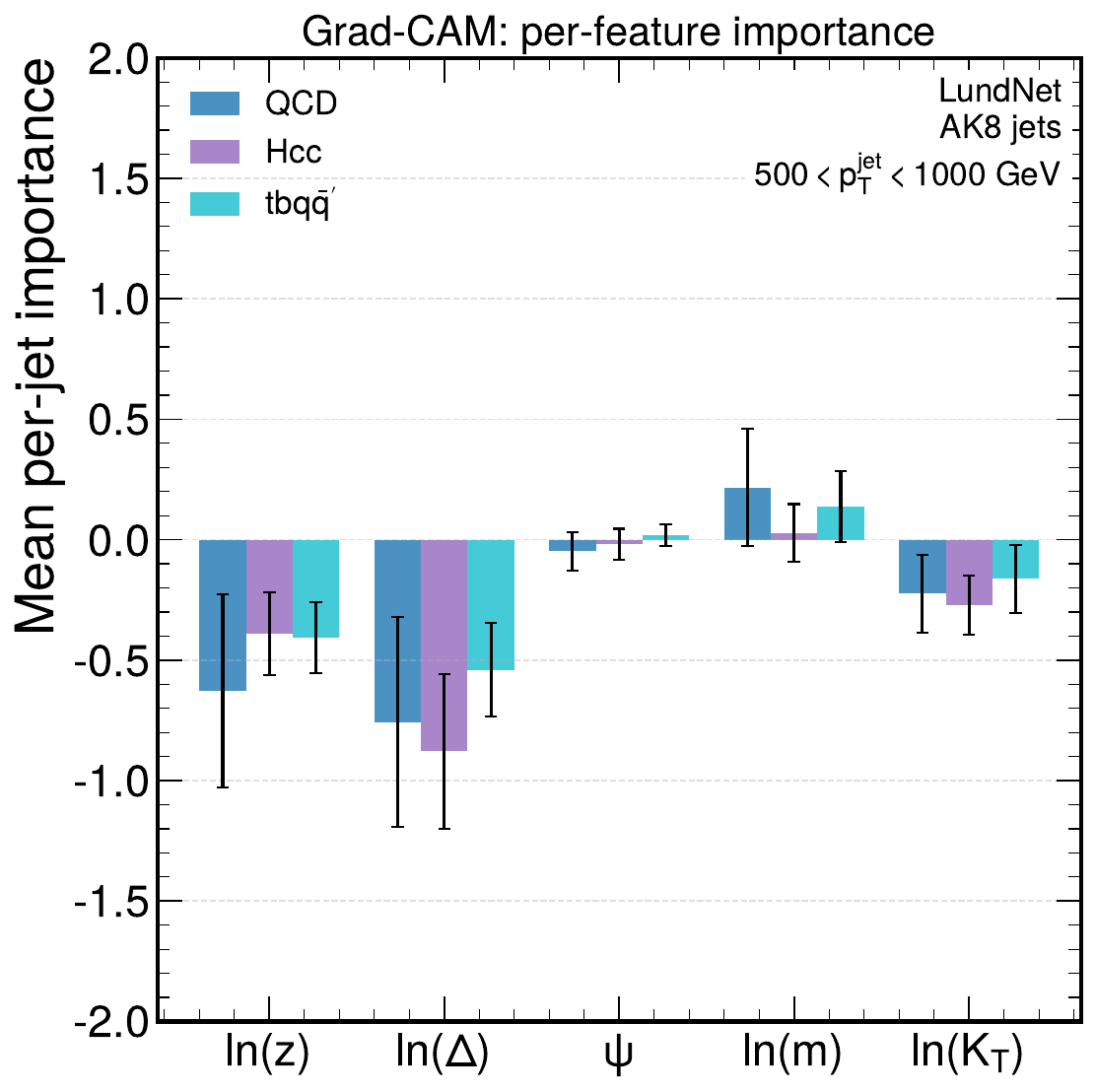}
    \caption{Weighted feature importance, as defined through 
    \Eq{eq:feat_avg}, is shown for all the five LJP features across the
    nine combinations of the neural architectures and explainability methods. It is evident that the natural logarithms of energy splitting fraction \(ln(z)\) and \(ln(\Delta)\) (both are negative quantities as in our problem \(z \le 1\) and \(\Delta \le 0.8\)) play a crucial role towards decision making, specially for QCD radiations.
    The \(ln(m)\) and \(ln(\kt)\) carry the relics of mult-prong jet substructure and hence plays a role of greater importance towards identification of 2-prong and 3-prong jets. Since the quantum dynamics associated with the parton fragmentation process has a symmetry over azimuthal angle of the emission, the variable \(\psi\) has a negligible role towards decision making for all the above cases.
    }
    \label{fig:feature_importance}
\end{figure*}

etc.) and three different explanation methods (\gnnshap{} etc.) ultimately returns a weight parameter \(w_i\) for each node \(i \in V_{\G}\), where \(V_{\G}\) is the set of nodes in the graph \(\G\)constructed in the LJP. The impact of the learned \(w_i\) is best demonstrated through a weighted average of the five LJP feature variables \( F_i \in \{\ln z,\,\ln\Delta,\,\psi,\,\ln m,\,\ln\kt\}  \). The weighted average is computed as 
\begin{equation}
  \label{eq:feat_avg}
  {\bar{F}_i} = \frac{\sum_{j \in V_{\G}} w_j F_{ij}}{\sum_{j \in V_{\G}} w_j}\,.
\end{equation}
Here \( F_{ij} \) is the \(i\)-th LJP feature associated with the \(j\)-th node. The distribution of the average feature importance are shown in \Fig{fig:feature_importance}.

It is quite evident that each combination of architecture and explanation method has it's own estimation of \(\bar{F}_i\) and thus it makes sense to study the correlation pattern of this quantity among all the nine combinations. Such a study gives us a numerical estimate about the degree of agreement between all the methods under consideration. This study is done in details \App{app:full_results} and the corresponding correlation coefficients are tabulated in the tables \Tab{tab:correlation_matrix_lund}, \Tab{tab:correlation_matrix_part} and \Tab{tab:correlation_matrix_parnet} respectively. 

\vspace{0.5cm}
\paragraph{Substructure rank correlation.}
For each jet, we rank Lund-plane nodes by their explainer-assigned
importance $w_i$ and, separately, by their contribution $c_i^{(\mathcal{O})}$
to a substructure observable $\mathcal{O}$ (e.g., for $\tauNN{2}{1}$,
$c_i$ is the $i$-th constituent's contribution to the $\tau_2$
numerator).  The Spearman rank correlation coefficient
\begin{equation}
  \label{eq:spearman}
  \rho_s(\mathcal{O})
  = 1 - \frac{6\sum_i d_i^2}{N(N^2-1)}\,,
  \quad d_i = \mathrm{rank}(w_i) - \mathrm{rank}(c_i^{(\mathcal{O})})\,,
\end{equation}
quantifies the agreement between the explainer's ranking and the
observable's ranking across $N$ nodes.

\FloatBarrier
\section{Results}
\label{sec:results}

The results presented in this section are organized around three
physics questions that the Lund Jet Plane representation makes
quantitatively addressable.  First, do the importance maps assigned
by the trained taggers populate the regions of phase space that QCD
factorization theorems identify as physically meaningful?  Second,
when interrogated through different XAI methods, do the networks
consistently emphasize the same parton-shower features, and how
does this consistency vary across the perturbative-versus-non-perturbative
boundary?  Third, are the learned importance scores quantitatively
aligned with the analytic substructure observables viz. $\tau_{21}$,
$\tau_{32}$, and the energy correlation function ratios $C_2$, $C_3$,
that decades of QCD theory have established as the canonical
discriminants for $N$-prong jets?

We address these questions in three subsections, working with the
nine combinations of three architectures (\part{},
\parnet{}, \lundnet{} ) and three explainability methods (\gnnexplainer,
\gnnshap, \gradcam), and stratifying every result by the transverse
momentum bins $500 < \pt < 700$~GeV (low) and
$800 < \pt < 1000$~GeV (high), in addition to the inclusive range.

\Fig{fig:raw_ljp_density} establishes the physical baseline. The
unweighted Lund-plane emission density $\rho(\kt,\Delta R)$ for
1-prong QCD jets is dominated by soft, wide-angle emissions
($\ln\kt \lesssim 0$, $\ln(R/\Delta R) \lesssim 2$), reaching peak
densities of $\rho \approx 1.6$.  This pattern is the direct
phase-space realization of the Sudakov double-logarithmic
distribution of QCD bremsstrahlung from a single light parton.
For 2-prong $H\to c\bar{c}$ and 3-prong $t\to bq\bar{q}'$ jets, the
density develops a pronounced overpopulation in the hard-collinear
region ($\ln\kt \gtrsim 3$, $\ln(R/\Delta R) \gtrsim 2$), with peak
values of $\rho \approx 3.0$ and $\rho \approx 4.0$ respectively.
This excess is the kinematic fingerprint of the parent resonance
mass: the hard splitting that releases the Higgs or top mass
populates a localized region of phase space that is parametrically
forbidden in pure QCD radiation.

The \gnnshap{} weighted maps in \Fig{fig:weighted_ljp_density_gnnshap_benchmark}
show that both the \part{} and \parnet{} correctly
identify these mass-induced excesses as the discriminating regions.
For $H\to c\bar{c}$ jets, the weighted density peaks sharply at
$\ln\kt \approx 4$--$6$ and $\ln(R/\Delta R) \approx 1$--$2$,
reaching $\rho_W \approx 0.35$ for the \part{} and
$\rho_w \approx 0.20$ for \parnet{}.  This is the phase-space
location where the Higgs mass-drop deposits the two charm prongs,
and the explainer has localised the network's attention precisely
there.  For top jets, the \gnnshap{}-weighted importance distributes
across two angular scales, i.e. a wider opening at
$\ln(R/\Delta R) \approx 3$--$4$ corresponding to the
$t\to bW$ splitting and a more collinear contribution at
$\ln(R/\Delta R) \approx 4$--$5$ corresponding to the
$W\to q\bar{q}'$ decay.  This bimodal structure in the weighted
density is the direct image of the $V$-$A$ decay cascade of the top
quark and constitutes a non-trivial physics validation of the
Shapley-based attribution.

\Fig{fig:weighted_ljp_density_part_benchmark}, which compares all three
XAI methods on the \part{}, brings into focus a result
that is implicit in the unweighted densities but only made explicit
through XAI: \emph{GNNShap is a near-perfect filter for the hard
collinear branching, while GNNExplainer and GradCAM additionally
weight low-$\kt$ fragmentation activity that is dominated by
non-perturbative dynamics}.  For $H\to c\bar{c}$, GNNShap concentrates
$\rho_w \approx 0.35$ in the hard-collinear strip and assigns
near-zero weight elsewhere; GNNExplainer and GradCAM, by contrast,
spread importance over the full Lund plane with peaks near
$\rho_w \approx 0.10$.  This finding has a clear physics interpretation.
Shapley values, by axiomatic construction, allocate the model's
predictive output across the most causally responsible edges; in
the Lund plane, those edges are the ancestors of the hard splitting.
Perturbation-based and gradient-based methods, by contrast, are
sensitive to the full message-passing pathway, and therefore also
expose the soft fragmentation tail that the network uses as
contextual confirmation.  Both signals are physical, but they probe
different aspects of the learned representation.

\subsection{Fidelity as a diagnostic of phase-space localisation}
\label{sec:results_fidelity}

The fidelity scores (\Fig{fig:gnnshap_fidplusminus} -- \Fig{fig:gnnshap_fidplus})
permit a quantitative test of whether the importance assignment
identifies a sufficient subgraph for classification.  For
$H\to c\bar{c}$ and $t\to bq\bar{q}'$ jets, GNNShap produces a Fid$^+$
distribution sharply peaked near unity: removing the \(w_{ij} > 0.5\)
edges identified as most important destroys the network's confidence
in the signal hypothesis, as expected when the explanation captures
the hard splitting that physically defines the signal.  The
corresponding Fid$^-$ distribution is peaked near zero, confirming
that the retained subgraph alone reproduces the full prediction.

A particularly informative anomaly arises for QCD jets, where the
Fid$^+$ distribution is itself peaked near zero rather than near
unity (\Fig{fig:gnnshap_fidplus}, right column).  At first
sight this appears to violate the expected behavior of a faithful
explainer; in fact it is a direct phase-space signature of
\emph{the absence of a discriminating subgraph}.  As \Fig{fig:raw_ljp_density}
and \Fig{fig:weighted_ljp_density_gnnshap_benchmark} make clear, QCD jet importance is
distributed quasi-uniformly across the Lund plane along the
soft-emission diagonal.  No sub-selection of \(w_{ij} > 0.5\) edges captures
a parametrically larger fraction of the discriminating signal than
its complement, because the network's QCD-tag decision is built from
the \emph{collective} pattern of soft, wide-angle radiation rather
than from any single hard splitting.  This is exactly the QCD
factorization expectation: the absence of a hard scale in 1-prong
jets means that the discriminant must integrate over the full Sudakov
double-logarithmic tail.  The fidelity anomaly is therefore a
physically meaningful negative result, and the contrast between
multi-prong and 1-prong jets in the Fid$^+$ distribution is itself
a phase-space-resolved discriminant of jet topology.

\Fig{fig:fidelityplus_pt_dependenc} confirms that this contrast sharpens
at high $\pt$.  In the $800 < \pt < 1000$~GeV bin, the GNNShap
Fid$^+$ for 2-prong and 3-prong jets is even more tightly peaked
near unity than in the inclusive range, while the QCD distribution
remains pinned near zero.  The increased boost compresses the hard
splittings into a smaller angular region and makes them
parametrically more separable from the soft background.  This
behavior is consistent with the analytic expectation that
substructure observables become better-resolved at higher transverse
momenta.

\subsection{Direct correlation with $\tau_{21}$, $\tau_{32}$, and
            energy correlation functions}
\label{sec:results_substructure}

The correlation tables (\Tab{tab:correlation_matrix_lund} -- \Tab{tab:C_corr_pt_C_3_parnet}) provide the most direct test
of whether the trained networks have discovered the same kinematic
structures that QCD theory has identified as discriminating.  We
discuss the findings observable by observable.

\paragraph{$\tau_{32}$ for top jets.}
The 3-prong substructure is best probed by $\tau_{32}$, whose
discriminating power derives from the additional resolved subjet
in $t \to bW \to b q \bar{q}'$ decays.  For LundNet
(\Tab{tab:tau_corr_pt_tau_32_lund}), GNNShap achieves
$\rho(\ln\Delta) = +0.46$, $\rho(\ln\kt) = -0.42$,
$\rho(\ln m) = -0.47$, and $\rho(\ln z) = +0.32$ in the inclusive
$\pt$ bin.  These four correlations form a physically coherent
picture: the network assigns greater importance to wide-angle hard
splittings with characteristic mass scales and asymmetric momentum
sharing---precisely the kinematic features that distinguish the
$W$ decay vertex from QCD background.  The negative correlation with
$\ln m$ is particularly informative: it indicates that the network
focuses on the moderate-mass splittings characteristic of the
$W$ decay rather than on the soft mass tail of QCD radiation.

For the \part{} (\Tab{tab:tau_corr_pt_tau_32_part}), the same
GNNShap analysis yields even stronger anti-correlations:
$\rho(\ln\kt) = -0.57$ and $\rho(\ln m) = -0.50$.  The attention
mechanism, by allowing every constituent to interact with every
other through learned pairwise biases, appears to encode a more
direct mapping to the hard-splitting kinematics than the local
EdgeConv message passing of LundNet does.  This is one of the
quantitative architecture-level findings of the present study.

ParticleNet displays a markedly different pattern: its strongest
$\tau_{32}$ correlations come not from GNNShap but from GradCAM,
with $\rho(\ln\kt) = -0.47$ and $\rho(\ln m) = -0.35$ in the
inclusive bin (\Tab{tab:tau_corr_pt_tau_32_parnet}).  We interpret
this as evidence that the dynamic $k$NN graph reconstruction of
ParticleNet relies more heavily on activation-level features than
on edge-level causal structure, so that the gradient-based method
captures the network's reasoning more faithfully than the
Shapley-based one in this architecture.

\paragraph{$\tau_{21}$ for $H\to c\bar{c}$.}
The 2-prong $H\to c\bar{c}$ decay is the canonical target for
$\tau_{21}$.  \Tab{tab:tau_corr_pt_tau_21_part} reveals the strongest result
of the entire study: for the \part{} with GNNShap,
$\rho(\ln\kt) = +0.65$ and $\rho(\ln m) = +0.63$ in the
$500 < \pt < 700$~GeV bin, and $\rho(\ln\Delta) = +0.58$ in the
$800 < \pt < 1000$~GeV bin.  These values approach the practical
upper bound for a Pearson correlation between a non-linear neural
network response and a single linear feature; the implication is
that the \part{} has learned a discriminant that is
functionally close to $\tau_{21}$ as a function of the dominant
Lund-plane variables.  The shift of the strongest correlation
from $\ln\kt$ at low boost to $\ln\Delta$ at high boost is
physically expected: as the boost increases, the angular separation
of the two charm prongs shrinks, and angular features become more
discriminating relative to absolute $\kt$.

For LundNet (\Tab{tab:tau_corr_pt_tau_21_lund}, \Tab{tab:tau_corr_pt_tau_32_lund}), the strongest
correlations come from GNNExplainer rather than GNNShap, with
$\rho(\ln z) = +0.41$ and $\rho(\ln\Delta) = +0.38$ in the low-$\pt$
bin.  The momentum-sharing variable $\ln z$ acquires importance
here because GNNExplainer's perturbation-based mask exposes the
near-symmetric splitting $z \approx 0.5$ characteristic of an
unpolarised Higgs decay, a feature that GNNShap's edge-level
attribution averages over.

\paragraph{Energy correlation functions $C_2$ and $C_3$.}
The $C_2$ and $C_3$ correlation tables
(\Tab{tab:C_corr_pt_C_2_lundnet} - \Tab{tab:C_corr_pt_C_2_part} yield two findings
that complement the $N$-subjettiness analysis.  First, GNNExplainer
shows a remarkably uniform positive correlation with $\ln\Delta$
across all jet types and architectures, with values clustering near
$\rho \approx +0.55$ to $+0.60$.  This near-universality reflects
the fact that the energy correlation functions, by construction,
sum over all pairs of constituents weighted by angular separations,
and the perturbation-based explainer is therefore sensitive to the
full angular extent of the radiation pattern rather than to a single
hard splitting.  Second, GNNShap's $C_2$ and $C_3$ correlations
\emph{strengthen at higher $\pt$} for top jets in LundNet:
$\rho(\ln\Delta)$ for $C_3$ rises from $+0.52$ at low boost to
$+0.59$ at high boost (\Tab{tab:C_corr_pt_C_3_lundnet}).  This is
consistent with the analytic result that $C_3$, which probes 3-point
energy correlations, becomes a more discriminating top-tagging
observable at higher boost where the three prongs are well-resolved.

For the \part{} with GradCAM (\Tab{tab:C_corr_pt_C_2_part}),
the $C_2$ correlations for QCD jets reveal a striking pattern:
$\rho(\ln m) = +0.50$ and $\rho(\ln z) = -0.45$.  The interpretation
is non-obvious: it suggests that GradCAM, when applied to the
\part{}, captures a feature in QCD jets that combines
mass and momentum-sharing information into the energy correlation
structure, something that has no direct $N$-subjettiness analogue
and may represent a learned feature beyond the canonical
substructure observables.

\paragraph{Behavior of the azimuthal angle $\psi$.}
Across all three architectures and all three explainers, the
azimuthal angle $\psi$ around the emitter axis exhibits correlations
consistent with zero ($|\rho| \lesssim 0.13$ in nearly all cases).
This is the strongest physics-validation signal in the entire
analysis: the approximate azimuthal symmetry of QCD radiation about
the jet axis demands that no $\psi$-dependent feature be useful for
classification, and the explainers correctly recover this expectation.
The result serves as an internal consistency check that the
correlation framework is calibrated: when a feature is genuinely
uninformative, the explainers report so.

\subsection{$\pt$-evolution of explainer agreement and the
            perturbative--non-perturbative boundary}
\label{sec:results_pt}

The pairwise inter-method correlations across $\pt$ bins
(\Tab{tab:correlation_matrix_lund}--\Tab{tab:correlation_matrix_part}) provide a final
diagnostic of whether the methods converge in the perturbative
regime, as theory expects.  For top jets in LundNet
(\Tab{tab:correlation_matrix_lund}), the GNNExplainer$\leftrightarrow$GNNShap
correlation for $\ln\Delta$ rises from $+0.386$ at low $\pt$ to
$+0.451$ at high $\pt$, indicating that the two methods identify
increasingly similar wide-angle features as the boost compresses
the hard splittings.  By contrast, the GradCAM$\leftrightarrow$GNNExplainer
correlation for $\ln m$ \emph{decreases} from $+0.432$ to $+0.338$,
because GradCAM continues to weight the soft mass tail at high
$\pt$ while GNNExplainer's mask migrates toward the hard splitting.

A more striking effect appears for $H\to c\bar{c}$ jets in the
\part{} (\Tab{tab:correlation_matrix_part}).  The
GNNExplainer$\leftrightarrow$GNNShap correlation for $\ln\kt$ drops
from $+0.654$ at low $\pt$ to $+0.494$ at high $\pt$, while the
analogous correlation for $\ln m$ falls from $+0.631$ to $+0.501$.
We attribute this to the increasing collimation of the 2-prong decay:
as the angular separation of the two charm prongs shrinks below the
characteristic resolution of the soft radiation field, the boundary
between hard signal splitting and contextual soft activity becomes
ambiguous, and the perturbation-based and Shapley-based methods
resolve this ambiguity differently.

This finding has direct consequences for high-luminosity LHC
analyses, where the most boosted jets sit precisely in this
collimated regime.  Explanation quality is not a monotonically
increasing function of $\pt$; it is sensitive to whether the boost
preserves or compresses the discriminating angular scale.
\section{Discussion}
\label{sec:discussion}

\subsection{What have the networks actually learned?}
\label{sec:disc_physics}

The combined evidence from the weighted density maps, the fidelity
distributions, and the substructure correlations supports a concrete
physical conclusion: \emph{all three architectures have, with varying
degrees of fidelity, rediscovered the QCD substructure observables
that decades of analytical work had identified as the canonical
discriminants for $N$-prong jets}.  The \part{} in
particular learns a discriminant that approaches functional
equivalence with $\tau_{21}$ and $\tau_{32}$ in the regimes where
those observables are themselves well-resolved.

This conclusion is significantly stronger than the standard claim
that ML jet taggers correlate with substructure variables.  Here
we have shown phase-space resolved correlations:  the explainer
weights at specific Lund-plane locations correlate with specific
substructure features in physically expected ways.  The wide-angle
hard splitting is associated with $\tau_{32}$ for top jets; the
mass-drop region is associated with $\tau_{21}$ for $H\to c\bar{c}$;
the soft Sudakov tail is associated with the absence of discriminating
substructure for QCD; and the azimuthal angle $\psi$ is correctly
identified as informationless across all categories.  The pattern
of correlations is itself a non-trivial physics result.

A related and more provocative finding concerns the energy correlation
function correlations.  The GradCAM importance for the Particle
Transformer on QCD jets shows a strong $C_2$ correlation pattern
($\rho(\ln m) = +0.50$, $\rho(\ln z) = -0.45$) that has no direct
analytical analogue in the canonical substructure literature.  This
may indicate that the transformer architecture has learned a feature
that combines kinematic information beyond what classical observables
encode---a candidate for a novel data-driven substructure variable.
A focused analytical study of this combined feature would be an
interesting direction for future work.

\subsection{Method-architecture pairings: there is no universally
            best XAI method}
\label{sec:disc_pairings}

A non-obvious lesson of this study is that the optimal explainability
method depends on the architecture being explained.  GNNShap performs
best for the \part{} and LundNet, where the
edge-level attention or message-passing weights map naturally onto
Shapley coalitions over edges.  GradCAM performs best for ParticleNet,
where the dynamic $k$NN graph reconstruction makes edge-level
attribution less directly meaningful but activation-level attribution
remains physically interpretable.  GNNExplainer is consistently the
most informative method for revealing the role of momentum-sharing
features ($\ln z$) and angular extent ($\ln\Delta$ for energy
correlation functions), because its perturbation-based mask exposes
the full radiation pattern rather than a sparse subgraph.

For physics analyses, this implies that no single XAI method is
adequate to characterise a tagger.  A multi-method evaluation,
similar to the one conducted here, should be the default protocol
when explainability claims are made.  This is a methodological
conclusion that we believe should propagate beyond jet tagging.

\subsection{The fidelity QCD anomaly is a feature, not a bug}
\label{sec:disc_anomaly}

The Fid$^+$ distribution for QCD jets, peaked near zero rather than
near unity, was initially counter-intuitive but has a clear physical
interpretation.  In a multi-prong signal jet, classification depends
on the presence of one or two hard splittings: removing those edges
collapses the signal hypothesis.  In a QCD jet, classification
depends on the absence of any single hard scale: removing any
particular subgraph of soft emissions still leaves the residual
soft pattern that the network uses for QCD identification.  The
fidelity anomaly is therefore a phase-space-resolved manifestation
of the QCD factorisation theorem: 1-prong jets cannot be reduced
to a small set of important emissions because their discriminating
power lies in the integrated soft tail.

This observation suggests a refinement to the standard fidelity
metric.  Rather than asking whether removing the top-$k$ important
edges destroys the prediction, one might ask whether removing
\emph{a typical random subset of size $k$} preserves the prediction.
For multi-prong jets, the random and importance-based removals
should differ markedly; for QCD jets, they should agree.  The
\emph{ratio} of these two fidelity measures is then a phase-space
discriminant in its own right.  We leave a quantitative implementation
to future work.

\subsection{Architecture comparison}
\label{sec:disc_architecture}

Quantitatively, the \part{} produces the strongest
substructure correlations of the three architectures
($|\rho| \approx 0.5$--$0.6$), followed by LundNet
($|\rho| \approx 0.4$--$0.5$), with ParticleNet showing weaker
edge-level correlations but compensating through stronger
gradient-based ones.  The interpretation is that attention-based
architectures, by allowing every constituent to interact with every
other, encode a more direct representation of pairwise kinematic
features---which $\tau_{N}$ and $C_N$ are explicit functions of.
Local message-passing architectures (LundNet, ParticleNet) require
multiple layers to assemble the same global features, and the
explanations distribute over those intermediate layers.

This finding is consistent with, but not implied by, the earlier
observation that \part{} outperforms the GNN-based
taggers on standard benchmarks.  The novel content here is that
the performance advantage is accompanied by a more direct internal
representation of the physical observables, not merely by a better
training landscape.

\subsection{Limitations}
\label{sec:disc_limitations}

Five limitations of the current analysis warrant explicit
acknowledgment.

\emph{First}, the substructure correlations are computed in the
Pearson (linear) framework.  The networks may exploit non-linear
combinations of features that are not captured by this metric.
The moderate values of $|\rho| \approx 0.4$--$0.6$ leave room for
unexplained variance that could reflect either non-linear effects
or genuine novel features beyond classical observables.  A
non-linear correlation analysis (e.g., distance correlation or
mutual information) would tighten the conclusions.

\emph{Second}, all three explainability methods are post-hoc.  They
characterise what the model \emph{has learned}, not what the physics
\emph{is}.  A strong correlation between explainer importance and
$\tau_{32}$ does not prove that the network internally computes
$\tau_{32}$; it may use a correlated but functionally distinct
discriminant.

\emph{Third}, the MC-truth ground-truth masks rely on the parton-level
information available in the generator record, which becomes
ambiguous after hadronisation, particularly at the boundary between
signal decay products and the underlying event.  A systematic study
using detector-level information alone would test the robustness
of the conclusions.

\emph{Fourth}, GradCAM's tendency to weight low-$\kt$ fragmentation
activity, while physically interpretable, can produce attributions
that are not amenable to direct physics validation against analytical
substructure observables.  We recommend cross-validating GradCAM
with at least one perturbation-based or game-theoretic method when
making claims about learned physics.

\emph{Fifth}, the analysis is performed on a single MC sample.
A Pythia \cite{Bierlich:2022pfr}-versus-Herwig \cite{Bahr:2008pv} comparison would test whether the explanation
patterns shift under generator change, providing a direct diagnostic
for simulation dependence --- a test that is essential before deploying
the methodology on real collision data.

\FloatBarrier
\section{Conclusion and Outlook}
\label{sec:outlook}

\subsection{Conclusion}
\label{sec:conclusions_summary}

We have presented the first systematic, multi-architecture, multi-method
explainability study of jet tagging in the Lund Jet Plane.  By
combining three explainability paradigms---perturbation-based
(GNNExplainer), Shapley-value-based (GNNShap), and gradient-based
(GradCAM)---and applying them to three state-of-the-art architectures
(Particle Transformer, ParticleNet, LundNet), we have constructed a
quantitative bridge between the learned representations of black-box
neural taggers and the analytical substructure observables that QCD
factorisation theorems identify as canonical.

The principal physics findings are:

\begin{enumerate}

\item \emph{All three architectures have rediscovered the canonical
      QCD substructure observables.}  The phase-space-resolved
      correlations between explainer importance and $\tau_{21}$,
      $\tau_{32}$, $C_2$, $C_3$ form a physically coherent pattern
      that matches the analytical expectations for 2-prong, 3-prong,
      and 1-prong jets.  The Particle Transformer in particular
      learns a discriminant that approaches functional equivalence
      with $\tau_{21}$ ($|\rho| \approx 0.65$) for $H\to c\bar{c}$
      tagging.

\item \emph{The fidelity anomaly for QCD jets is a manifestation of
      QCD factorisation.}  The peaking of Fid$^+$ near zero for
      1-prong jets reflects the absence of a discriminating subgraph,
      consistent with the integrated Sudakov double-logarithmic
      structure of QCD radiation.  This is a non-trivial physics
      validation of the methodology.

\item \emph{Attention-based architectures encode substructure
      observables more directly than message-passing GNNs.}  The
      stronger Pearson correlations achieved by the Particle
      Transformer ($|\rho| \approx 0.5$--$0.6$) compared to LundNet
      and ParticleNet ($|\rho| \approx 0.4$) suggest that pairwise
      attention biases are particularly well-suited for representing
      $N$-point energy correlation features.

\item \emph{The optimal explainability method depends on the
      architecture.}  GNNShap is most informative for the Particle
      Transformer and LundNet; GradCAM is most informative for
      ParticleNet.  No single method is universally best, and a
      multi-method evaluation should be the default protocol.

\item \emph{The azimuthal angle $\psi$ is correctly identified as
      informationless across all combinations}, providing an internal
      consistency check that the correlation framework is well-calibrated
      against a ground-truth physical symmetry.

\end{enumerate}

These findings collectively support the claim that the trained
neural taggers are not merely high-performance black boxes but are,
in a quantitatively verifiable sense, learning the substructure
physics that decades of theoretical work had identified.

\subsection{Outlook: applications in high-energy physics and beyond}

The methodology developed here opens several concrete avenues for
future investigation, both within high-energy physics and in
adjacent domains where graph-based machine learning intersects with
physically interpretable representations.

\paragraph{Robustness against generator and detector effects.}
The single most important next step is a Pythia-versus-Herwig
comparison.  If the explainer importance maps shift significantly
between generators, that shift directly localises the
generator-dependent components of the learned representation in
phase space.  The complement---phase-space regions where importance
is generator-stable---would identify the reliably perturbative
features that the network can safely exploit when deployed on real
data.  An analogous comparison between particle-level and
detector-level inputs would expose the detector-dependent
attributions, providing a quantitative tool for assessing
data--simulation transferability.

\paragraph{Application to real collision data.}
The Lund-plane density has been measured by both ATLAS and CMS at
$\sqrt{s} = 13$~TeV.  Computing the explainer importance maps on
the same datasets, and comparing them against the analytical
predictions matched to the experimental measurements, would for the
first time validate ML tagger explanations directly against data
rather than against simulation.  This is the natural endpoint of
the present work and would constitute a first-of-its-kind
data-driven physics validation of a deep tagger.

\paragraph{Anomaly detection and BSM searches.}
Lund-plane explainability is naturally suited to anomaly detection
tasks where no labelled signal exists.  An anomaly tagger trained
on QCD data alone, when interpreted through GNNShap, will indicate
which Lund-plane regions are unexpectedly populated in anomalous
events.  This provides a model-independent way to localise BSM
signatures in phase space, with direct application to dark sector
searches~\cite{Cohen:2023mya}, exotic Higgs decays, and unexpected
electroweak resonances.  The methodology turns XAI from a diagnostic
tool into a discovery instrument.

\paragraph{Fast-track tagger design through explanation feedback.}
The pattern of explanations identified here can directly inform
architecture design.  If GNNShap reveals that a particular subset
of Lund-plane regions carries the entire predictive signal, a
tagger with reduced input dimensionality---restricted to those
regions only---should preserve performance while reducing inference
cost.  Such explanation-guided architecture pruning could be
particularly impactful for HL-LHC trigger applications, where
inference latency is critical.

\paragraph{A general principle.}
The deeper insight of this study is that whenever a graph
representation has nodes with first-principles physical meaning,
explainability becomes more than a debugging tool: it becomes a
physical instrument that measures what the network has learned in
the language of the underlying theory.  The Lund Jet Plane is
unusually well-suited to this approach because every node corresponds
to a parton splitting whose properties are calculable from QCD.
But the principle generalises: any time first-principles theory
provides expectations for which features should matter, multi-method
XAI can quantify the extent to which a deep model has discovered
those expectations.  We expect this principle to be increasingly
important as machine learning becomes embedded in the analysis
pipelines of modern physics experiments.

\FloatBarrier
\section{Code Availability}
The analysis used JetClass public jet tagging dataset. The analysis codes will be made public upon publication.

\section{Acknowledgment}
SG is supported by the IIT-Kanpur faculty initiation grant (IITK /PHY /2023499) and Anusandhan National Research Foundation, Advanced Research Grant (ANRF /PHY /2025804). SG would like to thank the organizers
of "First Lund Jet Plane Workshop, July 2023", held at CERN TH department, from where many exciting ideas were obtained.  
\FloatBarrier
%
%

\appendix

\section{Description of used explainability methods}
\label{app:xai_math}

This appendix provides the full descriptions of all the explainability methods  that are
summarized in the main text (Section~\ref{sec:lund_plane_ml}).

\subsection{\gnnexplainer{}: Mutual information maximization}
\label{app:gnnexplainer}
Given a trained GNN $\Phi$ and an input graph
$\G = (V, E, \mathbf{X})$
with prediction $\hat{Y} = \Phi(\G)$, \gnnexplainer{} seeks a subgraph
$\Gs \subseteq \G$ that maximises the mutual information:
\begin{equation}
  \max_{\Gs}\; I(Y, \Gs)
  = H(Y) - H(Y \mid \Gs)\,,
\end{equation}
where $H(Y)$ is the entropy of the prediction (constant for a fixed
model and input).  Minimizing $H(Y \mid \Gs)$ is equivalent to finding
$\Gs$ such that the model's prediction is most certain when conditioned
on $\Gs$ alone.

Since optimizing over discrete subgraphs is combinatorialy intractable,
the method introduces a continuous relaxation via a learnable mask
$\mathbf{M} \in [0,1]^{|E|}$:
\begin{equation}
  \Gs = \bigl(V,\; E \odot \sigma(\mathbf{M}),\; \mathbf{X} \odot \sigma(\mathbf{F})\bigr)\,,
\end{equation}
where $\sigma$ is the sigmoid function and $\mathbf{F} \in \mathbb{R}^d$ is
a learnable feature mask.  The objective becomes:
\begin{align}
  \label{eq:gnnexplainer_full}
  \min_{\mathbf{M}, \mathbf{F}}\;
  &-\sum_{c=1}^{C} \mathbf{1}[\hat{Y}=c]\,
    \log P_\Phi\bigl(Y=c \mid \G \odot \sigma(\mathbf{M}),\,
    \mathbf{X}\odot\sigma(\mathbf{F})\bigr) \notag\\
  &+ \lambda_1\,\|\sigma(\mathbf{M})\|_1
   + \lambda_2\,H\bigl(\sigma(\mathbf{M})\bigr)\,,
\end{align}
where the first term is the cross-entropy loss encouraging the masked
graph to reproduce the original prediction, the second enforces sparsity,
and the third (element-wise entropy
$H(m) = -m\log m - (1-m)\log(1-m)$) encourages the mask values to be
near 0 or 1 rather than intermediate.

\subsection{\gnnshap{}: Shapley value computation}
\label{app:gnnshap}
GNNShap constructs its interpretability scores strictly upon the axioms of cooperative game theory, specifically utilizing Shapley values. Rather than focusing on nodes or features, GNNShap primarily treats the edges of the computational graph as the "players" in the game, providing a mathematically fair and fine-grained allocation of the model's predictive output.

Here is the step-by-step mathematical construction of how GNNShap computes these scores.
The key computational
challenge is that the sum runs over $2^{|\mathcal{E}|-1}$ subsets.
\gnnshap{} addresses this through:

\emph{(i) Computational graph pruning.}  For a target node $v$ in an
$L$-layer GNN, the computational graph $\G_L(v)$ contains all edges
whose messages reach $v$ within $L$ hops.  Edges in the $L$-hop induced
subgraph that do not actually carry messages to $v$ are pruned, reducing
$|\mathcal{E}|$ significantly.

\emph{(ii) Stratified sampling.}  Rather than sampling coalitions
uniformly, \gnnshap{} samples from all coalition sizes $|S| \in
\{0, 1, \ldots, |\mathcal{E}|-1\}$ with equal probability, then
applies importance weighting to recover unbiased Shapley estimates.

\emph{(iii) GPU batching.}  Multiple coalition evaluations are batched
on GPU, with the value function $f(S) = P_\Phi(Y = \hat{Y} \mid
\G_S)$ computed by zeroing out edges not in $S$.
The Shapley value for the \(i\)-th edge, \(\phi_i\) is computed by the formula 
\begin{equation}
\label{eqn:shap_val}
\phi_i = \sum_{S \subseteq E_c \setminus \{e_i\}} \frac{|S|! (|E_c| - |S| - 1)!}{|E_c|!} (v(S \cup \{e_i\}) - v(S))
\end{equation}
The resulting edge-level Shapley values $\phi_e$ satisfy the standard
axioms: efficiency ($\sum_e \phi_e = f(\mathcal{E}) - f(\emptyset)$),
symmetry, null player, and linearity. Finally we compute the average node level explainer 
weight for the \(i\)-th node is computed as : 
\begin{equation}
w_i = \frac{1}{|\mathcal{N}_i|} \sum_{j \in \mathcal{N}_i} e_{ij} \,,
\end{equation}
where \(\mathcal{N}_i\) is the neighborhood node set for the target node \(i\) and \(e_{ij}\) is the Shapley value computed through \Eq{eqn:shap_val} for the edge \((ij)\) which connects node \(j \rightarrow i\). 

\subsection{Graph \gradcam{}: Layer-wise activation mapping}
\label{app:gradcam}

For class $c$, \gradcam{} computes the importance weight for channel
$k$ at layer $\ell$ as:
\begin{equation}
\label{eqn:gradcam_alpha}
  \alpha_k^{c,(\ell)}
  = \frac{1}{N}\sum_{i=1}^{N}
    \frac{\partial y^c}{\partial H_{ik}^{(\ell)}}\,,
\end{equation}
where $N$ is the number of nodes and $H_{ik}^{(\ell)}$ is the activation
of node $i$ in channel $k$ at layer $\ell$.  The node-level importance is:
\begin{equation}
\label{eqn:gradcam_w}
  w_i^{c,(\ell)} = \mathrm{ReLU}\!\Bigl(
    \sum_{k=1}^{K_\ell} \alpha_k^{c,(\ell)}\,H_{ik}^{(\ell)}
  \Bigr)\,.
\end{equation}

When aggregating across layers, we take the element-wise maximum:
\begin{equation}
  w_i^c = \max_\ell\; w_i^{c,(\ell)}\,.
\end{equation}

The per-feature extension computes $\partial y^c / \partial x_{i,d}$
via backpropagation to the input layer, providing a $d$-dimensional
attribution vector for each node.

\section{Architecture Details}

\subsection{\lundnet{}  details}
\label{sec:lundnet}

\lundnet{}~\cite{Dreyer:2020brq} transforms the Lund-plane representation
into a graph suitable for GNN processing.  Each emission in the declustering
tree becomes a node $v_i$ with kinematic feature vector $\mathbf{x}_i$.  Two
variants are defined.  \lundnet{} constructs a $k$-nearest-neighbour graph
with $k=16$ in the five-dimensional feature space
$\mathbf{x}_i = (\ln z,\,\ln\Delta,\,\psi,\,\ln m,\,\ln\kt)_i$, where
$\psi$ is the azimuthal angle around the emitter axis and $m$ is the
invariant mass of the splitting. 

Message passing is performed through $L$ EdgeConv
layers~\cite{Wang:2019dgcnn}.  At layer $\ell$, the update for node $i$ is
\begin{equation}
  \label{eq:edgeconv}
  \mathbf{h}_i^{(\ell+1)}
  = \oplus_{j \in \mathcal{N}(i)}
    h_\Theta\!\bigl(\mathbf{h}_i^{(\ell)},\,
                     \mathbf{h}_j^{(\ell)} - \mathbf{h}_i^{(\ell)}\bigr)\,,
\end{equation}
where $\mathcal{N}(i)$ is the set of neighbours of node $i$,
$h_\Theta$ is a multi-layer perceptron (MLP), and $\oplus$ denotes
channel-wise permutation invariant pooling over the neighbourhood.  After $L$ layers, a
global aggregation (mean and max over all nodes) produces a fixed-length
vector that is passed to a fully connected classification head. The above EdgeConv operation is carried out on fixed static graph only.
Within the experimental collaborations, \lundnet{} was first studied by the ATLAS
collaboration \cite{ATLAS:2023ixc} for boosted-object tagging.
It achieved state-of-the-art performance while offering
an
order-of-magnitude improvement in computational speed over
previous
graph-based taggers \cite{Dreyer:2020brq}.

\begin{table}[h]
  \caption{\label{tab:architecture}%
    \lundnet{} architecture hyperparameters.}
  \begin{ruledtabular}
  \begin{tabular}{lc}
    Parameter & \lundnet{}  \\
    \hline
    Input features & 5 \\
    EdgeConv layers & 6\\
    MLP widths & (16,16), (32,32) (64,64)$\times$2, (128,128)$\times$2  \\
    $k$ (neighbours) & 16  \\
    Global pooling & Mean + Max  \\
    FC layers & 256, 128, $C$  \\
    Dropout & 0.1  \\
    Activation & ReLU \\
  \end{tabular}
  \end{ruledtabular}
\end{table}

\subsection{\parnet{}  details}
\label{sec:parnet}
The working principle of \parnet{} is similar to \lundnet{}, that of applying EdgeConvolution operation \Eq{eq:edgeconv}. The core difference is for \parnet{}, a graph is dynamically constructed, in the LJP,  at the input level of each EdgeConv layer. In this case we have used 3 EdgeConv layers.  

\subsection{\part{}  details}
\label{sec:part}
We applied the self-attention mechanism 
\begin{equation}
\mathrm{Attention}(Q,K,V) = \mathrm{softmax}\!\left(\frac{QK^\top}{\sqrt{d_k}} + U\right) V .
\end{equation}

on the input graph from LJP to adapt \part{}. We use three Embedding layers with latent space dimension \( (128, 128, 128)  \).

\section{Training details}
\label{app:training_details}

All the models are trained with AdamW optimizer with weight decay factor of 0.01 and learning rate = $10^{-4}$. 
Each of the model is trained to reach at least 86$\%$ signal tagging efficiency. The jets are taken from JetClass dataset. 

\section{Correlation between feature averages for different explainability methods across $\pt$ bins}
\label{app:full_results}
The feature average metric \(\bar{F}\), introduced in the 
\Sec{sec:evaluation}, was designed to capture the mutual 
compatibility between different explanation methods across $\pt$ 
ranges. In this section we study the pearson rank correlation 
coefficients in detail and report the obtained numbers in the 
tables \Tab{tab:correlation_matrix_lund}, 
\Tab{tab:correlation_matrix_part} and 
\Tab{tab:correlation_matrix_parnet} respectively.

It's evident from the numbers that \gnnexplainer{} and \gradcam{} has a higher mutual degree of correlation. As compared to \gnnshap{}. This behavior is readily understood by the fact that the former two explainer are more prone to soft emission \(ln(\kt) \le 0\). However, \gnnshap{} puts it emphasis on hard, wide-angled radiations corresponding to hard jet substructures.

\begin{table*}[!t]
\centering
\caption{Pearson correlation coefficients between  GNNExplainer (E), GNNShap (S), and Grad-CAM (G) for each jet process, individually for each node weight averaged Lund-plane feature. The correlation numbers are quoted individually for the low $\pt$ bin  $500 < p_T < 700$ GeV, the high $\pt$ bin $800 < p_T < 1000$ GeV and the inclusive $\pt$ range $500 < p_T < 1000$ GeV. The numbers tabulated are specific to the \lundnet{} architecture.}
\label{tab:correlation_matrix_lund}
\renewcommand{\arraystretch}{1.2}
\setlength{\tabcolsep}{6pt}
\scriptsize
\begin{tabular}{ll|rrr|rrr|rrr}
\hline
\multicolumn{2}{c|}{} & \multicolumn{3}{c|}{$500 < p_T < 700$ GeV} & \multicolumn{3}{c|}{$500 < p_T < 1000$ GeV} & \multicolumn{3}{c}{$800 < p_T < 1000$ GeV} \\
Process & Feature & E$\leftrightarrow$S & G$\leftrightarrow$E & G$\leftrightarrow$S & E$\leftrightarrow$S & G$\leftrightarrow$E & G$\leftrightarrow$S & E$\leftrightarrow$S & G$\leftrightarrow$E & G$\leftrightarrow$S \\
\hline\hline
\multirow{5}{*}{\textbf{QCD}} & $\ln\Delta$ & -0.204 & +0.353 & -0.238 & -0.225 & +0.342 & -0.236 & -0.293 & +0.365 & -0.215 \\
 & $\ln k_T$ & +0.004 & +0.363 & -0.102 & +0.005 & +0.356 & -0.098 & -0.001 & +0.346 & -0.031 \\
 & $\ln m$ & +0.097 & \textbf{+0.634} & +0.063 & +0.081 & \textbf{+0.619} & +0.043 & -0.007 & \textbf{+0.595} & +0.006 \\
 & $\ln z$ & +0.121 & \textbf{+0.536} & -0.040 & +0.105 & \textbf{+0.525} & -0.051 & +0.040 & +0.463 & -0.046 \\
 & $\psi$ & -0.075 & \textbf{+0.544} & -0.064 & -0.076 & \textbf{+0.541} & -0.065 & -0.043 & \textbf{+0.558} & -0.038 \\
\hline
\multirow{5}{*}{$\mathbf{H\!\to\!c\bar{c} }$} & $\ln\Delta$ & -0.276 & +0.302 & -0.160 & -0.287 & +0.312 & -0.180 & -0.343 & +0.326 & -0.239 \\
 & $\ln k_T$ & -0.035 & +0.115 & +0.076 & -0.034 & +0.103 & +0.082 & -0.046 & +0.052 & +0.091 \\
 & $\ln m$ & +0.137 & +0.290 & -0.068 & +0.119 & +0.264 & -0.072 & +0.127 & +0.088 & -0.110 \\
 & $\ln z$ & +0.277 & +0.094 & -0.088 & +0.248 & +0.078 & -0.092 & +0.179 & -0.026 & -0.057 \\
 & $\psi$ & -0.022 & +0.474 & -0.063 & -0.031 & +0.488 & -0.063 & -0.188 & \textbf{+0.504} & -0.169 \\
\hline
\multirow{5}{*}{$\mathbf{t \rightarrow bq\bar{q'} }$} & $\ln\Delta$ & +0.386 & +0.089 & -0.108 & +0.392 & +0.057 & -0.113 & +0.451 & +0.083 & +0.044 \\
 & $\ln k_T$ & +0.171 & +0.398 & +0.249 & +0.159 & +0.394 & +0.250 & +0.047 & +0.407 & +0.151 \\
 & $\ln m$ & +0.296 & +0.432 & +0.344 & +0.295 & +0.401 & +0.347 & +0.207 & +0.338 & +0.359 \\
 & $\ln z$ & +0.406 & +0.281 & +0.135 & +0.413 & +0.238 & +0.127 & +0.420 & +0.131 & +0.184 \\
 & $\psi$ & +0.279 & +0.472 & +0.154 & +0.267 & +0.464 & +0.145 & +0.128 & +0.395 & +0.178 \\
\hline
\end{tabular}
\label{tab:explainer_corr_lundnet}
\end{table*}
\begin{table*}[!t]
\centering
\caption{Pearson correlation coefficients between  GNNExplainer (E), GNNShap (S), and Grad-CAM (G) for each jet process, individually for each node weight averaged Lund-plane feature. The correlation numbers are quoted individually for the low $\pt$ bin  $500 < p_T < 700$ GeV, the high $\pt$ bin $800 < p_T < 1000$ GeV and the inclusive $\pt$ range $500 < p_T < 1000$ GeV. The numbers tabulated are specific to the \parnet{} architecture.}
\label{tab:correlation_matrix_parnet}
\renewcommand{\arraystretch}{1.2}
\setlength{\tabcolsep}{6pt}
\scriptsize
\begin{tabular}{ll|rrr|rrr|rrr}
\hline
\multicolumn{2}{c|}{} & \multicolumn{3}{c|}{$500 < p_T < 700$ GeV} & \multicolumn{3}{c|}{$500 < p_T < 1000$ GeV} & \multicolumn{3}{c}{$800 < p_T < 1000$ GeV} \\
Process & Feature & E$\leftrightarrow$S & G$\leftrightarrow$E & G$\leftrightarrow$S & E$\leftrightarrow$S & G$\leftrightarrow$E & G$\leftrightarrow$S & E$\leftrightarrow$S & G$\leftrightarrow$E & G$\leftrightarrow$S \\
\hline\hline
\multirow{5}{*}{\textbf{QCD}} & $\ln\Delta$ & +0.249 & +0.357 & +0.162 & +0.236 & +0.344 & +0.143 & +0.278 & +0.331 & +0.065 \\
 & $\ln k_T$ & -0.088 & +0.494 & -0.136 & -0.101 & +0.485 & -0.097 & -0.207 & +0.406 & +0.039 \\
 & $\ln m$ & -0.052 & \textbf{+0.807} & -0.068 & -0.053 & \textbf{+0.811} & -0.087 & -0.154 & \textbf{+0.806} & -0.207 \\
 & $\ln z$ & +0.003 & \textbf{+0.825} & -0.091 & +0.009 & \textbf{+0.829} & -0.097 & +0.125 & \textbf{+0.852} & -0.040 \\
 & $\psi$ & +0.198 & \textbf{+0.886} & +0.112 & +0.196 & \textbf{+0.886} & +0.099 & +0.183 & \textbf{+0.877} & -0.014 \\
\hline
\multirow{5}{*}{$\mathbf{H\!\to\!c\bar{c}}$} & $\ln\Delta$ & +0.133 & +0.326 & +0.108 & +0.144 & +0.312 & +0.091 & +0.040 & +0.173 & -0.005 \\
 & $\ln k_T$ & +0.231 & \textbf{+0.617} & +0.490 & +0.245 & \textbf{+0.609} & +0.489 & +0.247 & \textbf{+0.610} & \textbf{+0.524} \\
 & $\ln m$ & +0.237 & \textbf{+0.582} & +0.368 & +0.233 & \textbf{+0.577} & +0.348 & +0.258 & \textbf{+0.592} & +0.412 \\
 & $\ln z$ & +0.097 & \textbf{+0.504} & +0.238 & +0.066 & +0.485 & +0.204 & -0.027 & +0.361 & +0.175 \\
 & $\psi$ & +0.262 & +0.328 & \textbf{+0.902} & +0.245 & +0.317 & \textbf{+0.906} & +0.429 & +0.476 & \textbf{+0.922} \\
\hline
\multirow{5}{*}{$\mathbf{t \rightarrow bq\bar{q'} }$} & $\ln\Delta$ & +0.032 & +0.060 & +0.248 & +0.040 & +0.124 & +0.264 & +0.202 & +0.129 & +0.010 \\
 & $\ln k_T$ & +0.169 & +0.498 & +0.170 & +0.189 & \textbf{+0.503} & +0.175 & +0.229 & +0.440 & -0.015 \\
 & $\ln m$ & +0.162 & +0.442 & +0.352 & +0.155 & +0.451 & +0.343 & -0.010 & \textbf{+0.619} & +0.198 \\
 & $\ln z$ & +0.072 & +0.121 & \textbf{+0.501} & +0.078 & +0.134 & \textbf{+0.500} & +0.377 & \textbf{+0.531} & \textbf{+0.557} \\
 & $\psi$ & +0.202 & +0.245 & \textbf{+0.585} & +0.208 & +0.243 & \textbf{+0.612} & +0.271 & +0.279 & \textbf{+0.805} \\
\hline
\end{tabular}
\label{tab:explainer_corr_particlenet}
\end{table*}
\begin{table*}[!t]
\centering
\caption{Pearson correlation coefficients between  GNNExplainer (E), GNNShap (S), and Grad-CAM (G) for each jet process, individually for each node weight averaged Lund-plane feature. The correlation numbers are quoted individually for the low $\pt$ bin  $500 < p_T < 700$ GeV, the high $\pt$ bin $800 < p_T < 1000$ GeV and the inclusive $\pt$ range $500 < p_T < 1000$ GeV. The numbers tabulated are specific to the \part{} architecture.}
\label{tab:correlation_matrix_part}
\renewcommand{\arraystretch}{1.2}
\setlength{\tabcolsep}{6pt}
\scriptsize
\begin{tabular}{ll|rrr|rrr|rrr}
\hline
\multicolumn{2}{c|}{} & \multicolumn{3}{c|}{$500 < p_T < 700$ GeV} & \multicolumn{3}{c|}{$500 < p_T < 1000$ GeV} & \multicolumn{3}{c}{$800 < p_T < 1000$ GeV} \\
Process & Feature & E$\leftrightarrow$S & G$\leftrightarrow$E & G$\leftrightarrow$S & E$\leftrightarrow$S & G$\leftrightarrow$E & G$\leftrightarrow$S & E$\leftrightarrow$S & G$\leftrightarrow$E & G$\leftrightarrow$S \\
\hline\hline
\multirow{5}{*}{\textbf{QCD}} & $\ln\Delta$ & +0.121 & +0.104 & +0.160 & +0.120 & +0.122 & +0.148 & +0.128 & -0.080 & +0.001 \\
 & $\ln k_T$ & -0.190 & +0.235 & +0.020 & -0.168 & +0.261 & +0.031 & -0.096 & +0.309 & +0.071 \\
 & $\ln m$ & +0.056 & \textbf{+0.658} & -0.061 & +0.078 & \textbf{+0.661} & -0.050 & +0.217 & \textbf{+0.577} & +0.020 \\
 & $\ln z$ & +0.118 & \textbf{+0.756} & +0.047 & +0.122 & \textbf{+0.759} & +0.042 & +0.201 & \textbf{+0.778} & +0.045 \\
 & $\psi$ & +0.188 & \textbf{+0.517} & -0.002 & +0.181 & \textbf{+0.506} & -0.015 & +0.058 & \textbf{+0.516} & -0.048 \\
\hline
\multirow{5}{*}{$\mathbf{H\!\to\!c\bar{c}}$} & $\ln\Delta$ & -0.079 & +0.071 & +0.196 & -0.022 & +0.093 & +0.210 & +0.019 & -0.180 & +0.110 \\
 & $\ln k_T$ & \textbf{+0.654} & +0.263 & +0.186 & \textbf{+0.631} & +0.259 & +0.198 & +0.494 & +0.159 & +0.210 \\
 & $\ln m$ & \textbf{+0.631} & +0.403 & +0.250 & \textbf{+0.590} & +0.420 & +0.225 & \textbf{+0.501} & +0.455 & +0.200 \\
 & $\ln z$ & +0.374 & +0.324 & +0.128 & +0.348 & +0.324 & +0.103 & +0.415 & +0.392 & +0.158 \\
 & $\psi$ & +0.337 & \textbf{+0.618} & +0.231 & +0.323 & \textbf{+0.631} & +0.231 & +0.363 & \textbf{+0.622} & +0.250 \\
\hline
\multirow{5}{*}{$\mathbf{t \rightarrow bq\bar{q'} }$} & $\ln\Delta$ & +0.270 & +0.401 & +0.244 & +0.310 & +0.415 & +0.261 & -0.026 & +0.058 & +0.067 \\
 & $\ln k_T$ & +0.452 & +0.494 & +0.074 & +0.447 & +0.497 & +0.079 & +0.170 & +0.489 & -0.166 \\
 & $\ln m$ & +0.336 & +0.241 & +0.320 & +0.306 & +0.231 & +0.315 & +0.273 & \textbf{+0.611} & -0.068 \\
 & $\ln z$ & +0.029 & -0.077 & +0.035 & +0.021 & -0.105 & +0.047 & +0.304 & +0.315 & +0.065 \\
 & $\psi$ & +0.245 & \textbf{+0.683} & +0.245 & +0.230 & \textbf{+0.674} & +0.230 & +0.198 & \textbf{+0.623} & +0.114 \\
\hline
\end{tabular}
\label{tab:explainer_corr_particle_transformer}
\end{table*}


\begin{figure*}[!t]
    \includegraphics[width=0.32\textwidth]{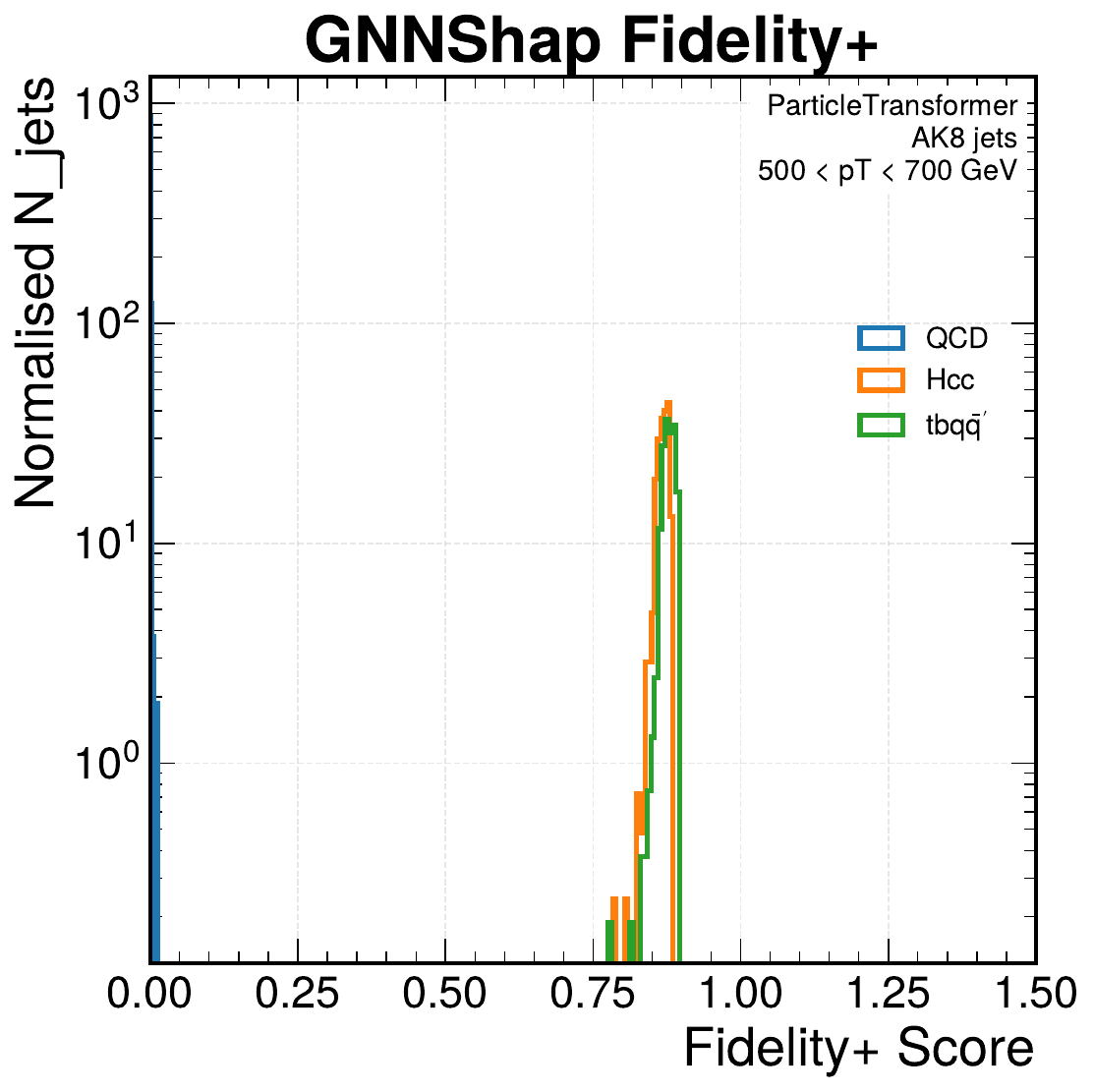}
    \includegraphics[width=0.32\textwidth]{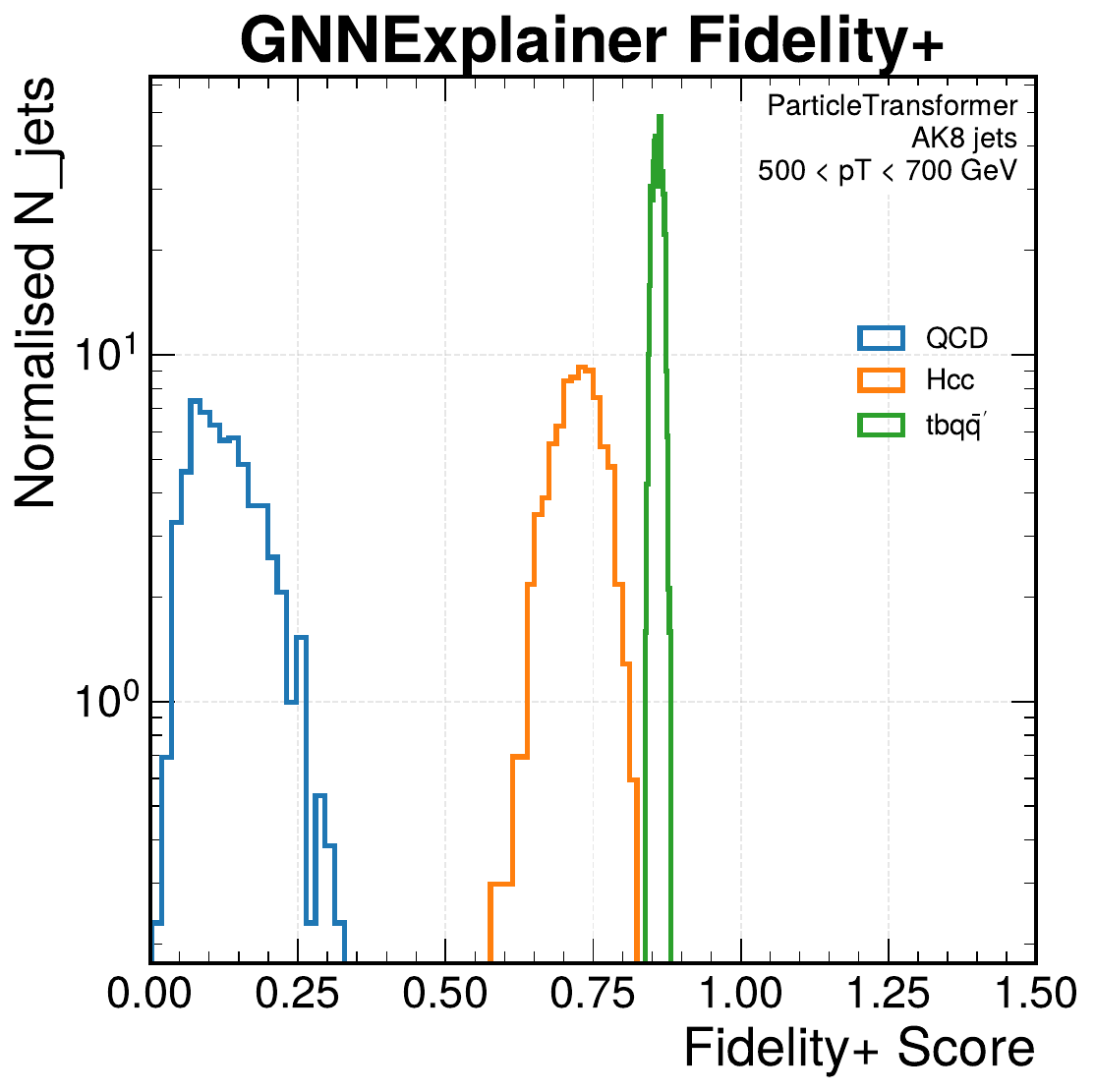}
    \includegraphics[width=0.32\textwidth]{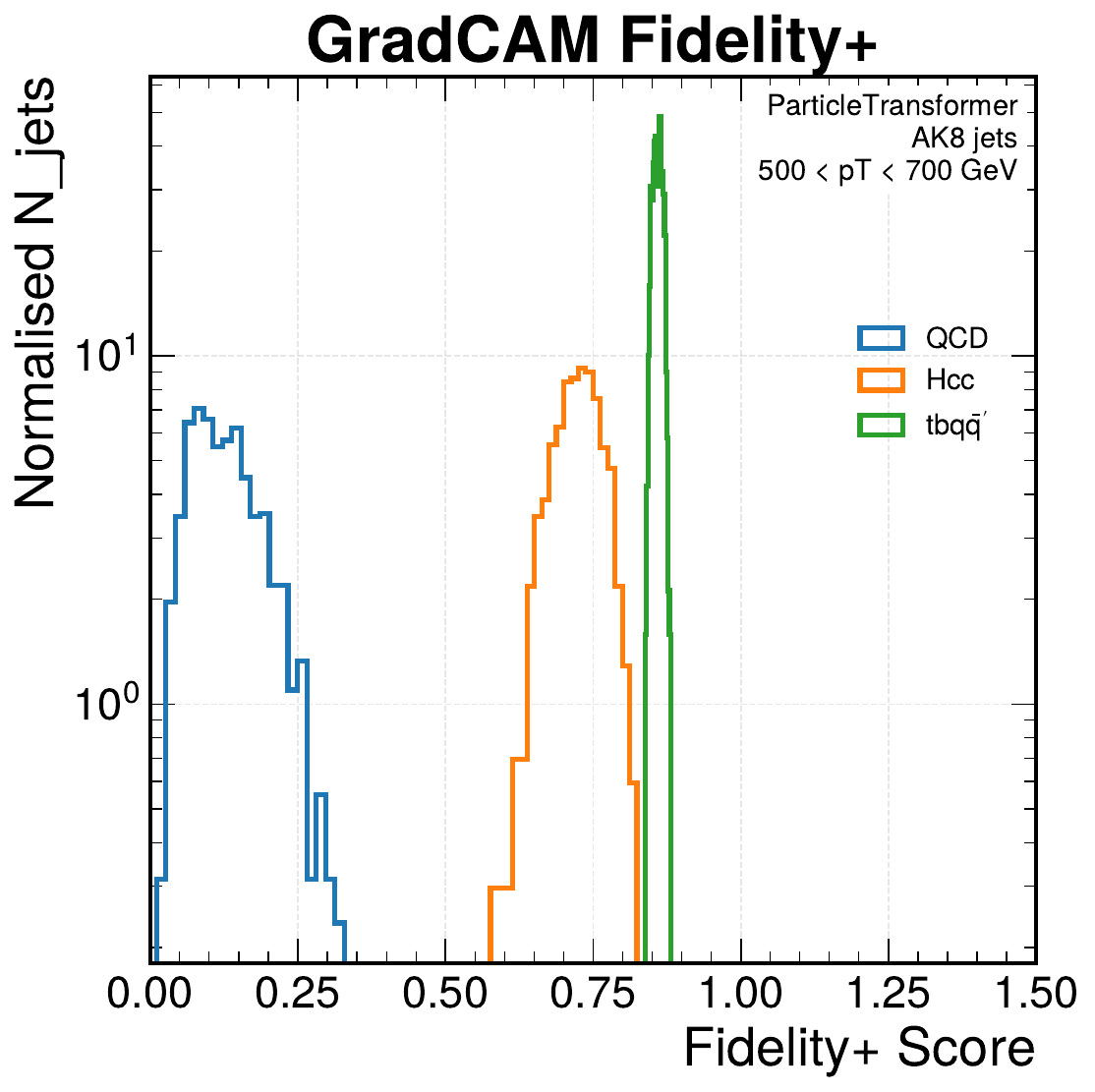}
    \includegraphics[width=0.32\textwidth]{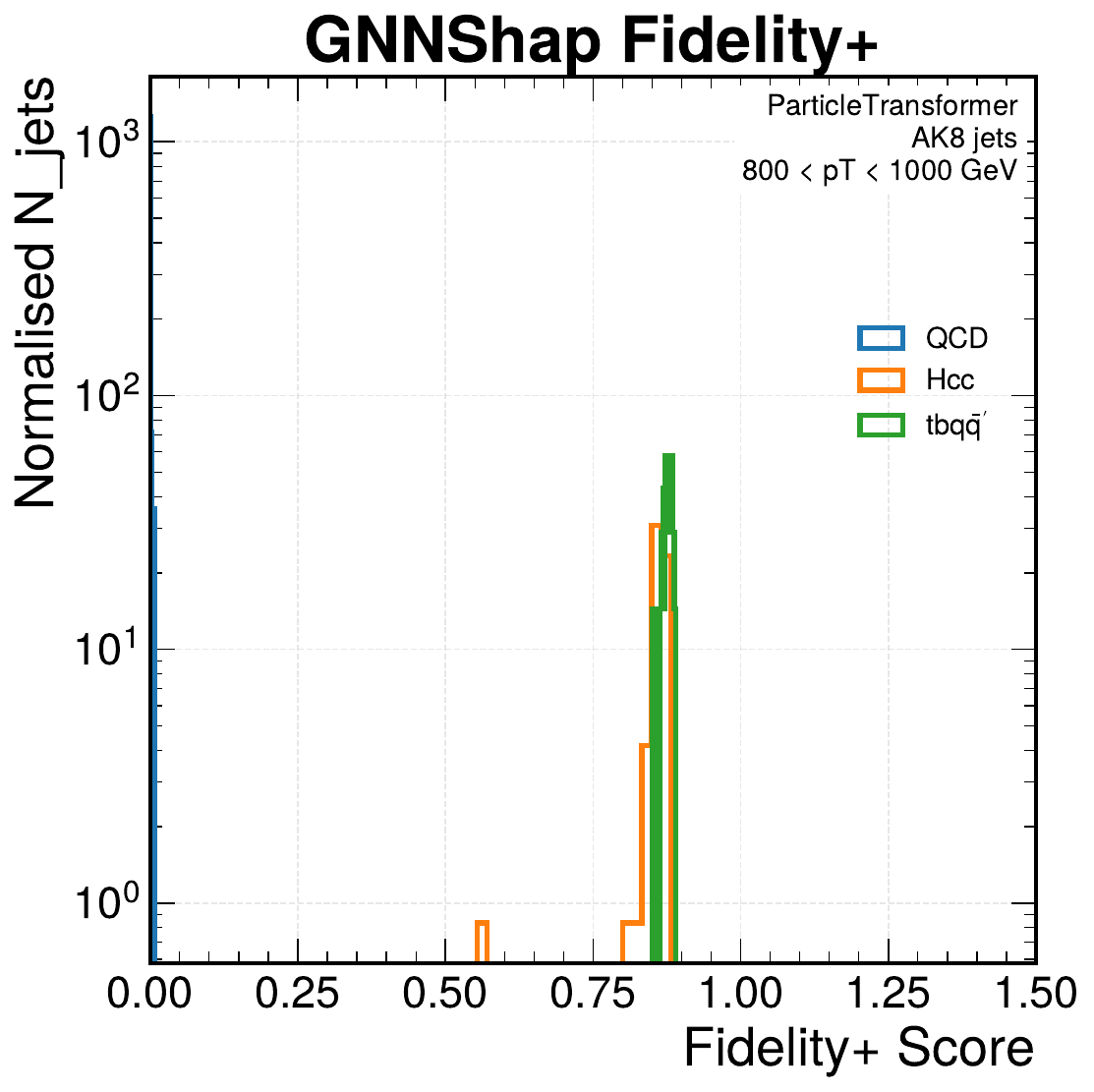}
    \includegraphics[width=0.32\textwidth]{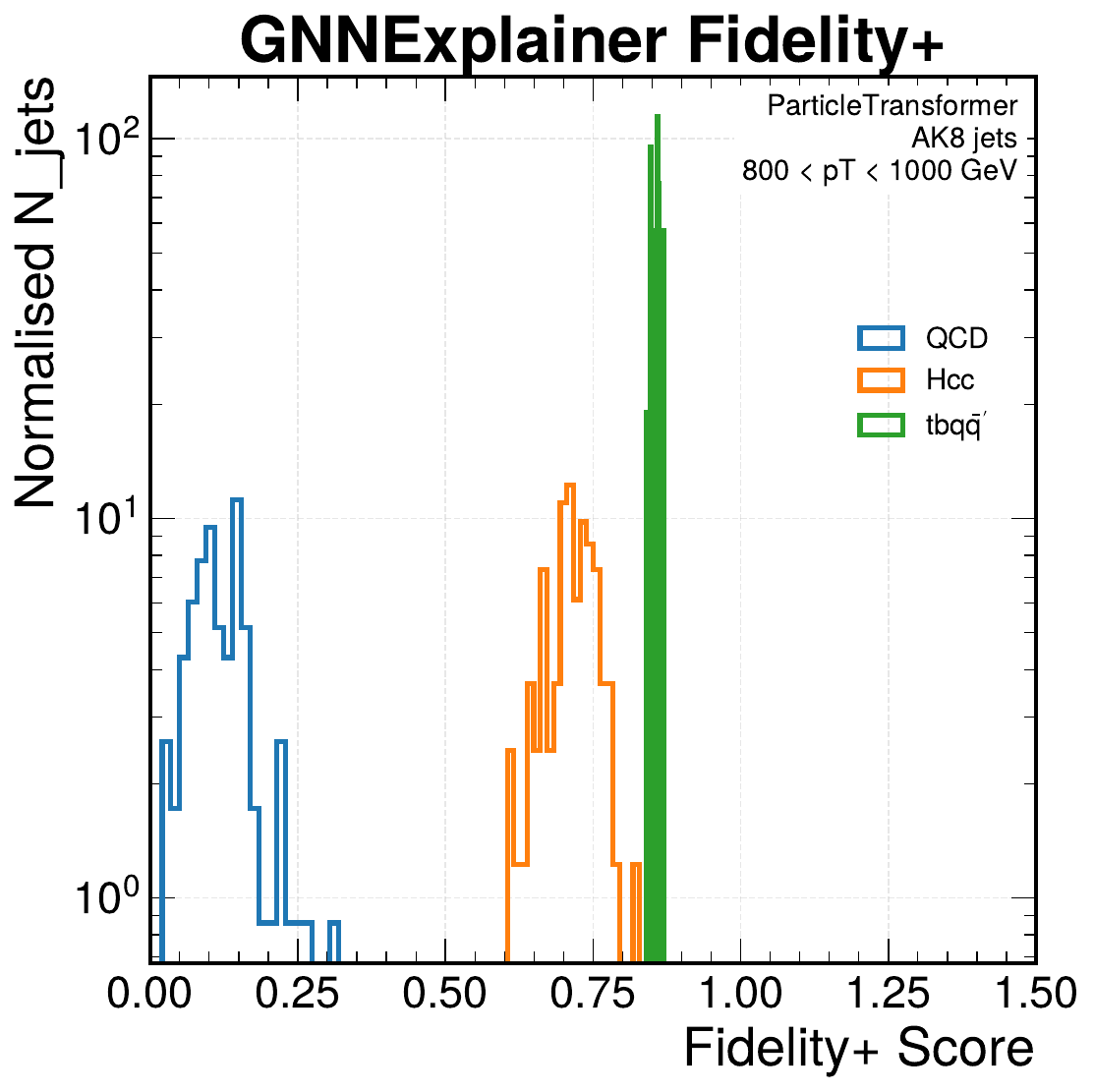}
    \includegraphics[width=0.32\textwidth]{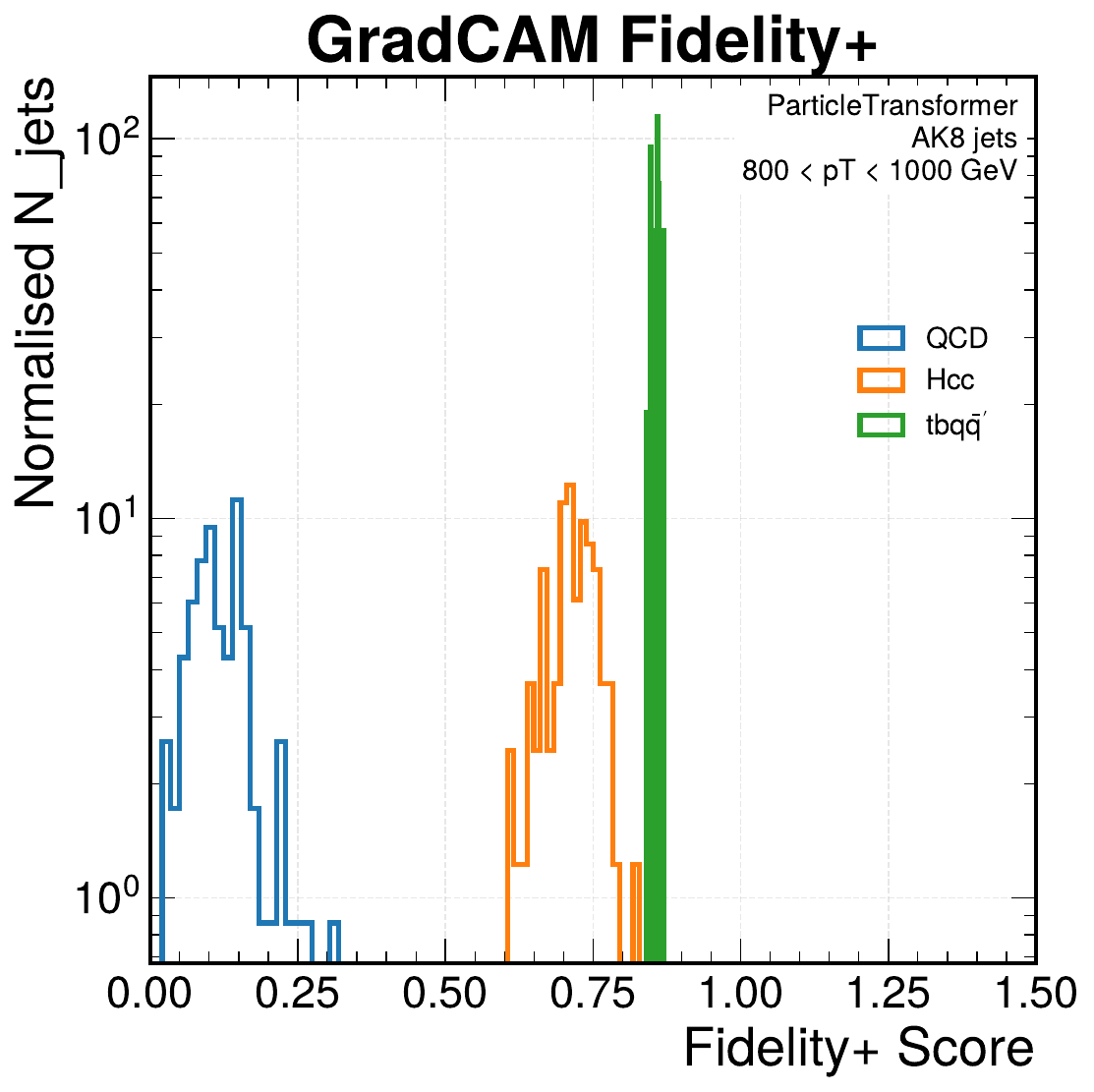}
    \caption{The distribution of the $\fidp$ score for the \part{} network along with \gnnexplainer{} and \gradcam{} explainability method. The reduction of the width of fidelity 
    scores at higher $\pt$ indicates that for a collimated jet, a subgraph captures a higher
    proportion of the core explanation weights.}
    \label{fig:fidelityplus_pt_dependenc}
\end{figure*}

\section{Correlation between jet-substructure observables and learned feature averages}
\label{app:weighted_scatter_tau}

In this section we study the pearson rank correlation 
coefficients between jet substructure observables \(\tau_{21}\,, \tau_{32}\,,C_2\,,C_3\) vs \(\bar{F}\) in detail and report the obtained 
numbers in the tables \Tab{tab:tau_corr_pt_tau_21_lund} - \Tab{tab:C_corr_pt_C_3_parnet}.




\begin{table*}[!t]
\centering
\caption{Pearson correlation coefficients $\rho$ between explainer importance scores and $\tau_{21}$ for each jet process, Lund-plane feature, and $p_T$ bin. The numbers quoted are for \lundnet{} architecture.}
\label{tab:tau_corr_pt_tau_21_lund}
\renewcommand{\arraystretch}{1.2}
\setlength{\tabcolsep}{6pt}
\scriptsize
\begin{tabular}{ll|rrr|rrr|rrr}
\hline
\multicolumn{2}{c|}{} & \multicolumn{3}{c|}{$500 < p_T < 700$ GeV} & \multicolumn{3}{c|}{$500 < p_T < 1000$ GeV} & \multicolumn{3}{c}{$800 < p_T < 1000$ GeV} \\
Process & Feature & GNNExpl. & GNN-SHAP & GradCAM & GNNExpl. & GNN-SHAP & GradCAM & GNNExpl. & GNN-SHAP & GradCAM \\
\hline\hline
\multirow{5}{*}{\textbf{QCD}} & $\ln\Delta$ & -0.04 & -0.02 & -0.13 & -0.05 & -0.04 & -0.13 & -0.06 & -0.07 & -0.20 \\
 & $\ln k_T$ & 0.02 & 0.24 & -0.05 & 0.01 & 0.22 & -0.06 & 0.05 & 0.11 & -0.08 \\
 & $\ln m$ & 0.10 & 0.23 & 0.20 & 0.08 & 0.23 & 0.18 & 0.07 & 0.11 & 0.19 \\
 & $\ln z$ & -0.16 & -0.15 & -0.22 & -0.15 & -0.16 & -0.21 & -0.11 & -0.09 & -0.22 \\
 & $\psi$ & 0.01 & -0.03 & -0.03 & 0.01 & -0.02 & -0.04 & -0.04 & 0.02 & -0.07 \\
\hline
\multirow{5}{*}{$\mathbf{H\!\to\!c\bar{c} }$} & $\ln\Delta$ & 0.38 & -0.10 & 0.16 & 0.33 & -0.14 & 0.18 & 0.37 & -0.36 & 0.33 \\
 & $\ln k_T$ & 0.02 & -0.17 & 0.16 & 0.01 & -0.14 & 0.16 & -0.08 & 0.11 & 0.17 \\
 & $\ln m$ & -0.22 & -0.12 & 0.11 & -0.20 & -0.09 & 0.09 & -0.16 & 0.17 & -0.05 \\
 & $\ln z$ & 0.41 & 0.07 & 0.04 & 0.37 & 0.03 & 0.07 & 0.19 & -0.20 & 0.19 \\
 & $\psi$ & 0.02 & -0.15 & -0.04 & 0.01 & -0.17 & -0.04 & -0.03 & -0.19 & 0.04 \\
\hline
\multirow{5}{*}{$\mathbf{t \rightarrow bq\bar{q'} }$} & $\ln\Delta$ & 0.21 & 0.17 & -0.03 & 0.20 & 0.18 & -0.03 & 0.21 & 0.20 & 0.14 \\
 & $\ln k_T$ & 0.01 & -0.10 & -0.15 & 0.01 & -0.12 & -0.15 & 0.02 & -0.17 & -0.02 \\
 & $\ln m$ & -0.13 & -0.12 & -0.23 & -0.13 & -0.14 & -0.24 & -0.10 & -0.19 & -0.16 \\
 & $\ln z$ & 0.24 & 0.13 & 0.15 & 0.25 & 0.14 & 0.16 & 0.27 & 0.15 & 0.20 \\
 & $\psi$ & -0.00 & 0.01 & 0.05 & 0.01 & 0.02 & 0.06 & 0.03 & 0.07 & 0.13 \\
\hline
\end{tabular}
\end{table*}
\begin{table*}[!t]
\centering
\caption{Pearson correlation coefficients $\rho$ between explainer importance scores and $\tau_{32}$ for each jet process, Lund-plane feature, and $p_T$ bin. The numbers quoted are for \lundnet{} architecture.}
\label{tab:tau_corr_pt_tau_32_lund}
\renewcommand{\arraystretch}{1.2}
\setlength{\tabcolsep}{6pt}
\scriptsize
\begin{tabular}{ll|rrr|rrr|rrr}
\hline
\multicolumn{2}{c|}{} & \multicolumn{3}{c|}{$500 < p_T < 700$ GeV} & \multicolumn{3}{c|}{$500 < p_T < 1000$ GeV} & \multicolumn{3}{c}{$800 < p_T < 1000$ GeV} \\
Process & Feature & GNNExpl. & GNN-SHAP & GradCAM & GNNExpl. & GNN-SHAP & GradCAM & GNNExpl. & GNN-SHAP & GradCAM \\
\hline\hline
\multirow{5}{*}{\textbf{QCD}} & $\ln\Delta$ & -0.04 & -0.02 & 0.02 & -0.04 & -0.02 & 0.01 & -0.13 & -0.04 & -0.09 \\
 & $\ln k_T$ & -0.15 & 0.01 & -0.07 & -0.15 & 0.04 & -0.07 & -0.22 & 0.16 & -0.19 \\
 & $\ln m$ & -0.16 & 0.01 & -0.07 & -0.15 & 0.04 & -0.06 & -0.18 & 0.20 & -0.09 \\
 & $\ln z$ & 0.10 & -0.02 & 0.04 & 0.09 & -0.03 & 0.02 & 0.08 & -0.16 & -0.01 \\
 & $\psi$ & 0.02 & -0.03 & 0.04 & 0.02 & -0.02 & 0.03 & -0.05 & -0.00 & -0.02 \\
\hline
\multirow{5}{*}{$\mathbf{H\!\to\!c\bar{c} }$} & $\ln\Delta$ & -0.02 & -0.10 & 0.06 & -0.01 & -0.09 & 0.05 & -0.06 & 0.01 & -0.01 \\
 & $\ln k_T$ & -0.03 & 0.23 & 0.02 & -0.04 & 0.21 & 0.02 & -0.07 & 0.13 & 0.01 \\
 & $\ln m$ & 0.06 & 0.23 & -0.07 & 0.04 & 0.21 & -0.07 & -0.01 & 0.15 & -0.08 \\
 & $\ln z$ & -0.17 & -0.18 & 0.07 & -0.15 & -0.17 & 0.07 & -0.10 & -0.10 & 0.07 \\
 & $\psi$ & -0.04 & 0.05 & 0.00 & -0.04 & 0.06 & 0.01 & -0.11 & 0.07 & -0.01 \\
\hline
\multirow{5}{*}{$\mathbf{t \rightarrow bq\bar{q'} }$} & $\ln\Delta$ & 0.31 & 0.46 & 0.04 & 0.30 & 0.46 & 0.03 & 0.45 & 0.48 & -0.12 \\
 & $\ln k_T$ & -0.00 & -0.44 & -0.15 & 0.01 & -0.42 & -0.14 & 0.07 & -0.21 & -0.11 \\
 & $\ln m$ & -0.18 & -0.48 & -0.32 & -0.17 & -0.47 & -0.30 & -0.17 & -0.24 & -0.09 \\
 & $\ln z$ & 0.35 & 0.34 & 0.28 & 0.33 & 0.32 & 0.26 & 0.25 & 0.16 & -0.05 \\
 & $\psi$ & -0.02 & -0.07 & 0.10 & -0.01 & -0.06 & 0.11 & -0.01 & 0.02 & 0.03 \\
\hline
\end{tabular}
\end{table*}

\begin{table*}[!t]
\centering
\caption{Pearson correlation coefficients $\rho$ between explainer importance scores and $\tau_{21}$ for each jet process, Lund-plane feature, and $p_T$ bin. The numbers quoted are for \parnet{} architecture.}
\label{tab:tau_corr_pt_tau_21_parnet}
\renewcommand{\arraystretch}{1.2}
\setlength{\tabcolsep}{6pt}
\scriptsize
\begin{tabular}{ll|rrr|rrr|rrr}
\hline
\multicolumn{2}{c|}{} & \multicolumn{3}{c|}{$500 < p_T < 700$ GeV} & \multicolumn{3}{c|}{$500 < p_T < 1000$ GeV} & \multicolumn{3}{c}{$800 < p_T < 1000$ GeV} \\
Process & Feature & GNNExpl. & GNN-SHAP & GradCAM & GNNExpl. & GNN-SHAP & GradCAM & GNNExpl. & GNN-SHAP & GradCAM \\
\hline\hline
\multirow{5}{*}{\textbf{QCD}} & $\ln\Delta$ & 0.01 & 0.14 & -0.06 & -0.01 & 0.13 & -0.09 & -0.10 & 0.21 & -0.33 \\
 & $\ln k_T$ & -0.05 & 0.14 & -0.21 & -0.07 & 0.14 & -0.23 & -0.06 & 0.12 & -0.42 \\
 & $\ln m$ & -0.03 & 0.06 & 0.06 & -0.04 & 0.04 & 0.08 & 0.05 & -0.19 & 0.28 \\
 & $\ln z$ & -0.03 & 0.02 & -0.13 & -0.03 & 0.04 & -0.15 & -0.17 & 0.25 & -0.36 \\
 & $\psi$ & 0.01 & 0.00 & 0.03 & -0.02 & 0.01 & 0.01 & -0.23 & -0.02 & -0.18 \\
\hline
\multirow{5}{*}{$\mathbf{H\!\to\!c\bar{c} }$} & $\ln\Delta$ & 0.31 & 0.19 & 0.39 & 0.22 & 0.15 & 0.36 & 0.31 & 0.04 & 0.43 \\
 & $\ln k_T$ & 0.11 & -0.27 & -0.40 & 0.10 & -0.26 & -0.41 & 0.10 & -0.21 & -0.42 \\
 & $\ln m$ & -0.15 & -0.22 & -0.44 & -0.16 & -0.19 & -0.46 & -0.09 & 0.03 & -0.39 \\
 & $\ln z$ & 0.39 & 0.05 & 0.30 & 0.37 & 0.01 & 0.29 & 0.21 & -0.26 & 0.10 \\
 & $\psi$ & 0.09 & 0.06 & -0.03 & 0.08 & 0.05 & -0.03 & 0.00 & 0.03 & -0.05 \\
\hline
\multirow{5}{*}{$\mathbf{t \rightarrow bq\bar{q'} }$} & $\ln\Delta$ & 0.10 & 0.06 & 0.08 & 0.13 & 0.06 & 0.07 & 0.37 & 0.05 & 0.26 \\
 & $\ln k_T$ & -0.00 & -0.06 & -0.18 & 0.02 & -0.06 & -0.18 & 0.08 & -0.11 & -0.34 \\
 & $\ln m$ & -0.07 & -0.07 & -0.13 & -0.06 & -0.08 & -0.14 & -0.08 & -0.19 & -0.34 \\
 & $\ln z$ & 0.11 & 0.06 & 0.02 & 0.11 & 0.07 & 0.02 & 0.26 & 0.27 & 0.27 \\
 & $\psi$ & 0.03 & -0.03 & -0.02 & 0.03 & -0.01 & -0.02 & -0.03 & 0.05 & -0.07 \\
\hline
\end{tabular}
\end{table*}
\begin{table*}[!t]
\centering
\caption{Pearson correlation coefficients $\rho$ between explainer importance scores and $\tau_{32}$ for each jet process, Lund-plane feature, and $p_T$ bin. The numbers quoted are for \parnet{} architecture.}
\label{tab:tau_corr_pt_tau_32_parnet}
\renewcommand{\arraystretch}{1.2}
\setlength{\tabcolsep}{6pt}
\scriptsize
\begin{tabular}{ll|rrr|rrr|rrr}
\hline
\multicolumn{2}{c|}{} & \multicolumn{3}{c|}{$500 < p_T < 700$ GeV} & \multicolumn{3}{c|}{$500 < p_T < 1000$ GeV} & \multicolumn{3}{c}{$800 < p_T < 1000$ GeV} \\
Process & Feature & GNNExpl. & GNN-SHAP & GradCAM & GNNExpl. & GNN-SHAP & GradCAM & GNNExpl. & GNN-SHAP & GradCAM \\
\hline\hline
\multirow{5}{*}{\textbf{QCD}} & $\ln\Delta$ & 0.02 & 0.11 & 0.21 & 0.04 & 0.12 & 0.22 & 0.18 & 0.19 & 0.22 \\
 & $\ln k_T$ & -0.09 & 0.18 & 0.02 & -0.08 & 0.20 & 0.04 & -0.10 & 0.29 & 0.05 \\
 & $\ln m$ & -0.15 & 0.17 & -0.16 & -0.16 & 0.16 & -0.16 & -0.25 & 0.09 & -0.13 \\
 & $\ln z$ & 0.15 & -0.10 & 0.16 & 0.16 & -0.09 & 0.17 & 0.15 & 0.01 & 0.11 \\
 & $\psi$ & 0.08 & -0.06 & 0.08 & 0.08 & -0.03 & 0.07 & -0.02 & -0.01 & -0.03 \\
\hline
\multirow{5}{*}{$\mathbf{H\!\to\!c\bar{c} }$} & $\ln\Delta$ & -0.06 & 0.11 & -0.13 & -0.02 & 0.11 & -0.12 & -0.18 & 0.02 & -0.15 \\
 & $\ln k_T$ & -0.08 & -0.06 & -0.05 & -0.08 & -0.07 & -0.05 & -0.17 & -0.12 & -0.02 \\
 & $\ln m$ & 0.05 & -0.09 & 0.05 & 0.04 & -0.10 & 0.07 & -0.12 & -0.16 & 0.09 \\
 & $\ln z$ & -0.19 & 0.07 & -0.17 & -0.18 & 0.06 & -0.18 & -0.00 & 0.10 & -0.10 \\
 & $\psi$ & -0.01 & 0.04 & 0.05 & -0.02 & 0.03 & 0.05 & -0.18 & 0.01 & 0.02 \\
\hline
\multirow{5}{*}{$\mathbf{t \rightarrow bq\bar{q'} }$} & $\ln\Delta$ & 0.24 & 0.22 & 0.32 & 0.20 & 0.20 & 0.28 & 0.42 & 0.28 & 0.43 \\
 & $\ln k_T$ & -0.00 & -0.09 & -0.47 & 0.00 & -0.09 & -0.47 & -0.01 & -0.13 & \textbf{-0.54} \\
 & $\ln m$ & -0.23 & -0.13 & -0.34 & -0.22 & -0.13 & -0.35 & -0.31 & -0.31 & -0.48 \\
 & $\ln z$ & 0.34 & 0.11 & -0.03 & 0.32 & 0.11 & -0.00 & 0.37 & 0.38 & 0.22 \\
 & $\psi$ & -0.03 & -0.01 & -0.05 & -0.03 & -0.01 & -0.06 & -0.06 & -0.24 & -0.12 \\
\hline
\end{tabular}
\end{table*}

\begin{table*}[!t]
\centering
\caption{Pearson correlation coefficients $\rho$ between explainer importance scores and $\tau_{21}$ for each jet process, Lund-plane feature, and $p_T$ bin. The numbers quoted are for \part{} architecture.}
\label{tab:tau_corr_pt_tau_21_part}
\renewcommand{\arraystretch}{1.2}
\setlength{\tabcolsep}{6pt}
\scriptsize
\begin{tabular}{ll|rrr|rrr|rrr}
\hline
\multicolumn{2}{c|}{} & \multicolumn{3}{c|}{$500 < p_T < 700$ GeV} & \multicolumn{3}{c|}{$500 < p_T < 1000$ GeV} & \multicolumn{3}{c}{$800 < p_T < 1000$ GeV} \\
Process & Feature & GNNExpl. & GNN-SHAP & GradCAM & GNNExpl. & GNN-SHAP & GradCAM & GNNExpl. & GNN-SHAP & GradCAM \\
\hline\hline
\multirow{5}{*}{\textbf{QCD}} & $\ln\Delta$ & 0.06 & 0.18 & 0.17 & 0.01 & 0.15 & 0.16 & -0.08 & 0.15 & 0.35 \\
 & $\ln k_T$ & -0.03 & 0.18 & -0.02 & -0.04 & 0.15 & -0.06 & -0.01 & 0.06 & -0.06 \\
 & $\ln m$ & -0.01 & -0.12 & 0.05 & -0.02 & -0.11 & 0.01 & 0.02 & -0.15 & -0.20 \\
 & $\ln z$ & -0.05 & 0.23 & -0.08 & -0.04 & 0.20 & -0.06 & -0.06 & 0.20 & 0.07 \\
 & $\psi$ & 0.01 & 0.04 & -0.00 & -0.01 & 0.05 & -0.03 & -0.10 & 0.13 & -0.18 \\
\hline
\multirow{5}{*}{$\mathbf{H\!\to\!c\bar{c} }$} & $\ln\Delta$ & 0.13 & 0.44 & 0.10 & 0.10 & 0.43 & 0.07 & 0.43 & \textbf{0.58} & 0.15 \\
 & $\ln k_T$ & -0.01 & -0.36 & -0.08 & 0.01 & -0.36 & -0.10 & 0.17 & -0.23 & -0.33 \\
 & $\ln m$ & -0.21 & -0.36 & -0.19 & -0.18 & -0.36 & -0.19 & -0.06 & -0.26 & -0.33 \\
 & $\ln z$ & 0.32 & 0.27 & 0.08 & 0.32 & 0.29 & 0.08 & 0.38 & 0.27 & 0.01 \\
 & $\psi$ & -0.04 & 0.02 & -0.05 & -0.03 & 0.01 & -0.06 & 0.13 & 0.01 & 0.06 \\
\hline
\multirow{5}{*}{$\mathbf{t \rightarrow bq\bar{q'} }$} & $\ln\Delta$ & 0.14 & 0.03 & 0.21 & 0.17 & 0.03 & 0.20 & \textbf{0.52} & -0.09 & 0.03 \\
 & $\ln k_T$ & 0.06 & -0.08 & 0.12 & 0.09 & -0.07 & 0.12 & 0.15 & -0.27 & 0.17 \\
 & $\ln m$ & -0.14 & -0.09 & -0.04 & -0.12 & -0.08 & -0.02 & -0.06 & -0.26 & 0.17 \\
 & $\ln z$ & 0.17 & 0.05 & 0.19 & 0.18 & 0.05 & 0.15 & 0.19 & 0.01 & -0.06 \\
 & $\psi$ & 0.03 & 0.05 & 0.02 & 0.03 & 0.06 & 0.02 & 0.22 & -0.08 & 0.03 \\
\hline
\end{tabular}
\end{table*}
\begin{table*}[!t]
\centering
\caption{Pearson correlation coefficients $\rho$ between explainer importance scores and $\tau_{32}$ for each jet process, Lund-plane feature, and $p_T$ bin. The numbers quoted are for \part{} architecture.}
\label{tab:tau_corr_pt_tau_32_part}
\renewcommand{\arraystretch}{1.2}
\setlength{\tabcolsep}{6pt}
\scriptsize
\begin{tabular}{ll|rrr|rrr|rrr}
\hline
\multicolumn{2}{c|}{} & \multicolumn{3}{c|}{$500 < p_T < 700$ GeV} & \multicolumn{3}{c|}{$500 < p_T < 1000$ GeV} & \multicolumn{3}{c}{$800 < p_T < 1000$ GeV} \\
Process & Feature & GNNExpl. & GNN-SHAP & GradCAM & GNNExpl. & GNN-SHAP & GradCAM & GNNExpl. & GNN-SHAP & GradCAM \\
\hline\hline
\multirow{5}{*}{\textbf{QCD}} & $\ln\Delta$ & -0.05 & 0.31 & 0.15 & -0.03 & 0.31 & 0.17 & 0.03 & 0.38 & 0.17 \\
 & $\ln k_T$ & -0.16 & 0.26 & -0.07 & -0.14 & 0.26 & -0.06 & -0.15 & 0.19 & -0.09 \\
 & $\ln m$ & -0.19 & -0.15 & -0.09 & -0.18 & -0.16 & -0.08 & -0.18 & -0.21 & -0.17 \\
 & $\ln z$ & 0.15 & 0.33 & 0.07 & 0.14 & 0.33 & 0.07 & 0.06 & 0.32 & 0.10 \\
 & $\psi$ & 0.08 & -0.01 & 0.01 & 0.08 & 0.01 & -0.00 & -0.08 & 0.03 & -0.11 \\
\hline
\multirow{5}{*}{$\mathbf{H\!\to\!c\bar{c} }$} & $\ln\Delta$ & -0.11 & -0.01 & -0.11 & -0.09 & -0.03 & -0.10 & -0.23 & -0.23 & -0.22 \\
 & $\ln k_T$ & -0.06 & -0.12 & 0.07 & -0.07 & -0.13 & 0.08 & -0.13 & -0.17 & 0.24 \\
 & $\ln m$ & 0.12 & -0.07 & 0.20 & 0.10 & -0.05 & 0.20 & 0.09 & -0.03 & 0.13 \\
 & $\ln z$ & -0.25 & -0.00 & -0.08 & -0.24 & -0.04 & -0.08 & -0.29 & -0.15 & -0.01 \\
 & $\psi$ & 0.01 & 0.00 & 0.00 & 0.01 & -0.01 & 0.01 & -0.15 & -0.04 & -0.14 \\
\hline
\multirow{5}{*}{$\mathbf{t \rightarrow bq\bar{q'} }$} & $\ln\Delta$ & -0.11 & 0.15 & 0.18 & -0.11 & 0.14 & 0.14 & 0.08 & 0.27 & 0.02 \\
 & $\ln k_T$ & -0.17 & \textbf{-0.58} & 0.06 & -0.16 & \textbf{-0.57} & 0.04 & -0.03 & -0.37 & -0.19 \\
 & $\ln m$ & -0.15 & \textbf{-0.50} & -0.17 & -0.14 & \textbf{-0.50} & -0.18 & -0.14 & -0.33 & -0.11 \\
 & $\ln z$ & -0.02 & 0.07 & 0.32 & -0.00 & 0.09 & 0.31 & 0.41 & 0.09 & 0.25 \\
 & $\psi$ & -0.01 & -0.04 & -0.03 & -0.01 & -0.05 & -0.03 & -0.02 & -0.11 & -0.01 \\
\hline
\end{tabular}
\end{table*}

\label{app:nsubjettiness_scatter_ecf}
\begin{table*}[!t]
\centering
\caption{Pearson correlation coefficients $\rho$ between explainer importance scores and $C_2$ for each jet process, Lund-plane feature, and $p_T$ bin. The numbers quoted are for \lundnet{} architecture.}
\label{tab:C_corr_pt_C_2_lundnet}
\renewcommand{\arraystretch}{1.2}
\setlength{\tabcolsep}{6pt}
\scriptsize
\begin{tabular}{ll|rrr|rrr|rrr}
\hline
\multicolumn{2}{c|}{} & \multicolumn{3}{c|}{$500 < p_T < 700$ GeV} & \multicolumn{3}{c|}{$500 < p_T < 1000$ GeV} & \multicolumn{3}{c}{$800 < p_T < 1000$ GeV} \\
Process & Feature & GNNExpl. & GNN-SHAP & GradCAM & GNNExpl. & GNN-SHAP & GradCAM & GNNExpl. & GNN-SHAP & GradCAM \\
\hline\hline
\multirow{5}{*}{\textbf{QCD}} & $\ln\Delta$ & \textbf{0.59} & -0.07 & 0.28 & \textbf{0.59} & -0.09 & 0.27 & \textbf{0.60} & -0.13 & 0.24 \\
 & $\ln k_T$ & 0.20 & -0.14 & 0.25 & 0.19 & -0.16 & 0.24 & 0.17 & -0.31 & 0.22 \\
 & $\ln m$ & 0.12 & -0.19 & 0.10 & 0.11 & -0.20 & 0.10 & 0.11 & -0.32 & 0.08 \\
 & $\ln z$ & -0.00 & 0.17 & 0.06 & -0.00 & 0.17 & 0.05 & -0.03 & 0.26 & 0.04 \\
 & $\psi$ & 0.02 & 0.04 & -0.05 & 0.00 & 0.03 & -0.05 & -0.08 & -0.01 & -0.02 \\
\hline
\multirow{5}{*}{$\mathbf{H\!\to\!c\bar{c} }$} & $\ln\Delta$ & \textbf{0.54} & -0.26 & 0.24 & \textbf{0.57} & -0.29 & 0.26 & \textbf{0.56} & -0.41 & 0.32 \\
 & $\ln k_T$ & 0.10 & -0.02 & 0.23 & 0.09 & 0.05 & 0.23 & 0.01 & 0.33 & 0.16 \\
 & $\ln m$ & -0.08 & 0.06 & 0.19 & -0.08 & 0.13 & 0.16 & 0.02 & 0.37 & -0.10 \\
 & $\ln z$ & 0.25 & -0.10 & 0.01 & 0.23 & -0.16 & 0.04 & 0.01 & -0.35 & 0.18 \\
 & $\psi$ & 0.04 & -0.15 & -0.07 & 0.02 & -0.14 & -0.06 & -0.00 & -0.16 & 0.08 \\
\hline
\multirow{5}{*}{$\mathbf{t \rightarrow bq\bar{q'} }$} & $\ln\Delta$ & 0.47 & 0.43 & 0.01 & \textbf{0.51} & 0.44 & -0.01 & \textbf{0.59} & \textbf{0.52} & 0.09 \\
 & $\ln k_T$ & 0.13 & -0.20 & 0.00 & 0.13 & -0.20 & 0.02 & 0.18 & -0.31 & 0.17 \\
 & $\ln m$ & -0.04 & -0.22 & -0.07 & -0.06 & -0.22 & -0.03 & -0.03 & -0.26 & 0.12 \\
 & $\ln z$ & 0.23 & 0.15 & 0.09 & 0.26 & 0.15 & 0.05 & 0.21 & 0.08 & -0.00 \\
 & $\psi$ & -0.04 & 0.03 & -0.05 & -0.03 & 0.04 & -0.07 & 0.09 & 0.06 & 0.03 \\
\hline
\end{tabular}
\end{table*}
\begin{table*}[!t]
\centering
\caption{Pearson correlation coefficients $\rho$ between explainer importance scores and $C_3$ for each jet process, Lund-plane feature, and $p_T$ bin. The numbers quoted are for \lundnet{} architecture.}
\label{tab:C_corr_pt_C_3_lundnet}
\renewcommand{\arraystretch}{1.2}
\setlength{\tabcolsep}{6pt}
\scriptsize
\begin{tabular}{ll|rrr|rrr|rrr}
\hline
\multicolumn{2}{c|}{} & \multicolumn{3}{c|}{$500 < p_T < 700$ GeV} & \multicolumn{3}{c|}{$500 < p_T < 1000$ GeV} & \multicolumn{3}{c}{$800 < p_T < 1000$ GeV} \\
Process & Feature & GNNExpl. & GNN-SHAP & GradCAM & GNNExpl. & GNN-SHAP & GradCAM & GNNExpl. & GNN-SHAP & GradCAM \\
\hline\hline
\multirow{5}{*}{\textbf{QCD}} & $\ln\Delta$ & \textbf{0.57} & -0.06 & 0.28 & \textbf{0.56} & -0.08 & 0.27 & \textbf{0.56} & -0.13 & 0.24 \\
 & $\ln k_T$ & 0.14 & -0.15 & 0.24 & 0.15 & -0.16 & 0.23 & 0.16 & -0.18 & 0.23 \\
 & $\ln m$ & 0.08 & -0.20 & 0.06 & 0.08 & -0.19 & 0.06 & 0.09 & -0.20 & 0.03 \\
 & $\ln z$ & 0.04 & 0.15 & 0.09 & 0.03 & 0.14 & 0.07 & -0.00 & 0.13 & 0.07 \\
 & $\psi$ & 0.01 & 0.02 & -0.05 & -0.00 & 0.02 & -0.05 & -0.10 & -0.04 & 0.04 \\
\hline
\multirow{5}{*}{$\mathbf{H\!\to\!c\bar{c} }$} & $\ln\Delta$ & \textbf{0.54} & -0.32 & 0.25 & \textbf{0.57} & -0.34 & 0.25 & \textbf{0.56} & -0.39 & 0.20 \\
 & $\ln k_T$ & 0.15 & 0.16 & 0.22 & 0.14 & 0.22 & 0.20 & 0.10 & 0.40 & 0.09 \\
 & $\ln m$ & 0.17 & 0.22 & 0.21 & 0.17 & 0.28 & 0.16 & 0.22 & 0.44 & -0.08 \\
 & $\ln z$ & -0.11 & -0.24 & -0.01 & -0.11 & -0.28 & 0.01 & -0.26 & -0.41 & 0.08 \\
 & $\psi$ & 0.02 & -0.11 & -0.04 & 0.02 & -0.08 & -0.02 & -0.03 & 0.00 & 0.09 \\
\hline
\multirow{5}{*}{$\mathbf{t \rightarrow bq\bar{q'} }$} & $\ln\Delta$ & \textbf{0.55} & \textbf{0.52} & 0.04 & \textbf{0.56} & \textbf{0.52} & 0.03 & \textbf{0.68} & \textbf{0.59} & -0.02 \\
 & $\ln k_T$ & 0.12 & -0.26 & -0.05 & 0.13 & -0.25 & -0.04 & 0.21 & -0.23 & 0.03 \\
 & $\ln m$ & -0.03 & -0.31 & -0.14 & -0.02 & -0.30 & -0.11 & -0.00 & -0.26 & 0.08 \\
 & $\ln z$ & 0.24 & 0.25 & 0.14 & 0.23 & 0.24 & 0.10 & 0.17 & 0.14 & -0.10 \\
 & $\psi$ & -0.04 & -0.02 & 0.01 & -0.03 & -0.01 & 0.01 & 0.10 & -0.00 & 0.07 \\
\hline
\end{tabular}
\end{table*}

\begin{table*}[!t]
\centering
\caption{Pearson correlation coefficients $\rho$ between explainer importance scores and $C_2$ for each jet process, Lund-plane feature, and $p_T$ bin. The numbers quoted are for \parnet{} architecture.}
\label{tab:C_corr_pt_C_2_parnet}
\renewcommand{\arraystretch}{1.2}
\setlength{\tabcolsep}{6pt}
\scriptsize
\begin{tabular}{ll|rrr|rrr|rrr}
\hline
\multicolumn{2}{c|}{} & \multicolumn{3}{c|}{$500 < p_T < 700$ GeV} & \multicolumn{3}{c|}{$500 < p_T < 1000$ GeV} & \multicolumn{3}{c}{$800 < p_T < 1000$ GeV} \\
Process & Feature & GNNExpl. & GNN-SHAP & GradCAM & GNNExpl. & GNN-SHAP & GradCAM & GNNExpl. & GNN-SHAP & GradCAM \\
\hline\hline
\multirow{5}{*}{\textbf{QCD}} & $\ln\Delta$ & \textbf{0.59} & 0.29 & 0.12 & \textbf{0.59} & 0.30 & 0.11 & \textbf{0.61} & 0.39 & 0.05 \\
 & $\ln k_T$ & 0.29 & 0.03 & 0.02 & 0.29 & 0.01 & 0.00 & 0.26 & -0.14 & -0.13 \\
 & $\ln m$ & 0.18 & -0.18 & 0.03 & 0.17 & -0.20 & 0.02 & 0.08 & -0.33 & -0.03 \\
 & $\ln z$ & 0.00 & 0.25 & -0.01 & 0.01 & 0.27 & -0.01 & 0.06 & 0.37 & -0.02 \\
 & $\psi$ & 0.02 & -0.04 & 0.00 & 0.00 & -0.00 & -0.02 & -0.03 & 0.08 & -0.13 \\
\hline
\multirow{5}{*}{$\mathbf{H\!\to\!c\bar{c} }$} & $\ln\Delta$ & \textbf{0.54} & 0.36 & 0.33 & \textbf{0.56} & 0.39 & 0.31 & \textbf{0.54} & 0.42 & 0.30 \\
 & $\ln k_T$ & 0.09 & -0.49 & -0.49 & 0.09 & -0.49 & -0.47 & 0.06 & \textbf{-0.54} & \textbf{-0.50} \\
 & $\ln m$ & -0.11 & -0.37 & -0.43 & -0.10 & -0.38 & -0.37 & 0.05 & -0.37 & -0.29 \\
 & $\ln z$ & 0.26 & 0.05 & 0.21 & 0.24 & 0.06 & 0.15 & 0.01 & -0.05 & -0.03 \\
 & $\psi$ & 0.03 & 0.08 & -0.01 & 0.03 & 0.06 & -0.02 & 0.05 & 0.08 & -0.04 \\
\hline
\multirow{5}{*}{$\mathbf{t \rightarrow bq\bar{q'} }$} & $\ln\Delta$ & 0.42 & 0.20 & 0.35 & 0.47 & 0.23 & 0.40 & \textbf{0.52} & 0.37 & 0.29 \\
 & $\ln k_T$ & 0.05 & -0.31 & -0.19 & 0.03 & -0.30 & -0.15 & 0.10 & -0.45 & -0.17 \\
 & $\ln m$ & -0.09 & -0.24 & -0.10 & -0.13 & -0.26 & -0.08 & 0.05 & -0.42 & -0.09 \\
 & $\ln z$ & 0.18 & 0.03 & -0.14 & 0.20 & 0.07 & -0.12 & 0.01 & 0.19 & -0.16 \\
 & $\psi$ & -0.03 & 0.02 & 0.03 & -0.03 & 0.02 & 0.03 & 0.11 & 0.05 & -0.04 \\
\hline
\end{tabular}
\end{table*}
\begin{table*}[!t]
\centering
\caption{Pearson correlation coefficients $\rho$ between explainer importance scores and $C_3$ for each jet process, Lund-plane feature, and $p_T$ bin. The numbers quoted are for \parnet{} architecture.}
\label{tab:C_corr_pt_C_3_parnet}
\renewcommand{\arraystretch}{1.2}
\setlength{\tabcolsep}{6pt}
\scriptsize
\begin{tabular}{ll|rrr|rrr|rrr}
\hline
\multicolumn{2}{c|}{} & \multicolumn{3}{c|}{$500 < p_T < 700$ GeV} & \multicolumn{3}{c|}{$500 < p_T < 1000$ GeV} & \multicolumn{3}{c}{$800 < p_T < 1000$ GeV} \\
Process & Feature & GNNExpl. & GNN-SHAP & GradCAM & GNNExpl. & GNN-SHAP & GradCAM & GNNExpl. & GNN-SHAP & GradCAM \\
\hline\hline
\multirow{5}{*}{\textbf{QCD}} & $\ln\Delta$ & \textbf{0.58} & 0.25 & 0.27 & \textbf{0.58} & 0.26 & 0.26 & \textbf{0.58} & 0.30 & 0.21 \\
 & $\ln k_T$ & 0.22 & 0.05 & 0.14 & 0.23 & 0.04 & 0.13 & 0.24 & 0.01 & 0.08 \\
 & $\ln m$ & 0.14 & -0.03 & -0.03 & 0.14 & -0.04 & -0.03 & 0.08 & -0.11 & -0.06 \\
 & $\ln z$ & 0.04 & 0.07 & 0.10 & 0.03 & 0.08 & 0.09 & 0.04 & 0.16 & 0.08 \\
 & $\psi$ & 0.01 & -0.05 & -0.00 & 0.00 & -0.02 & -0.02 & -0.03 & 0.11 & -0.13 \\
\hline
\multirow{5}{*}{$\mathbf{H\!\to\!c\bar{c} }$} & $\ln\Delta$ & \textbf{0.53} & 0.30 & 0.13 & \textbf{0.55} & 0.32 & 0.12 & \textbf{0.52} & 0.26 & 0.12 \\
 & $\ln k_T$ & 0.15 & -0.36 & -0.20 & 0.14 & -0.37 & -0.18 & 0.11 & -0.42 & -0.23 \\
 & $\ln m$ & 0.17 & -0.34 & -0.06 & 0.17 & -0.35 & -0.03 & 0.25 & -0.38 & -0.04 \\
 & $\ln z$ & -0.12 & 0.14 & -0.07 & -0.13 & 0.16 & -0.11 & -0.28 & 0.13 & -0.16 \\
 & $\psi$ & 0.00 & 0.04 & 0.01 & -0.00 & 0.03 & 0.01 & 0.01 & 0.04 & 0.01 \\
\hline
\multirow{5}{*}{$\mathbf{t \rightarrow bq\bar{q'} }$} & $\ln\Delta$ & 0.47 & 0.26 & 0.44 & 0.49 & 0.29 & 0.44 & \textbf{0.50} & 0.41 & 0.32 \\
 & $\ln k_T$ & 0.00 & -0.36 & -0.41 & 0.01 & -0.37 & -0.37 & 0.07 & -0.47 & -0.28 \\
 & $\ln m$ & -0.10 & -0.37 & -0.35 & -0.09 & -0.40 & -0.32 & 0.13 & \textbf{-0.52} & -0.25 \\
 & $\ln z$ & 0.18 & 0.21 & 0.14 & 0.15 & 0.23 & 0.14 & -0.13 & 0.36 & 0.12 \\
 & $\psi$ & -0.05 & -0.01 & -0.02 & -0.03 & 0.00 & -0.02 & 0.09 & 0.05 & -0.07 \\
\hline
\end{tabular}
\end{table*}

\begin{table*}[!t]
\centering
\caption{Pearson correlation coefficients $\rho$ between explainer importance scores and $C_2$ for each jet process, Lund-plane feature, and $p_T$ bin. The numbers quoted are for \part{} architecture.}
\label{tab:C_corr_pt_C_2_part}
\renewcommand{\arraystretch}{1.2}
\setlength{\tabcolsep}{6pt}
\scriptsize
\begin{tabular}{ll|rrr|rrr|rrr}
\hline
\multicolumn{2}{c|}{} & \multicolumn{3}{c|}{$500 < p_T < 700$ GeV} & \multicolumn{3}{c|}{$500 < p_T < 1000$ GeV} & \multicolumn{3}{c}{$800 < p_T < 1000$ GeV} \\
Process & Feature & GNNExpl. & GNN-SHAP & GradCAM & GNNExpl. & GNN-SHAP & GradCAM & GNNExpl. & GNN-SHAP & GradCAM \\
\hline\hline
\multirow{5}{*}{\textbf{QCD}} & $\ln\Delta$ & \textbf{0.57} & 0.21 & 0.28 & \textbf{0.55} & 0.19 & 0.26 & \textbf{0.52} & 0.04 & 0.14 \\
 & $\ln k_T$ & 0.17 & 0.12 & 0.11 & 0.17 & 0.11 & 0.11 & 0.18 & 0.00 & 0.04 \\
 & $\ln m$ & 0.32 & -0.23 & 0.49 & 0.32 & -0.22 & \textbf{0.50} & 0.26 & -0.21 & 0.46 \\
 & $\ln z$ & -0.44 & 0.26 & -0.44 & -0.42 & 0.25 & -0.45 & -0.39 & 0.10 & -0.46 \\
 & $\psi$ & -0.01 & 0.01 & -0.01 & -0.03 & 0.02 & -0.03 & -0.18 & 0.02 & -0.10 \\
\hline
\multirow{5}{*}{$\mathbf{H\!\to\!c\bar{c} }$} & $\ln\Delta$ & 0.28 & 0.47 & 0.10 & 0.34 & \textbf{0.50} & 0.13 & \textbf{0.57} & \textbf{0.54} & 0.03 \\
 & $\ln k_T$ & 0.03 & -0.31 & -0.14 & 0.04 & -0.28 & -0.19 & 0.22 & -0.11 & -0.31 \\
 & $\ln m$ & -0.07 & -0.28 & -0.12 & -0.09 & -0.24 & -0.17 & 0.11 & -0.11 & -0.18 \\
 & $\ln z$ & 0.18 & 0.20 & -0.01 & 0.21 & 0.16 & -0.00 & 0.16 & 0.08 & -0.14 \\
 & $\psi$ & -0.00 & 0.01 & -0.02 & -0.00 & -0.01 & -0.03 & 0.20 & 0.13 & 0.10 \\
\hline
\multirow{5}{*}{$\mathbf{t \rightarrow bq\bar{q'} }$} & $\ln\Delta$ & \textbf{0.56} & 0.33 & 0.44 & \textbf{0.62} & 0.38 & 0.44 & 0.28 & -0.11 & -0.00 \\
 & $\ln k_T$ & 0.34 & -0.21 & 0.37 & 0.36 & -0.14 & 0.34 & 0.21 & -0.03 & 0.10 \\
 & $\ln m$ & -0.20 & -0.11 & 0.28 & -0.24 & -0.05 & 0.32 & 0.39 & 0.05 & 0.24 \\
 & $\ln z$ & 0.49 & -0.18 & 0.06 & \textbf{0.54} & -0.18 & -0.04 & -0.30 & -0.36 & -0.30 \\
 & $\psi$ & 0.00 & 0.04 & -0.05 & 0.02 & 0.04 & -0.02 & 0.16 & 0.40 & 0.22 \\
\hline
\end{tabular}
\end{table*}
\begin{table*}[!t]
\centering
\caption{Pearson correlation coefficients $\rho$ between explainer importance scores and $C_3$ for each jet process, Lund-plane feature, and $p_T$ bin. The numbers quoted are for \part{} architecture.}
\label{tab:C_corr_pt_C_3_part}
\renewcommand{\arraystretch}{1.2}
\setlength{\tabcolsep}{6pt}
\scriptsize
\begin{tabular}{ll|rrr|rrr|rrr}
\hline
\multicolumn{2}{c|}{} & \multicolumn{3}{c|}{$500 < p_T < 700$ GeV} & \multicolumn{3}{c|}{$500 < p_T < 1000$ GeV} & \multicolumn{3}{c}{$800 < p_T < 1000$ GeV} \\
Process & Feature & GNNExpl. & GNN-SHAP & GradCAM & GNNExpl. & GNN-SHAP & GradCAM & GNNExpl. & GNN-SHAP & GradCAM \\
\hline\hline
\multirow{5}{*}{\textbf{QCD}} & $\ln\Delta$ & \textbf{0.56} & 0.35 & 0.30 & \textbf{0.56} & 0.33 & 0.31 & \textbf{0.57} & 0.17 & 0.18 \\
 & $\ln k_T$ & 0.08 & 0.26 & 0.05 & 0.11 & 0.25 & 0.08 & 0.13 & 0.15 & 0.01 \\
 & $\ln m$ & 0.25 & -0.27 & \textbf{0.50} & 0.26 & -0.26 & \textbf{0.51} & 0.25 & -0.15 & \textbf{0.56} \\
 & $\ln z$ & -0.33 & 0.38 & -0.43 & -0.33 & 0.36 & -0.43 & -0.41 & 0.17 & \textbf{-0.52} \\
 & $\psi$ & -0.02 & -0.01 & -0.03 & -0.04 & -0.00 & -0.05 & -0.19 & 0.02 & -0.13 \\
\hline
\multirow{5}{*}{$\mathbf{H\!\to\!c\bar{c} }$} & $\ln\Delta$ & 0.31 & 0.07 & -0.10 & 0.36 & 0.08 & -0.09 & 0.47 & 0.01 & -0.41 \\
 & $\ln k_T$ & 0.11 & -0.04 & -0.14 & 0.10 & -0.02 & -0.16 & 0.20 & 0.04 & -0.14 \\
 & $\ln m$ & 0.30 & 0.03 & 0.10 & 0.27 & 0.05 & 0.06 & 0.37 & 0.13 & 0.09 \\
 & $\ln z$ & -0.29 & -0.08 & -0.16 & -0.26 & -0.11 & -0.16 & -0.29 & -0.24 & -0.33 \\
 & $\psi$ & 0.02 & -0.00 & 0.01 & 0.00 & 0.01 & 0.01 & -0.03 & 0.07 & -0.07 \\
\hline
\multirow{5}{*}{$\mathbf{t \rightarrow bq\bar{q'} }$} & $\ln\Delta$ & 0.29 & 0.27 & 0.30 & 0.32 & 0.29 & 0.29 & 0.36 & 0.13 & -0.11 \\
 & $\ln k_T$ & 0.11 & -0.42 & 0.17 & 0.13 & -0.38 & 0.17 & 0.29 & -0.28 & 0.18 \\
 & $\ln m$ & -0.07 & -0.36 & 0.04 & -0.07 & -0.33 & 0.08 & 0.33 & -0.34 & 0.35 \\
 & $\ln z$ & 0.17 & 0.10 & 0.14 & 0.18 & 0.10 & 0.08 & -0.24 & 0.17 & -0.22 \\
 & $\psi$ & 0.01 & -0.03 & -0.05 & 0.03 & -0.02 & -0.02 & 0.29 & 0.19 & 0.45 \\
\hline
\end{tabular}
\end{table*}

\FloatBarrier
\clearpage
\bibliography{bibliography}

@article{Dreyer:2018nbf,
    author = "Dreyer, Fr{\'e}d{\'e}ric A. and Salam, Gavin P. and Soyez, Gr{\'e}gory",
    title = "{The Lund Jet Plane}",
    eprint = "1807.04758",
    archivePrefix = "arXiv",
    primaryClass = "hep-ph",
    reportNumber = "CERN-TH-2018-151",
    doi = "10.1007/JHEP12(2018)064",
    journal = "JHEP",
    volume = "12",
    pages = "064",
    year = "2018"
}

@inproceedings{Thais:2022iok,
    author = "Thais, Savannah and Calafiura, Paolo and Chachamis, Grigorios and DeZoort, Gage and Duarte, Javier and Ganguly, Sanmay and Kagan, Michael and Murnane, Daniel and Neubauer, Mark S. and Terao, Kazuhiro",
    title = "{Graph Neural Networks in Particle Physics: Implementations, Innovations, and Challenges}",
    booktitle = "{Snowmass 2021}",
    eprint = "2203.12852",
    archivePrefix = "arXiv",
    primaryClass = "hep-ex",
    month = "3",
    year = "2022"
}

@article{Salam:2010nqg,
    author = "Salam, Gavin P.",
    title = "{Towards Jetography}",
    eprint = "0906.1833",
    archivePrefix = "arXiv",
    primaryClass = "hep-ph",
    doi = "10.1140/epjc/s10052-010-1314-6",
    journal = "Eur. Phys. J. C",
    volume = "67",
    pages = "637--686",
    year = "2010"
}

@article{Larkoski:2017jix,
    author = "Larkoski, Andrew J. and Moult, Ian and Nachman, Benjamin",
    title = "{Jet Substructure at the Large Hadron Collider: A Review of Recent Advances in Theory and Machine Learning}",
    eprint = "1709.04464",
    archivePrefix = "arXiv",
    primaryClass = "hep-ph",
    doi = "10.1016/j.physrep.2019.11.001",
    journal = "Phys. Rept.",
    volume = "841",
    pages = "1--63",
    year = "2020"
}

@article{Kogler:2018hem,
    author = "Kogler, Roman and others",
    title = "{Jet Substructure at the Large Hadron Collider: Experimental Review}",
    eprint = "1803.06991",
    archivePrefix = "arXiv",
    primaryClass = "hep-ex",
    reportNumber = "FERMILAB-PUB-18-123-PPD",
    doi = "10.1103/RevModPhys.91.045003",
    journal = "Rev. Mod. Phys.",
    volume = "91",
    number = "4",
    pages = "045003",
    year = "2019"
}

@book{Marzani:2019hun,
    author = "Marzani, Simone and Soyez, Gregory and Spannowsky, Michael",
    title = "{Looking inside jets: an introduction to jet substructure and boosted-object phenomenology}",
    eprint = "1901.10342",
    archivePrefix = "arXiv",
    primaryClass = "hep-ph",
    doi = "10.1007/978-3-030-15709-8",
    publisher = "Springer",
    volume = "958",
    year = "2019"
}

@article{Kasieczka:2019dbj,
    author = "Butter, Anja and others",
    editor = "Kasieczka, Gregor and Plehn, Tilman",
    title = "{The Machine Learning landscape of top taggers}",
    eprint = "1902.09914",
    archivePrefix = "arXiv",
    primaryClass = "hep-ph",
    doi = "10.21468/SciPostPhys.7.1.014",
    journal = "SciPost Phys.",
    volume = "7",
    pages = "014",
    year = "2019"
}

@techreport{CMS-PAS-HIG-23-012,
      collaboration = "CMS",
      title         = "{Search for highly energetic double Higgs boson production
                       in the two bottom quark and two vector boson all-hadronic
                       final state}",
      institution   = "CERN",
      reportNumber  = "CMS-PAS-HIG-23-012",
      address       = "Geneva",
      year          = "2024",
      url           = "https://cds.cern.ch/record/2904879",
}

@techreport{ATLAS:2026vyw,
    collaboration = "ATLAS",
    title = "{GN3: Multi-task, Multi-modal Transformers for Jet Flavour Tagging in ATLAS}",
    reportNumber = "ATL-PHYS-PUB-2026-001",
    year = "2026"
}

@article{deOliveira:2015xxd,
    author = "de Oliveira, Luke and Kagan, Michael and Mackey, Lester and Nachman, Benjamin and Schwartzman, Ariel",
    title = "{Jet-images {\textemdash} deep learning edition}",
    eprint = "1511.05190",
    archivePrefix = "arXiv",
    primaryClass = "hep-ph",
    doi = "10.1007/JHEP07(2016)069",
    journal = "JHEP",
    volume = "07",
    pages = "069",
    year = "2016"
}

@article{Komiske:2016rsd,
    author = "Komiske, Patrick T. and Metodiev, Eric M. and Schwartz, Matthew D.",
    title = "{Deep learning in color: towards automated quark/gluon jet discrimination}",
    eprint = "1612.01551",
    archivePrefix = "arXiv",
    primaryClass = "hep-ph",
    reportNumber = "MIT-CTP-4866, MIT-CTP 4866",
    doi = "10.1007/JHEP01(2017)110",
    journal = "JHEP",
    volume = "01",
    pages = "110",
    year = "2017"
}

@article{Kasieczka:2017nvn,
    author = "Kasieczka, Gregor and Plehn, Tilman and Russell, Michael and Schell, Torben",
    title = "{Deep-learning Top Taggers or The End of QCD?}",
    eprint = "1701.08784",
    archivePrefix = "arXiv",
    primaryClass = "hep-ph",
    reportNumber = "MCNET-17-07",
    doi = "10.1007/JHEP05(2017)006",
    journal = "JHEP",
    volume = "05",
    pages = "006",
    year = "2017"
}

@article{Qu:2019gqs,
    author = "Qu, Huilin and Gouskos, Loukas",
    title = "{ParticleNet: Jet Tagging via Particle Clouds}",
    eprint = "1902.08570",
    archivePrefix = "arXiv",
    primaryClass = "hep-ph",
    doi = "10.1103/PhysRevD.101.056019",
    journal = "Phys. Rev. D",
    volume = "101",
    number = "5",
    pages = "056019",
    year = "2020"
}

@article{Moreno:2019neq,
    author = "Moreno, Eric A. and Nguyen, Thong Q. and Vlimant, Jean-Roch and Cerri, Olmo and Newman, Harvey B. and Periwal, Avikar and Spiropulu, Maria and Duarte, Javier M. and Pierini, Maurizio",
    title = "{Interaction networks for the identification of boosted $H \rightarrow b\overline{b}$ decays}",
    eprint = "1909.12285",
    archivePrefix = "arXiv",
    primaryClass = "hep-ex",
    reportNumber = "FERMILAB-PUB-19-492-CMS-E",
    doi = "10.1103/PhysRevD.102.012010",
    journal = "Phys. Rev. D",
    volume = "102",
    number = "1",
    pages = "012010",
    year = "2020"
}

@article{Gong:2022lye,
    author = "Gong, Shiqi and Meng, Qi and Zhang, Jue and Qu, Huilin and Li, Congqiao and Qian, Sitian and Du, Weitao and Ma, Zhi-Ming and Liu, Tie-Yan",
    title = "{An efficient Lorentz equivariant graph neural network for jet tagging}",
    eprint = "2201.08187",
    archivePrefix = "arXiv",
    primaryClass = "hep-ph",
    doi = "10.1007/JHEP07(2022)030",
    journal = "JHEP",
    volume = "07",
    pages = "030",
    year = "2022"
}

@misc{Qu:2022mxj,
      title={Particle Transformer for Jet Tagging}, 
      author={Huilin Qu and Congqiao Li and Sitian Qian},
      year={2024},
      eprint={2202.03772},
      archivePrefix={arXiv},
      primaryClass={hep-ph},
      url={https://arxiv.org/abs/2202.03772}, 
}

@article{Shlomi:2020gdn,
   title={Graph neural networks in particle physics},
   volume={2},
   ISSN={2632-2153},
   url={http://dx.doi.org/10.1088/2632-2153/abbf9a},
   DOI={10.1088/2632-2153/abbf9a},
   number={2},
   journal={Machine Learning: Science and Technology},
   publisher={IOP Publishing},
   author={Shlomi, Jonathan and Battaglia, Peter and Vlimant, Jean-Roch},
   year={2021},
   month=jan, pages={021001} }

@article{Piacquadio:2008zza,
    author = "Piacquadio, Giacinto and Weiser, Christian",
    editor = "Sobie, Randall and Tafirout, Reda and Thomson, Jana",
    title = "{A new inclusive secondary vertex algorithm for b-jet tagging in ATLAS}",
    doi = "10.1088/1742-6596/119/3/032032",
    journal = "J. Phys. Conf. Ser.",
    volume = "119",
    pages = "032032",
    year = "2008"
}

@article{Shlomi:2020ufi,
    author = "Shlomi, Jonathan and Ganguly, Sanmay and Gross, Eilam and Cranmer, Kyle and Lipman, Yaron and Serviansky, Hadar and Maron, Haggai and Segol, Nimrod",
    title = "{Secondary vertex finding in jets with neural networks}",
    eprint = "2008.02831",
    archivePrefix = "arXiv",
    primaryClass = "hep-ex",
    doi = "10.1140/epjc/s10052-021-09342-y",
    journal = "Eur. Phys. J. C",
    volume = "81",
    number = "6",
    pages = "540",
    year = "2021"
}

@article{ATLAS:2025dkv,
    author = "Aad, Georges and others",
    collaboration = "ATLAS",
    title = "{Transforming jet flavour tagging at ATLAS}",
    eprint = "2505.19689",
    archivePrefix = "arXiv",
    primaryClass = "hep-ex",
    reportNumber = "CERN-EP-2025-103",
    doi = "10.1038/s41467-025-65059-6",
    journal = "Nature Commun.",
    volume = "17",
    number = "1",
    pages = "541",
    year = "2026"
}

@article{Thaler:2010cxa,
    author = "Thaler, Jesse and Van Tilburg, Ken",
    title = "{Identifying Boosted Objects with N-subjettiness}",
    eprint = "1011.2268",
    archivePrefix = "arXiv",
    primaryClass = "hep-ph",
    reportNumber = "MIT-CTP-4191",
    doi = "10.1007/JHEP03(2011)015",
    journal = "JHEP",
    volume = "03",
    pages = "015",
    year = "2011"
}

@article{Thaler:2011gf,
    author = "Thaler, Jesse and Van Tilburg, Ken",
    title = "{Maximizing Boosted Top Identification by Minimizing N-subjettiness}",
    eprint = "1108.2701",
    archivePrefix = "arXiv",
    primaryClass = "hep-ph",
    reportNumber = "MIT-CTP-4287",
    doi = "10.1007/JHEP02(2012)093",
    journal = "JHEP",
    volume = "02",
    pages = "093",
    year = "2012"
}

@article{Larkoski:2013eya,
    author = "Larkoski, Andrew J. and Salam, Gavin P. and Thaler, Jesse",
    title = "{Energy Correlation Functions for Jet Substructure}",
    eprint = "1305.0007",
    archivePrefix = "arXiv",
    primaryClass = "hep-ph",
    reportNumber = "MIT-CTP-4446, CERN-PH-TH-2013-066, LPN13-026",
    doi = "10.1007/JHEP06(2013)108",
    journal = "JHEP",
    volume = "06",
    pages = "108",
    year = "2013"
}

@article{Larkoski:2024uoc,
    author = "Larkoski, Andrew J.",
    title = "{QCD masterclass lectures on jet physics and machine learning}",
    eprint = "2407.04897",
    archivePrefix = "arXiv",
    primaryClass = "hep-ph",
    doi = "10.1140/epjc/s10052-024-13341-0",
    journal = "Eur. Phys. J. C",
    volume = "84",
    number = "10",
    pages = "1117",
    year = "2024"
}

@inproceedings{Salam:2010zt,
    author = "Salam, Gavin P.",
    title = "{Elements of QCD for hadron colliders}",
    booktitle = "{2009 European School of High-Energy Physics}",
    eprint = "1011.5131",
    archivePrefix = "arXiv",
    primaryClass = "hep-ph",
    month = "11",
    year = "2010"
}

@article{Cacciari:2008gp,
    author = "Cacciari, Matteo and Salam, Gavin P. and Soyez, Gregory",
    title = "{The anti-$k_t$ jet clustering algorithm}",
    eprint = "0802.1189",
    archivePrefix = "arXiv",
    primaryClass = "hep-ph",
    reportNumber = "LPTHE-07-03",
    doi = "10.1088/1126-6708/2008/04/063",
    journal = "JHEP",
    volume = "04",
    pages = "063",
    year = "2008"
}

@article{Moult:2016cvt,
    author = "Moult, Ian and Necib, Lina and Thaler, Jesse",
    title = "{New Angles on Energy Correlation Functions}",
    eprint = "1609.07483",
    archivePrefix = "arXiv",
    primaryClass = "hep-ph",
    reportNumber = "MIT-CTP-4825",
    doi = "10.1007/JHEP12(2016)153",
    journal = "JHEP",
    volume = "12",
    pages = "153",
    year = "2016"
}

@article{Larkoski:2014wba,
    author = "Larkoski, Andrew J. and Marzani, Simone and Soyez, Gregory and Thaler, Jesse",
    title = "{Soft Drop}",
    eprint = "1402.2657",
    archivePrefix = "arXiv",
    primaryClass = "hep-ph",
    reportNumber = "MIT-CTP-4531, DCPT-14-24, IPPP-14-12",
    doi = "10.1007/JHEP05(2014)146",
    journal = "JHEP",
    volume = "05",
    pages = "146",
    year = "2014"
}

@ARTICLE{Wetzel:2025review,
       author = {{Wetzel}, Sebastian Johann and {Ha}, Seungwoong and {Iten}, Raban and {Klopotek}, Miriam and {Liu}, Ziming},
        title = "{Interpretable Machine Learning in Physics: A Review}",
      journal = {arXiv e-prints},
         year = 2025,
        month = mar,
          eid = {arXiv:2503.23616},
        pages = {arXiv:2503.23616},
          doi = {10.48550/arXiv.2503.23616},
archivePrefix = {arXiv},
       eprint = {2503.23616},
 primaryClass = {physics.comp-ph},
       adsurl = {https://ui.adsabs.harvard.edu/abs/2025arXiv250323616W}
}

@article{Dreyer:2020brq,
    author = "Dreyer, Fr{\'e}d{\'e}ric A. and Qu, Huilin",
    title = "{Jet tagging in the Lund plane with graph networks}",
    eprint = "2012.08526",
    archivePrefix = "arXiv",
    primaryClass = "hep-ph",
    reportNumber = "OUTP-20-15P",
    doi = "10.1007/JHEP03(2021)052",
    journal = "JHEP",
    volume = "03",
    pages = "052",
    year = "2021"
}

@article{Barnard:2016qma,
    author = "Barnard, James and Dawe, Edmund Noel and Dolan, Matthew J. and Rajcic, Nina",
    title = "{Parton Shower Uncertainties in Jet Substructure Analyses with Deep Neural Networks}",
    eprint = "1609.00607",
    archivePrefix = "arXiv",
    primaryClass = "hep-ph",
    doi = "10.1103/PhysRevD.95.014018",
    journal = "Phys. Rev. D",
    volume = "95",
    number = "1",
    pages = "014018",
    year = "2017"
}

@article{Grojean:2022mef,
    author = {Grojean, Christophe and Paul, Ayan and Qian, Zhuoni and Str{\"u}mke, Inga},
    title = "{Lessons on interpretable machine learning from particle physics}",
    eprint = "2203.08021",
    archivePrefix = "arXiv",
    primaryClass = "hep-ph",
    reportNumber = "DESY-22-038, HU-EP-21/46, DESY-22-038, HU-EP-21/46",
    doi = "10.1038/s42254-022-00456-0",
    journal = "Nature Rev. Phys.",
    volume = "4",
    number = "5",
    pages = "284--286",
    year = "2022"
}

@article{Murdoch:2019pnas,
       author = {{Murdoch}, W. James and {Singh}, Chandan and {Kumbier}, Karl and {Abbasi-Asl}, Reza and {Yu}, Bin},
        title = "{Interpretable machine learning: definitions, methods, and applications}",
      journal = {arXiv e-prints},
         year = 2019,
        month = jan,
          eid = {arXiv:1901.04592},
        pages = {arXiv:1901.04592},
          doi = {10.48550/arXiv.1901.04592},
archivePrefix = {arXiv},
       eprint = {1901.04592},
 primaryClass = {stat.ML},
       url = {https://ui.adsabs.harvard.edu/abs/2019arXiv190104592M}
}

@article{BarredoArrieta:2020xaif,
       author = {{Barredo Arrieta}, Alejandro and {D{\'\i}az-Rodr{\'\i}guez}, Natalia and {Del Ser}, Javier and {Bennetot}, Adrien and {Tabik}, Siham and {Barbado}, Alberto and {Garc{\'\i}a}, Salvador and {Gil-L{\'o}pez}, Sergio and {Molina}, Daniel and {Benjamins}, Richard and {Chatila}, Raja and {Herrera}, Francisco},
        title = "{Explainable Artificial Intelligence (XAI): Concepts, Taxonomies, Opportunities and Challenges toward Responsible AI}",
      journal = {arXiv e-prints},
         year = 2019,
        month = oct,
          eid = {arXiv:1910.10045},
        pages = {arXiv:1910.10045},
          doi = {10.48550/arXiv.1910.10045},
archivePrefix = {arXiv},
       eprint = {1910.10045},
 primaryClass = {cs.AI},
       url = {https://ui.adsabs.harvard.edu/abs/2019arXiv191010045B}
}

@inproceedings{Wang:2024rup,
    author = "Wang, Aaron and Gandrakota, Abhijith and Ngadiuba, Jennifer and Sahu, Vivekanand and Bhatnagar, Priyansh and Khoda, Elham E. and Duarte, Javier",
    title = "{Interpreting Transformers for Jet Tagging}",
    eprint = "2412.03673",
    archivePrefix = "arXiv",
    primaryClass = "hep-ph",
    month = "12",
    year = "2024"
}

@article{Bogatskiy:2023nnw,
    author = "Bogatskiy, Alexander and Hoffman, Timothy and Miller, David W. and Offermann, Jan T. and Liu, Xiaoyang",
    title = "{Explainable equivariant neural networks for particle physics: PELICAN}",
    eprint = "2307.16506",
    archivePrefix = "arXiv",
    primaryClass = "hep-ph",
    doi = "10.1007/JHEP03(2024)113",
    journal = "JHEP",
    volume = "03",
    pages = "113",
    year = "2024"
}

@article{Islam:2025kjf,
    author = "Islam, Md Raqibul and Khan, Adrita and Hossain, Mir Sazzat and Siddiqui, Choudhury Ben Yamin and Hossan, Md. Zakir and Khan, Tanjib and Momen, M. Arshad and Ali, Amin Ahsan and Rahman, AKM Mahbubur",
    title = "{E-PCN: Jet Tagging with Explainable Particle Chebyshev Networks Using Kinematic Features}",
    eprint = "2512.07420",
    archivePrefix = "arXiv",
    primaryClass = "hep-ph",
    journal = "",
    month = "12",
    year = "2025"
}

@ARTICLE{Yuan:2022taxonomy,
author={Yuan, Hao and Yu, Haiyang and Gui, Shurui and Ji, Shuiwang},
journal={ IEEE Transactions on Pattern Analysis \& Machine Intelligence },
title={{ Explainability in Graph Neural Networks: A Taxonomic Survey }},
year={2023},
volume={45},
number={05},
ISSN={1939-3539},
pages={5782-5799},
doi={10.1109/TPAMI.2022.3204236},
url = {https://doi.ieeecomputersociety.org/10.1109/TPAMI.2022.3204236},
publisher={IEEE Computer Society},
address={Los Alamitos, CA, USA},
month=may}

@INPROCEEDINGS{Li:2022graphxai,
       author = {{Agarwal}, Chirag and {Queen}, Owen and {Lakkaraju}, Himabindu and {Zitnik}, Marinka},
        title = "{Evaluating explainability for graph neural networks}",
    booktitle = {Nature Scientific Data},
         year = 2023,
       series = {Nature Scientific Data},
       volume = {10},
        month = mar,
          eid = {144},
        pages = {144},
          doi = {10.1038/s41597-023-01974-x},
archivePrefix = {arXiv},
       eprint = {2208.09339},
 primaryClass = {cs.LG},
       adsurl = {https://ui.adsabs.harvard.edu/abs/2023NatSD..10..144A}
}

@ARTICLE{Ying:2019gnnexplainer,
       author = {{Ying}, Rex and {Bourgeois}, Dylan and {You}, Jiaxuan and {Zitnik}, Marinka and {Leskovec}, Jure},
        title = "{GNNExplainer: Generating Explanations for Graph Neural Networks}",
      journal = {arXiv e-prints},
         year = 2019,
        month = mar,
          eid = {arXiv:1903.03894},
        pages = {arXiv:1903.03894},
          doi = {10.48550/arXiv.1903.03894},
archivePrefix = {arXiv},
       eprint = {1903.03894},
 primaryClass = {cs.LG},
       adsurl = {https://ui.adsabs.harvard.edu/abs/2019arXiv190303894Y}
}

@inproceedings{Akkas:2024gnnshap,
author = {Akkas, Selahattin and Azad, Ariful},
title = {GNNShap: Scalable and Accurate GNN Explanation using Shapley Values},
year = {2024},
isbn = {9798400701719},
publisher = {Association for Computing Machinery},
address = {New York, NY, USA},
url = {https://doi.org/10.1145/3589334.3645599},
doi = {10.1145/3589334.3645599},
booktitle = {Proceedings of the ACM Web Conference 2024},
pages = {827–838},
numpages = {12},
location = {Singapore, Singapore},
series = {WWW '24}
}

@INPROCEEDINGS{Pope:2019gradcam,
  author={Pope, Phillip E. and Kolouri, Soheil and Rostami, Mohammad and Martin, Charles E. and Hoffmann, Heiko},
  booktitle={2019 IEEE/CVF Conference on Computer Vision and Pattern Recognition (CVPR)}, 
  title={Explainability Methods for Graph Convolutional Neural Networks}, 
  year={2019},
  volume={},
  number={},
  pages={10764-10773},
  doi={10.1109/CVPR.2019.01103}}

@ARTICLE{Amara:2022graphframex,
       author = {{Amara}, Kenza and {Ying}, Rex and {Zhang}, Zitao and {Han}, Zhihao and {Shan}, Yinan and {Brandes}, Ulrik and {Schemm}, Sebastian and {Zhang}, Ce},
        title = "{GraphFramEx: Towards Systematic Evaluation of Explainability Methods for Graph Neural Networks}",
      journal = {arXiv e-prints},
         year = 2022,
        month = jun,
          eid = {arXiv:2206.09677},
        pages = {arXiv:2206.09677},
          doi = {10.48550/arXiv.2206.09677},
archivePrefix = {arXiv},
       eprint = {2206.09677},
 primaryClass = {cs.LG},
       adsurl = {https://ui.adsabs.harvard.edu/abs/2022arXiv220609677A}
}

@INPROCEEDINGS{Agarwal:2023evaluating,
       author = {{Agarwal}, Chirag and {Queen}, Owen and {Lakkaraju}, Himabindu and {Zitnik}, Marinka},
        title = "{Evaluating explainability for graph neural networks}",
    booktitle = {Nature Scientific Data},
         year = 2023,
       series = {Nature Scientific Data},
       volume = {10},
        month = mar,
          eid = {144},
        pages = {144},
          doi = {10.1038/s41597-023-01974-x},
archivePrefix = {arXiv},
       eprint = {2208.09339},
 primaryClass = {cs.LG},
       adsurl = {https://ui.adsabs.harvard.edu/abs/2023NatSD..10..144A}
}

@article{Dokshitzer:1997in,
    author = "Dokshitzer, Yuri L. and Leder, G. D. and Moretti, S. and Webber, B. R.",
    title = "{Better jet clustering algorithms}",
    eprint = "hep-ph/9707323",
    archivePrefix = "arXiv",
    reportNumber = "CAVENDISH-HEP-97-06",
    doi = "10.1088/1126-6708/1997/08/001",
    journal = "JHEP",
    volume = "08",
    pages = "001",
    year = "1997"
}

@inproceedings{Wobisch:1998wt,
    author = "Wobisch, M. and Wengler, T.",
    title = "{Hadronization corrections to jet cross-sections in deep inelastic scattering}",
    booktitle = "{Workshop on Monte Carlo Generators for HERA Physics (Plenary Starting Meeting)}",
    eprint = "hep-ph/9907280",
    archivePrefix = "arXiv",
    reportNumber = "PITHA-99-16",
    pages = "270--279",
    month = "4",
    year = "1998"
}

@article{Lifson:2020gua,
    author = "Lifson, Andrew and Salam, Gavin P. and Soyez, Gregory",
    title = "{Calculating the primary Lund Jet Plane density}",
    eprint = "2007.06578",
    archivePrefix = "arXiv",
    primaryClass = "hep-ph",
    doi = "10.1007/JHEP10(2020)170",
    journal = "JHEP",
    volume = "10",
    pages = "170",
    year = "2020"
}

@article{ATLAS:2020bbn,
    author = "Aad, Georges and others",
    collaboration = "ATLAS",
    title = "{Measurement of the Lund Jet Plane Using Charged Particles in 13 TeV Proton-Proton Collisions with the ATLAS Detector}",
    eprint = "2004.03540",
    archivePrefix = "arXiv",
    primaryClass = "hep-ex",
    reportNumber = "CERN-EP-2020-030",
    doi = "10.1103/PhysRevLett.124.222002",
    journal = "Phys. Rev. Lett.",
    volume = "124",
    number = "22",
    pages = "222002",
    year = "2020"
}

@ARTICLE{Wang:2019dgcnn,
       author = {{Wang}, Yue and {Sun}, Yongbin and {Liu}, Ziwei and {Sarma}, Sanjay E. and {Bronstein}, Michael M. and {Solomon}, Justin M.},
        title = "{Dynamic Graph CNN for Learning on Point Clouds}",
      journal = {arXiv e-prints},
         year = 2018,
        month = jan,
          eid = {arXiv:1801.07829},
        pages = {arXiv:1801.07829},
          doi = {10.48550/arXiv.1801.07829},
archivePrefix = {arXiv},
       eprint = {1801.07829},
 primaryClass = {cs.CV},
       adsurl = {https://ui.adsabs.harvard.edu/abs/2018arXiv180107829W}
}

@article{ATLAS:2023ixc,
    author = "Malara, Andrea",
    collaboration = "ATLAS, CMS",
    title = "{Exploring jets: substructure and flavour tagging in CMS and ATLAS}",
    eprint = "2410.14330",
    archivePrefix = "arXiv",
    primaryClass = "hep-ex",
    reportNumber = "CMS-CR-2024-177",
    doi = "10.22323/1.478.0150",
    journal = "PoS",
    volume = "LHCP2024",
    pages = "150",
    year = "2025"
}

@article{Adams:2015hiv,
    author = "Adams, D. and others",
    title = "{Towards an Understanding of the Correlations in Jet Substructure}",
    eprint = "1504.00679",
    archivePrefix = "arXiv",
    primaryClass = "hep-ph",
    reportNumber = "FERMILAB-PUB-15-670-CMS, SLAC-PUB-16703",
    doi = "10.1140/epjc/s10052-015-3587-2",
    journal = "Eur. Phys. J. C",
    volume = "75",
    number = "9",
    pages = "409",
    year = "2015"
}

@article{Dasgupta:2013ihk,
    author = "Dasgupta, Mrinal and Fregoso, Alessandro and Marzani, Simone and Salam, Gavin P.",
    title = "{Towards an understanding of jet substructure}",
    eprint = "1307.0007",
    archivePrefix = "arXiv",
    primaryClass = "hep-ph",
    reportNumber = "CERN-PH-TH-2013-145, DCPT-13-86, IPPP-13-43, LPN13-036, MAN-HEP-2013-12",
    doi = "10.1007/JHEP09(2013)029",
    journal = "JHEP",
    volume = "09",
    pages = "029",
    year = "2013"
}

@article{CMS:2023lpp,
    author = "Hayrapetyan, Aram and others",
    collaboration = "CMS",
    title = "{Measurement of the primary Lund jet plane density in proton-proton collisions at $ \sqrt{\textrm{s}} $ = 13 TeV}",
    eprint = "2312.16343",
    archivePrefix = "arXiv",
    primaryClass = "hep-ex",
    reportNumber = "CMS-SMP-22-007, CERN-EP-2023-282",
    doi = "10.1007/JHEP05(2024)116",
    journal = "JHEP",
    volume = "05",
    pages = "116",
    year = "2024"
}

@article{CMS:2025eyd,
    author = "Hayrapetyan, Aram and others",
    collaboration = "CMS",
    title = "{A method for correcting the substructure of multiprong jets using the Lund jet plane}",
    eprint = "2507.07775",
    archivePrefix = "arXiv",
    primaryClass = "hep-ex",
    reportNumber = "CMS-JME-23-001, CERN-EP-2025-128",
    doi = "10.1007/JHEP11(2025)038",
    journal = "JHEP",
    volume = "11",
    pages = "038",
    year = "2025"
}

@article{Cohen:2023mya,
    author = "Cohen, Timothy and Roloff, Jennifer and Scherb, Christiane",
    title = "{Dark sector showers in the Lund jet plane}",
    eprint = "2301.07732",
    archivePrefix = "arXiv",
    primaryClass = "hep-ph",
    reportNumber = "CERN-TH-2023-007",
    doi = "10.1103/PhysRevD.108.L031501",
    journal = "Phys. Rev. D",
    volume = "108",
    number = "3",
    pages = "L031501",
    year = "2023"
}

@article{Vent:2025ddm,
    author = "Vent, Sophia and Winterhalder, Ramon and Plehn, Tilman",
    title = "{The Physics Behind ML-based Quark-Gluon Taggers}",
    eprint = "2507.21214",
    archivePrefix = "arXiv",
    primaryClass = "hep-ph",
    reportNumber = "TIF-UNIMI-2025-16",
    doi = "10.21468/SciPostPhys.20.3.084",
    journal = "SciPost Phys.",
    volume = "20",
    pages = "084",
    year = "2026"
}

@article{Konar:2025vts,
    author = "Konar, Partha and Ngairangbam, Vishal S. and Spannowsky, Michael and Srivastava, Deepanshu",
    title = "{Stable and interpretable jet physics with IRC-safe equivariant feature extraction}",
    eprint = "2509.22059",
    archivePrefix = "arXiv",
    primaryClass = "hep-ph",
    reportNumber = "IPPP/25/58",
    doi = "10.1007/JHEP03(2026)219",
    journal = "JHEP",
    volume = "03",
    pages = "219",
    year = "2026"
}

@article{Bierlich:2022pfr,
    author = "Bierlich, Christian and others",
    title = "{A comprehensive guide to the physics and usage of PYTHIA 8.3}",
    eprint = "2203.11601",
    archivePrefix = "arXiv",
    primaryClass = "hep-ph",
    reportNumber = "LU-TP 22-16, MCNET-22-04, FERMILAB-PUB-22-227-SCD",
    doi = "10.21468/SciPostPhysCodeb.8",
    journal = "SciPost Phys. Codeb.",
    volume = "2022",
    pages = "8",
    year = "2022"
}

@article{Bahr:2008pv,
    author = "Bahr, M. and others",
    title = "{Herwig++ Physics and Manual}",
    eprint = "0803.0883",
    archivePrefix = "arXiv",
    primaryClass = "hep-ph",
    reportNumber = "CERN-PH-TH-2008-038, CAVENDISH-HEP-08-03, KA-TP-05-2008, DCPT-08-22, IPPP-08-11, CP3-08-05",
    doi = "10.1140/epjc/s10052-008-0798-9",
    journal = "Eur. Phys. J. C",
    volume = "58",
    pages = "639--707",
    year = "2008"
}

\end{document}